\documentclass[english]{article}
\usepackage{geometry}
\usepackage{footnote}
\usepackage{babel}
\usepackage{float}
\usepackage{rotfloat}
\usepackage{amsthm}
\usepackage{epsfig,color,amsmath,amssymb,euscript}
\usepackage{graphicx}
\usepackage{subfig}
\usepackage{setspace}
\usepackage{url}
\usepackage{hyperref}
\usepackage[sort,comma,numbers, authoryear]{natbib}
\usepackage{authblk}
\usepackage{multirow}
\usepackage{textcomp}
\usepackage{comment}
\usepackage{xcolor}

\makeatletter

\makeatletter

\newtheorem{theorem}{Theorem}

\newtheorem{lemma}{Lemma}

\newtheorem{remark}[theorem]{Remark}

\def\red{\color{red}}

\def\T{{ \mathrm{\scriptscriptstyle T} }}

\newcommand{\norm}[1]{\big\| #1 \big\|}

\def\Q{\mathbf{Q}}
\def\Z{\mathbf{Z}}
\def\R{\mathbf{R}}
\def\P{\mathbf{P}}
\def\E{\mathbf{E}}
\def\1{\mathbf{1}}        
\def\U{\mathbf{U}}

\def\I{\mathbf{I}}
\def\D{\mathbf{D}}
\def\H{\mathbf{H}}
\def\y{\mathbf{y}}
\def\0{\mathbf{0}}
\def\C{\mathbf{C}}
\def\W{\mathbf{W}}
\def\w{\mathbf{w}}
\def\A{\mathbf{A}}

\def\G{\mathbf{G}}
\def\z{\mathbf{z}}
\def\B{\mathbf{B}}
\def\V{\mathbf{V}}
\def\x{\mathbf{x}}
\def\e{\mathbf{e}}
\def\F{\mathbf{F}}
\def\M{\mathbf{M}}

\def\q{\mathbf{q}}

\def\bTheta{\boldsymbol{\Theta}}
\def\mat{\text{mat}}
\def\Amk{\mathbf{A}_{\text{-}k}}
\def\qmk{\mathbf{q}_{\text{-}k}}
\def\cmk{\mathbf{c}_{\text{-}k}}

\def\a{\mathbf{a}}

\def\c{\mathbf{c}}
\def\X{\mathbf{X}}
\def \bbeta{\boldsymbol{\beta}}

\def \S{\mathbf{S}}
\def \I{\mathbf{I}}
\def \A{\mathbf{A}}

\def\bepsilon{\boldsymbol{\epsilon}}
\def\bSigma{\boldsymbol{\Sigma}}

\def\bPsi{\boldsymbol{\Psi}}

\def\bxi{\boldsymbol{\xi}}

\def\vec{\text{vec}}
\def\tr{\text{tr}}

\def\diag{\text{diag}}

\def\var{\text{var}}
\def\cov{\text{cov}}

\def\f{\mathbf{f}}

\def\wt{\widetilde}
\def\wh{\widehat}

\def\cF{\mathcal{F}}

\def\cA{\mathcal{A}}
\def\cX{\mathcal{X}}

\def\cS{\mathcal{S}}

\def\u{\mathbf{u}}
\def\cC{\mathcal{C}}
\def\cE{\mathcal{E}}
\def\Amk{\A_{\text{-}k}}
\def\cN{\mathcal{N}}
\def\gmkhat{\hat{g}_{\text{-}k}}

\def\Umk{\U_{\text{-}k}}
\def\Gmk{\G_{\text{-}k}}
\def\gmkcheck{\check{g}_{\text{-}k}}
\def\qmkcheck{\check{\q}_{\text{-}k}}
\def\Umkpre1{\hat{\U}_{\text{-}k,pre,(1)}}
\def\Umkc1{\U_{\text{-}k,(1)}}
\def\Qm{\mathbf{Qm}}

\def\dmk{d_{\text{-}k}}
\def\rmk{r_{\text{-}k}}

\def\red{\color{red}}

\renewcommand{\hat}{\widehat}
\renewcommand{\tilde}{\widetilde}

\begin{document}
	\setlength{\parindent}{18pt}
	\doublespacing
	\begin{titlepage}
		
		\title{\bf Rank and Factor Loadings Estimation in Time Series Tensor Factor Model by Pre-averaging}
		\author{Weilin Chen\thanks{Weilin Chen is PhD student, Department of Statistics, London School of Economics. Email: w.chen56@lse.ac.uk}}
				\author{Clifford Lam\thanks{Clifford Lam is Professor, Department of Statistics, London School of Economics. Email: C.Lam2@lse.ac.uk}}
		
		\affil{Department of Statistics, London School of Economics and Political Science}
		
		\date{}
		
		\maketitle
		
\begin{abstract}
Tensor time series data appears naturally in a lot of fields, including finance and economics. As a major dimension reduction tool, similar to its factor model counterpart, the idiosyncratic components of a tensor time series factor model can exhibit serial correlations, especially in financial and economic applications. This rules out a lot of state-of-the-art methods that assume white idiosyncratic components, or even independent/Gaussian data.
While the traditional higher order orthogonal iteration (HOOI) is proved to be convergent to a set of factor loading matrices, the closeness of them to the true underlying factor loading matrices are in general not established, or only under some strict circumstances like having i.i.d. Gaussian noises \citep{ZhangXia2018}. Under the presence of serial and cross-correlations in the idiosyncratic components and time series variables with only bounded fourth order moments, we propose a pre-averaging method that accumulates information from tensor fibres for better estimating all the factor loading spaces. The estimated directions corresponding to the strongest factors are then used for projecting the data for a potentially improved re-estimation of the factor loading spaces themselves, with theoretical guarantees and rate of convergence spelt out. We also propose a new rank estimation method which utilizes correlation information from the projected data, in the same spirit as \cite{Fanetal2022} for factor models with independent data. Extensive simulation results reveal competitive performance of our rank and factor loading estimators relative to other state-of-the-art or traditional alternatives. A set of matrix-valued portfolio return data is also analyzed.

\end{abstract}
		
		\bigskip
		\bigskip


\noindent

	\end{titlepage}
	
	\setcounter{page}{2}
	
	\section{Introduction}\label{sec:intro}
	\label{intro}
Thanks to the advancement of the internet and general computing power, the collection and analysis of panel data are made ever easier over the past decade. Toolboxes in high dimensional vector time series analysis play increasingly important roles in extracting useful information from high dimensional time series data. Time series factor modelling is a major dimension reduction tool for such data, allowing insights into the common dynamics of different observed time series. For instance, when considering many macroeconomic time series for forecasting \citep{StockWatson2002}, the estimation and forecasting through the common factors can give more accurate results overall, and allowing for the interpretation of the factors (e.g., potential grouping of macroeconomic time series as factors) at the same time.

To improve the accuracy of forecasting, one can add the time series of macroeconomic indicators from other countries, and stack all observed time series into one high dimensional vector time series. The problem in doing this is that we are now ignoring the natural structure of the data, namely, all macroeconomic time series are now categorized by countries. Moreover, stacking all time series into a long vector can create curse of dimensionality (e.g., when the stacked length is too much larger than the sample size), leading to inaccurate estimation and predictions.

A more natural approach is to consider the country-categorized macroeconomic time series a matrix-valued time series (i.e., an {\em order-2 tensor}), with different countries by row and different macroeconomic time series by columns. \cite{Wangetal2019} describes a factor model for such matrix-valued time series, and provides estimation methods together with theoretical results. Their work is extended to a general order-$K$ tensor $\{\cX_t\}$ in \cite{Chenetal2022}, where
\[\cX_t = \cC_t + \cE_t,\]
with $\cC_t$ the common component and $\cE_t$ the noise tensor, assuming that the elements in each $\cE_t$ are sub-Gaussian, with each $\cE_t$ independent of each other. Based on the above, \cite{Hanetal2020} analyzes iterative projection procedures iTOPUP and iTIPUP for estimation, while \cite{Hanetal2022} proposes core rank (or multilinear tensor rank) estimators of $\cC_t$ based on information criterion and eigen-ratio criterion that are intertwined with iTIPUP and iTOPUP. Core rank is similar to the number of factors, and will be explained in Section \ref{sec:pre-averaging} (see Section \ref{sec:BasicTensorManipulations} as well).

In other recent developments, \cite{ZhangXia2018} proposes a similar model for an order-3 tensor, with the tensor noise elements being i.i.d. normal having a common variance, and develops minimax theoretical guarantees for their estimators. With the same tensor noise assumption, \cite{Yokotaetal2017} proposes a core rank estimator for $\cC_t$ for a general order-$K$ tensor $\cX_t$ based on a BIC-like criterion, while \cite{Liuetal2022} proposes a tensor SVD method for estimation under a CP decomposition of $\cC_t$. \cite{Chenetal2020} proposes a semiparametric model with $\cC_t$ taking covariates under the assumption of i.i.d. sub-Gaussian elements in $\cE_t$, which are themselves independent of each other.

All the tensor factor modelling works mentioned above assumed at least independent noise tensor series $\{\cE_t\}$ with sub-Gaussian elements. The i.i.d. assumption for the elements in $\cE_t$ in many of them is also considered a standard assumption for statistical analysis. However, if we have applications in economics and finance for instance, it is very easy that (weak) serial correlations exist in $\{\cE_t\}$, representing any serial correlations in $\cX_t$ not captured by the common components $\cC_t$ (some time series in $\cX_t$ have ``unique'' company or macroeconomic characteristics, for example).
The {\em Approximate factor model} of \cite{BaiNg2002} allows for such weak serial correlations (as well as weak cross-correlations) in the idiosyncratic noise series $\{\cE_t\}$. When $\cE_t$ has a higher order tensor structure, allowing for weak-serial and cross-correlations becomes even more essential as there could be even more potentially intricate serial and cross-correlations in $\{\cE_t\}$.

In this paper, we make three important contributions to the literature of tensor time series factor modelling. The first one is to allow for (weak) serial correlations and cross-correlations in the noise tensor series $\{\cE_t\}$, allowing for elements in both $\cC_t$ and $\cE_t$ to be driven by different general linear processes, and without the need to restrict the variables in $\cE_t$ to be sub-Gaussian. Our methods utilize covariance information, which are more natural to apply to financial return data for example as opposed to methods that utilize only autocovariance information (see \cite{Wangetal2019} or \cite{Chenetal2022} for example). Due to market efficiency, population autocovariances of the data can be close to zero and methods that only utilize autocovariance information can have low signal-to-noise ratio.

Secondly, a spectrum of different factor strengths are allowed in our settings, which is a generalization to \cite{Lametal2011} when static vector time series factor model is concerned. While \cite{Hanetal2020} has two parameters $\delta_0$ and $\delta_1$ controlling the factor strengths, they are less easily interpretable compared to our $\alpha_{k,j}$, $j\in[r_k]$, $k\in[K]$, which has the $j$th diagonal entry of $\A_k^\T\A_k \asymp d_k^{\alpha_{k,j}}$
(see Assumption (L1) as well). Hence if the $j$th column of $\A_k$ is dense (a pervasive factor), then $\alpha_{k,j} = 1$. If there are only finitely many non-zeros in the  $j$th column of $\A_k$, then it is a very weak factor, and $\alpha_{k,j} = 0$. \cite{Freyaldenhoven2022} allows for these weaker factors in its vector time series factor model, and called them ``local factors''. With relaxed assumptions for wider applications, and allowing for a spectrum of factors with different strengths, our final contribution is to provide a ``pre-averaging'' and iterative projection estimators for our model, with theoretical analysis provided and rate of convergence spelt out. To complete the paper, we also provide estimators of the core tensor rank through correlation analysis, which is inspired by \cite{Fanetal2022}, but we provide a bootstrap method for tuning parameter selection as well.

The rest of the paper is organized as follows. Section \ref{sec:basics} reviews some basic notations we use throughout the paper. Section \ref{sec:pre-averaging} presents the idea of pre-averaging, together with important assumptions on our model. Discussions and theory on choosing the ``best'' samples for aggregating results are presented, together with rate of convergence for our pre-averaging estimator for the strongest factors spelt out. Section \ref{sec:re-estimation} utilizes the pre-averaging estimator as the ideal initial estimator for re-estimating the projection direction by iterations, and presents the key theoretical results on the iterative projection estimators. Section \ref{sec:rank_estimation} presents theoretical justifications for using correlation analysis in finding the rank of the core tensor, and provides a fibre bootstrapping technique in determining the tuning parameter of the procedure. Section \ref{sec:simulation} presents our simulation studies on a number of different settings and compare to other benchmarks or state-of-the-art estimators. A set of matrix-valued portfolio return data is also analyzed in this section. Section \ref{sec:BasicTensorManipulations} is a review of basic tensor manipulations for unfamiliar readers.
All proofs are relegated to Section \ref{sec:Proofs}.

\section{Notations}\label{sec:basics}
\setcounter{equation}{0}

In this paper, we use $a \asymp b$ to denote $a = O(b)$ and $b = O(a)$ (also $a\asymp_P b$ for $a=O_P(b)$ and $b = O_P(a)$), while $a\succeq b$ is equivalent to $b = O(a)$, and $a \succ b$ is equivalent to $b = o(a)$. We also use $\norm{\cdot}$ to denote the $L_2$ norm (of a vector or a matrix), and $\norm{\cdot}_F$ to denote the Frobenius norm, while $\norm{\cdot}_{\max}$ represents the maximum element (of a vector or a matrix). We also use $\norm{\A}_{\infty} = \max_{i}\sum_{j}|a_{ij}|$ and $\norm{\A}_{1} = \max_{j}\sum_{i}|a_{ij}|$ to denote the $L_{\infty}$ and $L_1$ norm of a matrix $\A$ respectively. The notation $\vec(\cdot)$ represents the vectorization of a matrix, stacking columns of the matrix from left to right. We also use $\1_m$ to represent a vector of ones with length $m$, whereas $\I_m$ is the identity matrix with size $m$. The notation $\diag(\A)$ of a square matrix $\A$ is the diagonal matrix with only the diagonal elements of $\A$ remain, and everything else set to 0. This notation is also used to represent a block diagonal matrix. For instance, $\diag(\A_1,\ldots,\A_n)$ is the block diagonal matrix with diagonal block matrices $\A_1,\ldots,\A_n$. For a positive integer $m$, we define
$[m] := \{1,\ldots,m\}$.

Please refer to the Appendix in Section \ref{sec:BasicTensorManipulations} for some basic tensor manipulations related to this paper.

\section{Estimation of Strongest Factors by Pre-averaging}\label{sec:pre-averaging}
\setcounter{equation}{0}
The tensor factor model for each $\cX_t \in \mathbb{R}^{d_1\times\cdots\times d_K}$, $t\in[T]$, is
\begin{equation}\label{eqn:tensorfactormodel}
  \cX_t = \mu + \cC_t + \cE_t = \mu + \cF_t\times_1\A_1\times_2\cdots\times_K\A_K + \cE_t,
\end{equation}
where $\cF_t\in\mathbb{R}^{r_1\times\cdots\times r_K}$ is the core tensor, and $\A_k \in \mathbb{R}^{d_k\times r_k}$, $k\in[K]$, is called a mode-$k$ factor loading matrix. The product $\times_k$ is the tensor $k$-mode product.
We develop a summing method to estimate the factor loading space of each $\A_k$. The basic idea is, in estimating the mode-$k$ factor loading matrix $\A_k$, we can write the tensor factor model (\ref{eqn:tensorfactormodel}) as
\begin{equation*}
\cX_t = \mu+ \cF_{t,\text{-}k}  \times_k \A_k + \cE_t,
\end{equation*}
where $\cF_{t,\text{-}k} = \cF_t\times_1\A_1\times_2\cdots \times_{k-1}\A_{k-1} \times_{k+1}\A_{k+1} \times_{k+2} \cdots \times_K\A_K$, and we have
\begin{equation}\label{eqn:mode-k-flattening}
    \mat_k(\cX_t) = \mat_k(\mu) + \A_k\mat_k(\cF_{t,\text{-}k}) + \mat_k(\cE_t),
\end{equation}
where $\mat_k(\cdot)$ is the mode-$k$ unfolding matrix of a tensor.
Let $\x_{t,\text{-}k,i}, \; i\in[d_{\text{-}k}]$ be the $i$-th mode-$k$ fiber of $\cX_t$. In other words, $\x_{t,\text{-}k,i}$ is the $i$-th column vector of $\mat_k(\cX_t)$. Then we have,
\begin{equation}\label{eqn:vectordecomposition}
    \x_{t,\text{-}k,i} = \mu_{\text{-}k,i} +  \A_k \f_{t,\text{-}k,i} + \e_{t,\text{-}k,i}, \;\;\; i\in[d_{\text{-}k}],
\end{equation}
where the $\f_{t,\text{-}k,i}$'s, $\e_{t,\text{-}k,i}$'s and $\mu_{\text{-}k,i}$'s are the mode-$k$ fibers of $\cF_{t,\text{-}k}$, $\cE_t$ and $\mu$ respectively. The decomposition (\ref{eqn:vectordecomposition}) can be seen as a factor model for vector time series, as \cite{BaiNg2002} describes. For tensor time series $\cX_t$, such decomposition implies there are totally $d_{\text{-}k}$ vector time series in the form of (\ref{eqn:vectordecomposition}), which all share the same factor loading matrix $\A_k$. Our purpose is to combine these $d_{\text{-}k}$ models in a certain way to obtain one single estimator of $\A_k$. An intuitive idea is to sum over some or all of the $d_{\text{-}k}$ vector time series. Note that for any subset $S\subseteq [d_{\text{-}k}]$,
\begin{equation}\label{eqn:sumvectordecomposition}
    \sum_{i \in S} \x_{t,\text{-}k,i} = \sum_{i \in S} \mu_{\text{-}k,i} +  \A_k \sum_{i \in S} \f_{t,\text{-}k,i} + \sum_{i \in S} \e_{t,\text{-}k,i}.
\end{equation}
Hence, we can carry out usual estimations of vector factor models, for instance, using PCA on sample covariance matrix based on (\ref{eqn:sumvectordecomposition})  \citep{BaiNg2002}, to obtain an estimator of $\A_k$. For the choice of $S$, one simple option is to use $S = [d_{\text{-}k}]$, i.e., summing over all $d_{\text{-}k}$ mode-$k$ fibers of $\cX_t$. Using $S = [\dmk]$ first, we present the corresponding model assumptions and theoretical properties in the subsequent sections. We discuss more choices of $S$ to improve estimation performance and the theoretical properties of the corresponding estimators in Section \ref{subsec:Partialsumfibres}.

\subsection{Assumptions}

We present assumptions for the tensor factor model (\ref{eqn:tensorfactormodel}).

\subsubsection{Assumptions on the errors}\label{subsubsec:errorassumptions}
We present assumptions (E1) - (E2) below with explanations.
\begin{itemize}
\item[(E1)] (Decomposition of error) {\em Assume that
\begin{align}
  \text{\em mat}_k(\cE_t) &= (\bxi_{t,1}^{(k)},\ldots, \bxi_{t,\dmk}^{(k)}), \;\;\text{ where}\notag\\
  \bxi_{t,\ell}^{(k)} &:=  \bPsi_{\ell}^{(k)}\e_t^{(k)} + (\bSigma_{\epsilon,\ell}^{(k)})^{1/2}\bepsilon_{t,\ell}^{(k)}, \label{eqn:unfoldeddecomposition}
\end{align}
with $E(\e_t^{(k)})=\0$, $E(\bxi_{t,\ell}^{(k)})=\0$, $\e_t^{(k)} \in \mathbb{R}^{r_e}$ independent of $\bepsilon_{s,\ell}^{(k)}$, $\bepsilon_{t,\ell}^{(k)}$ independent of $\bepsilon_{t,m}^{(k)}$ for $\ell\neq m$, $\text{\em var}(\e_t^{(k)}) = \I_{r_e}$ and $\text{\em var}(\bepsilon_{t,\ell}^{(k)}) = \I_{d_k}$ for each $s,t \in [T]$, $\ell,m\in[\dmk]$, $k\in[K]$. Also, each $\bSigma_{\epsilon,\ell}^{(k)}$ has non-vanishing diagonals with $\norm{\bSigma_{\epsilon,\ell}^{(k)}}_{\max} = O(1)$ and $\text{\em tr}(\bSigma_{\epsilon,\ell}^{(k)}) = O(d_k)$, where $\text{\em tr}(\cdot)$ is the trace of a square matrix. Moreover,
define $\bPsi^{(k)} := \sum_{\ell=1}^{d_{\text{-}k}}\bPsi_l^{(k)}$ and  $\bSigma_{\epsilon}^{(k)} := \sum_{\ell=1}^{\dmk}\bSigma_{\epsilon,\ell}^{(k)}$. Then we assume
$\norm{\bPsi^{(k)} \bPsi^{(k)\T}} = O(\dmk)$ and $\norm{\bSigma_{\epsilon}^{(k)}} = O(\dmk)$.}
\end{itemize}
Hence with (E1), each mode-$k$ fibre of the error tensor $\cE_t$ is a sum of two independent parts. The first part is $\bPsi_\ell^{(k)}\e_t^{(k)}$, which is similar to a common component in a factor model because $\e_t^{(k)}$ is independent of the index $\ell$, so that it appears in every mode-$k$ fibre at time $t$. However, we assume $\bPsi_\ell^{(k)}$ is sparse or approximately sparse (such that $\norm{\bPsi^{(k)} \bPsi^{(k)\T} } = O(\dmk)$), meaning that this common component is too weak to be detected. We allow the existence of this part in our model because this allows (weak) cross-correlations among the fibres. For $\ell,m\in [\dmk]$,
\begin{align*}
  \cov(\bxi_{t,\ell}^{(k)}, \bxi_{t,m}^{(k)}) = \bPsi_{\ell}^{(k)}\bPsi_{m}^{(k)\T}.
\end{align*}
For instance, if $\cX_t \in \mathbb{R}^{10\times 10}$ is a matrix-valued time series ($K=2$) of 100 asset returns, with rows denoting 10 different size levels and columns denoting 10 different book-to-market levels, then the error tensor series $\cE_t$ can exhibit weak cross-correlations, especially if the assets are all within the same market and some are closely related (two assets with the same parent company, for instance). The inclusion of the term $\bPsi_\ell^{(k)}\e_t^{(k)}$ for each mode-$k$ fibre facilitates such weak cross-correlations.
For the last part $\norm{\bSigma_{\epsilon}^{(k)} } = O(\dmk)$, as long as $\bSigma_{\epsilon}^{(k)}$ is sparse enough with the majority of $\bSigma_{\epsilon,\ell}^{(k)}$'s well-conditioned, the condition will be satisfied. In fact, this includes the scenario when each $\bSigma_{\epsilon,\ell}^{(k)}$ is dense, but the sum $\bSigma_{\epsilon}^{(k)}$ is approximately sparse (relative to the diagonal entries now with order $\dmk$). When  the off-diagonals are generated randomly with mean 0 and finite variances, we can easily see that if $d_k = O(\dmk^{1/2})$, then the $L_1$ norm, and thus the $L_2$ norm of $\bSigma_{\epsilon}^{(k)}$ will be $O_P(\dmk)$. This is true when $K\geq 3$ and $d_1\asymp\cdots\asymp d_K$.

\begin{itemize}
\item[(E2)] (Time series){\em The elements in $\e_t^{(k)} = (e_{t,j}^{(k)})$ and $\bepsilon_{t,\ell}^{(k)} = (\epsilon_{t,\ell,j}^{(k)})$ are following weakly stationary general linear processes, such that with $\ell\in[\dmk]$, $t\in[T]$ and $k\in[K]$,
    \begin{equation*}
    \begin{split}\label{eqn:timeseriesE2}
      e_{t,j}^{(k)} &= \sum_{q\geq 0} a_{e,q}z_{e,t-q,j}^{(k)}, \;\;\; j\in[r_e],\\
      \epsilon_{t,\ell,j}^{(k)} &= \sum_{q\geq 0}a_{\epsilon,q}z_{\epsilon,t-q,\ell,j}^{(k)}, \;\;\; j\in[d_k],
    \end{split}
    \end{equation*}
    where the coefficients $a_{e,q}$ and $a_{\epsilon,q}$ are such that $\sum_{q\geq 0}a_{e,q}^2 = \sum_{q\geq 0}a_{\epsilon,q}^2=1$ and $\sum_{q\geq 0}|a_{e,q}| \leq C$, $\sum_{q\geq 0}|a_{\epsilon,q}| \leq C$ for some constant C. For each $k\in[K]$, the series of random variables $\{z_{e,t,j}^{(k)}\}$ and $\{z_{\epsilon,t,\ell,j}^{(k)}\}$ are independent of each other, with i.i.d. elements having mean 0 and variance 1.
    }
\end{itemize}
With this assumption, the error variables are serially correlated in general. Together with (E1), (weak) serial and cross-sectional dependence within and among fibres are allowed for the errors.

\subsubsection{Assumptions on the factors}\label{subsubsec:factorassumptions}
Similar to (E2), the factors in $\cF_t$ are assumed to follow general linear processes.
\begin{itemize}
  \item[(F1)] {\em Let $\f_{t,\ell}^{(k)} = (f_{t,\ell,j}^{(k)})$ be the $\ell$-th column vector in $\text{\em mat}_k(\cF_t)$, $\ell\in[r_{\text{-}k}]$, where $r_{\text{-}k} := \prod_{\ell\neq k}r_\ell$. We assume that $\text{\em var}(\f_{t,\ell}^{(k)}) = \I_{r_k}$ (the identity matrix with size $r_k$), and $\text{\em cov}(\f_{t,\ell_1}^{(k)}, \f_{t,\ell_2}^{(k)})=\0$ for $\ell_1\neq \ell_2$.

      Then we can write
      \begin{equation*}\label{eqn:timeseriesF1}
         f_{t,\ell,j}^{(k)} = \sum_{q\geq 0}a_{f,q}z_{f,t-q,\ell,j}^{(k)}, \;\;\; j\in[r_k],
      \end{equation*}
      where we have $\sum_{q\geq 0}a_{f,q}^2 = 1$ and $\sum_{q\geq 0}|a_{f,q}| \leq C$ for some constant C. For each $k\in[K]$, the series of random variables $\{z_{f,t,\ell,j}^{(k)}\}$ has i.i.d. elements having zero mean and variance 1.
  }
\end{itemize}
Note the series of coefficients $\{a_{e,q}\}$, $\{a_{\epsilon,q}\}$ and $\{a_{f,q}\}$ are not necessarily equal.

\subsubsection{Assumptions on the model parameters}\label{subsubsec:Assumption_modelparameters}
We present the assumptions needed for the factor loading matrices $\A_k$, $k\in[K]$, and other model parameters.
\begin{itemize}
  \item[(L1)] (Factor Strength) {\em We assume that, for $k\in[K]$, $\A_k$ is of full rank, $r_k = o(T^{1/3})$, and as $d_k \rightarrow \infty$,
  \begin{align}\label{eqn:factorstrengthassumption}
  \D_k^{-1/2} \A_k^\T \A_k \D_k^{-1/2} \rightarrow \Sigma_{\A,k},
  \end{align}
  where $\D_k = \diag{(\A_k^\T \A_k)}$ is a diagonal matrix consisting of the diagonal elements of $\A_k^\T \A_k$, and $\Sigma_{\A,k}$ is positive definite with all eigenvalues bounded away from 0 and infinity. Let $(\D_k)_j$ be the $j$-th diagonal element of $\D_k$, then we assume $(\D_k)_j \asymp d_k^{\alpha_{k,j}}$ for $j \in [r_k]$, and $0 < \alpha_{k,r_k} \leq \dots \leq \alpha_{k,2} \leq \alpha_{k,1} \leq 1$.

  }


  \item[(L2)] (Signal Cancellation) {\em For $k\in[K]$ and $j \in [r_k]$, define the growth rate $\kappa_{k,j} \in [0,1]$ of column sum to be such that $\sum_{i=1}^{d_k}(\A_k)_{ij} \asymp d_k^{\kappa_{k,j}}$ (or the smallest $\kappa_{k,j}$ such that $\mathbb{P}(c_1 d_k^{\kappa_{k,j}} \leq \sum_{i=1}^{d_k}(\A_k)_{ij} \leq c_2 d_k^{\kappa_{k,j}}) = 1$ for constants $c_1, c_2$ when $\A_k$ has random entries). It measures how close the column sum of $\A_k$ is to 0. For each $k\in [K]$, let $s_k = \sum_{j=1}^{r_k} \left( \sum_{i=1}^{d_k}(\A_k)_{ij} \right)^2 \asymp \sum_{j=1}^{r_k}d_k^{2\kappa_{k,j}}$, and $s_{\text{-}k} := \prod_{l=1;  l \neq k}^{K} s_l$. We assume for $k \in [K]$ and $z_k\leq r_k$ (see Theorem \ref{thm:sumestimatorrate_newmethod} and the explanations thereafter as well),
  \begin{align}\label{scalecondition}
      \frac{d_{\text{-}k}}{s_{\text{-}k}}\left(1+\frac{d_k}{T}\right) = o\left(d_k^{\alpha_{k,z_k}}\right).
  \end{align}
  }

  \item[(R1)] {\em The elements in $\{z_{f,t,\ell,j}^{(k)}\}$ from Assumption (F1), and those in  $\{z_{e,t,j}^{(k)}\}$ and $\{z_{\epsilon,t,\ell,j}^{(k)}\}$ from Assumption (E2)
       have uniformly bounded fourth moments. }

\end{itemize}
 Assumption (L1) states that the factors can have different strengths. When $K = 1$ and $\alpha_{1,j} = \alpha$ for $j \in [r_1]$, (\ref{eqn:factorstrengthassumption}) reduces to the assumption of (approximate) vector factor model with the same strengths, which is discussed in \cite{BaiNg2021}. Hence, our assumption is a generalization of \cite{BaiNg2021} to a tensor time series factor model setting with different factor strengths. In addition, we do not assume the orthogonality of $\A_k$ as \cite{Freyaldenhoven2022} did, since this would be incompatible with the expression of factor strength and signal cancellation in terms of the norm and column sum of $\A_k$. The concept of a pervasive factor, for instance, depends on a column of $\A_k$ being dense. However, such an interpretation can be lost completely under the assumption of orthogonal columns in $\A_k$.

 Assumption (L2) indicates that the column sum of each $\A_k$ is not too close to 0, so that signal cancellation due to sum of fibers is not likely to occur. A sufficient condition for (\ref{scalecondition}) is $d_k = O(T)$ and $\kappa_{k,j} \geq 0.5$ for $k \in [K]$ and $j \in [r_k]$, which implies $d_k / s_k = O(1)$, but this is not a necessary condition.

Assumption (R1) relaxes the need for Gaussian or sub-Gaussian random variables (see \cite{ZhangXia2018} and \cite{Chenetal2022} for example), with only bounded fourth order moments required. This allows for substantially more types of data to be analyzed. For instance, financial returns data over more volatile periods where we do not usually want to assume moments beyond order four exist.

For convenience of further theoretical analysis, we define $\Q_k = \A_k \D_k^{-1/2}$. Then the subspace spanned by the columns of $\Q_k$ and $\A_k$ are the same. Since  $\Q_k^\T \Q_k \rightarrow \Sigma_{A,k}$, we can regard $\Q_k$ as a re-normalized version of $\A_k$ by normalising its factor strength. In addition, we apply the singular value decomposition of $\A_k$ as
\begin{align}
    \A_k = \U_k \G_k^{1/2} \V_k^\T, \label{eqn:SVD_Ak}
\end{align}
where $\U_k \in \mathbb{R}^{d_k \times r_k}$ has orthogonal columns such that $\U_k^\T \U_k = \I_{r_k}$, $\G_k \in \mathbb{R}^{r_k \times r_k}$ is diagonal and consists of the eigenvalues of $\A_k^\T \A_k$ in decreasing order, and $\V_k\in \mathbb{R}^{r_k \times r_k}$ is an orthogonal matrix. The subspaces spanned by the columns of $\U_k$, $\Q_k$ and $\A_k$ are the same, and hence it is equivalent to estimate $\U_k$ (or $\Q_k$) and $\A_k$, and the columns of $\U_k$ form an orthonormal basis for the column space spanned by $\Q_k$ (or $\A_k$). We will estimate $\U_k$ (or $\Q_k$) instead of $\A_k$ in the sections that follow. We need another regularity condition on the singular values on $\G_k$. This can be relaxed at the expense of lengthier explanations involving factor loading spaces in all subsequent theorems.
\begin{itemize}
  \item[(L1')] The singular values on $\G_k$ are distinct.
\end{itemize}

 We use the following notations for the remainder of the paper. Let $\lambda_j(B)$ to be the $j$-th largest eigenvalue of matrix $B$. Denote $\tilde{\x}_{t,k} = \sum_{i \in [d_{\text{-}k}]} \x_{t,\text{-}k,i}$, $\; \tilde{\f}_{t,k} = \sum_{i \in [d_{\text{-}k}]} \f_{t,\text{-}k,i}$, $\tilde{\e}_{t,k} = \sum_{i \in [d_{\text{-}k}]} \e_{t,\text{-}k,i}$ and $\tilde{\mu}_{k} = \sum_{i \in [d_{\text{-}k}]} \mu_{\text{-}k,i}$. Define $\ddot{\f}_{t,k} = \tilde{\f}_{t,k} / s_{\text{-}k}^{\frac{1}{2}}$, write $\ddot{\F}_k = \left[\ddot{\f}_{1,k}, \dots, \ddot{\f}_{T,k} \right] \in \mathbb{R}^{r_k \times T}$ and $\tilde{\F}_k = \left[\tilde{\f}_{1,k}, \dots, \tilde{\f}_{T,k} \right]$. Similarly, denote $\ddot{\x}_{t,k} = \tilde{\x}_{t,k} / s_{\text{-}k}^{\frac{1}{2}}$ and $\ddot{\X}_k = (\ddot{\x}_{1,k},...,\ddot{\x}_{T,k})^\T \in \mathbb{R}^{T \times d_k}$, $\tilde{\X}_k = (\tilde{\x}_{1,k},...,\tilde{\x}_{T,k})^\T$. Also let $\ddot{\e}_{t,k} = \tilde{\e}_{t,k} / s_{\text{-}k}^{\frac{1}{2}}$, $\ddot{\E}_k = \left( \ddot{\e}_{1,k}, ..., \ddot{\e}_{T,k} \right)^\T$ and $\tilde{\E}_k = \left( \tilde{\e}_{1,k}, ..., \tilde{\e}_{T,k} \right)^\T$. Finally, let $\ddot{\mu}_{k} = \tilde{\mu}_{k} /  s_{\text{-}k}^{\frac{1}{2}}$. Then,
\begin{align}
    \ddot{\X}_k =  \1_T \ddot{\mu}_{k}^\T +  \ddot{\F}_k^\T \A_k^\T + \ddot{\E}_k, \label{eqn:X_k_double_dot}
\end{align}
where $\1_T$ is a column vector of $T$ ones.

\subsection{Estimator based on summing all fibers}

Under Assumptions (E1), (E2), (F1), (L1), (L2), (R1), we can obtain estimators of $\Q_k$ or $\U_k$, $k \in [K]$ by performing PCA on the data covariance matrix of the sum of all mode-$k$ fibers of $\cX_t$. In other words, consider the vector time series  $\tilde{\x}_{t,k} = \sum_{i \in [d_{\text{-}k}]} \x_{t,\text{-}k,i}$. We perform PCA on the ``pseudo data covariance matrix'' of $\tilde{\x}_{t,k}$, which is
\begin{align*}
    \hat{\Sigma}_{\tilde{\x}_{k}} = \frac{\tilde{\X}_{k}^\T \left( \I_T - \frac{1}{T} \1_T \1_T^\T \right)\tilde{\X}_{k}}{T},
\end{align*}
where $\I_{T}$ is a $T \times T$ identity matrix.
Suppose we do not know the true number of factor $r_k$, and start with an arbitrary $z_k$ with $z_k \leq r_k$, then we can obtain $\hat{\Q}_{k,(z_k)}$ as the $z_k$ largest eigenvectors of $\hat{\Sigma}_{\tilde{\x}_{k}}$, with normalization $\hat{\Q}_{k,(z_k)}^\T \hat{\Q}_{k,(z_k)} = \I_{z_k}$. Let $\tilde{\V}_k$ be the $z_k \times z_k$ diagonal matrix of the first $z_k$ largest eigenvalues of $\hat{\Sigma}_{\tilde{\x}_{k}}$ in decreasing order, and $\ddot{\V}_k := \tilde{\V}_k / s_{\text{-}k}$. We present the theoretical properties of $\hat{\Q}_{k,(z_k)}$ in the following theorem.

\begin{theorem}\label{thm:sumestimatorrate_newmethod}
Under Assumption (E1), (E2), (F1), (L1), (L2), (R1), for $k \in [K]$, let $c_k := \frac{d_{\text{-}k}^2}{s_{\text{-}k}^2}\left(1+\frac{d_k^2}{T^2}\right) + \frac{d_{\text{-}k}}{s_{\text{-}k}}  d_k^{\alpha_{k,1}} \min{\left\{1+\frac{d_k}{T},\frac{r_kd_k}{T}\right\}}$, then
\begin{align}\label{eqn:sumestimatorrate_newmethod_newrate}
     \norm{ \hat{\Q}_{k,(z_k)} - \Q_k\ddot{\H}_k }^2 &= O_p\left( d_k^{-2\alpha_{k,z_k}} c_k\right),
\end{align}
where $\ddot{\H}_k = \frac{\D_k^{\frac{1}{2}} \ddot{\F}_k \left( \I_T - \frac{1}{T} \1_T \1_T^\T \right) \ddot{\F}_k^\T \A_k^\T\hat{\Q}_{k,(z_k)}\ddot{\V}_k^{-1}}{T}$ has $rank(\ddot{\H}_k) = z_k$. Moreover, further assuming (L1'), there exists $\hat\U_{k,(z_k)}$ with $\hat\U_{k,(z_k)}^\T \hat\U_{k,(z_k)} = \I_{z_k}$ such that $\hat\Q_{k,(z_k)} = \hat\U_{k,(z_k)} \P_{k,(z_k)}$ with $\P_{k,(z_k)}$ an orthogonal matrix, and
\begin{align}\label{eqn:sumestimatorrate_lemma3_Lam}
    \norm{\hat\U_{k,(z_k)} - \U_{k,(z_k)} }^2 = O_p \left(d_k^{-2\alpha_{k,z_k}} \left[d_k^{2\alpha_{k,1}}\frac{r_k}{T} + c_k \right]\right),
\end{align}
where $\U_{k,(z_k)}$ is the matrix consisting of the first $z_k$ columns of $\U_k$.
\end{theorem}
Theorem \ref{thm:sumestimatorrate_newmethod} holds for any $z_k \leq r_k$. When $z_k = r_k$, it suggests that the convergence rate of the estimator of the factor loading space corresponding to $\A_k$ depends on the strengths of the strongest and weakest factors in $\A_k$.

Note that the meanings for (\ref{eqn:sumestimatorrate_newmethod_newrate}) and (\ref{eqn:sumestimatorrate_lemma3_Lam}) are different. When $z_k < r_k$, (\ref{eqn:sumestimatorrate_newmethod_newrate}) suggests that the estimated directions $\hat\Q_{k,(z_k)}$ will lie in the subspace spanned by columns of $\Q_k$ (or $\U_k$), but it may not be ``close'' the directions correspond to the strongest $z_k$ factors. However, with (\ref{eqn:sumestimatorrate_lemma3_Lam}), we can conclude that $\hat\U_{k,(z_k)}$ (same column space of $\hat\Q_{k,(z_k)}$) will be ``close'' the directions which correspond to the strongest $z_k$ factors. As a compromise, (\ref{eqn:sumestimatorrate_lemma3_Lam}) involves an extra rate  $d_k^{2(\alpha_{k,1}-\alpha_{k,z_k})}\frac{r_k}{T}$ as compared to (\ref{eqn:sumestimatorrate_newmethod_newrate}). Such a difference is especially notable when we set $z_k = 1$ and perform the iterative projection in Section 4. Also, note that when $z_k = 1$, both (\ref{eqn:sumestimatorrate_newmethod_newrate}) and (\ref{eqn:sumestimatorrate_lemma3_Lam}) achieve the best rate of convergence among all $z_k$, with
\begin{align*}
    \norm{ \hat{\Q}_{k,(1)} - \Q_k\ddot{\H}_k }^2 = O_p\left(d_k^{-2\alpha_{k,1}} c_k \right), \\
    \norm{\hat\U_{k,(1)} - \U_{k,(1)} }^2 = O_p \left(d_k^{-2\alpha_{k,1}}c_k +  \frac{r_k}{T}\right).
\end{align*}
In this case, $\wh\Q_{k,(1)}$ (and hence $\wh\U_{k,(1)}$) is a vector that serves as an initial estimator for a factor loading direction corresponding to the strongest factor, which can be utilized in re-estimating the factor loading matrices through projection in Section \ref{sec:re-estimation}, and core rank estimation
in Section \ref{sec:rank_estimation}.

\subsection{Estimator based on partial sum of fibers}\label{subsec:Partialsumfibres}

In the previous section, we sum over all $d_{\text{-}k}$ mode-$k$ fibers of $\cX_t$ to obtain an estimator of $\hat{\Q}_{k,(z_k)}$. From Theorem \ref{thm:sumestimatorrate_newmethod}, the convergence rate of the factor loading space depend on the term $\frac{d_{\text{-}k}}{s_{\text{-}k}}$, where $s_{\text{-}k}$ defined in Assumption (L2) is a measure of the scale of the product of column sums of $\A_l$ for $l \neq k$. However, if some or all of the column sums of $\A_l$ are close to 0, $s_{\text{-}k}$ may then be too small, potentially violating Assumption (L2) and affecting the convergence rate of the factor loading space. Therefore, instead of summing all mode-$k$ fibers, we can consider only summing part of them for enhancing performances. In this section, we discuss two possible estimators of $\Q_k$ (or $\U_k$) based on partial sum of fibers.

\subsubsection{Maximum eigenvalue ratio estimator}\label{subsubsec:MaximumEigenvaluesRatioEstimators}

Since we want to sum a part of the mode-$k$ fibers, a natural way is to do random sampling. Suppose we perform $M_0$ random sampling of the $\dmk$ mode-$k$ fibres, corresponding to randomly choosing an index set $\cS_m \subseteq [d_{\text{-}k}]$ for the $m$-th sample. {(In practice, we can sample $\cS_m \subseteq [d_{\text{-}k}]$ directly. However, for theoretical analysis and practical performance control, we may need to first sample $\cS_{l,m} \subseteq [d_{l}] $ for $l \neq k$, and then let $\cS_m := \prod_{l \in [K] \setminus \{ k\}}  \cS_{l, m}$ to be the Cartesian product, to facilitate the definition of $s_{\text{-}k,m}$. See (\ref{eqn:MER_estimator_convergencerate}) and the explanations thereafter.)} Then, for each random sample, we can obtain an estimator $\hat{\Q}_{k,m,(z_k)}$ based on summing the $\cS_m$ mode-$k$ fibers of $\cX_t$. From Theorem \ref{thm:sumestimatorrate_newmethod}, if Assumptions (E1) -- (R1) are satisfied for the corresponding sample, then
\begin{align}
    \norm{ \hat{\Q}_{k,m,(z_k)} - \Q_k\ddot{\H}_{k,m} }^2 &= O_p\left( d_k^{-2\alpha_{k,z_k}} \left[\frac{d_{\text{-}k,m}^2}{s_{\text{-}k,m}^2} \left( 1+ \frac{d_k^2}{T^2} \right) +  d_k^{\alpha_{k,1}}\frac{d_{\text{-}k,m}}{s_{\text{-}k,m}} \min{\left\{ 1 + \frac{d_k}{T},\frac{r_kd_k}{T}\right\}} \right] \right), \label{eqn:MER_estimator_convergencerate}\\
     \norm{\hat\U_{k,m,(z_k)} - \U_{k,(z_k)} }^2 &=  O_p\left( d_k^{-2\alpha_{k,z_k}} \left[\frac{d_{\text{-}k,m}^2}{s_{\text{-}k,m}^2} \left( 1+ \frac{d_k^2}{T^2} \right) +  d_k^{\alpha_{k,1}}\frac{d_{\text{-}k,m}}{s_{\text{-}k,m}} \min{\left\{ 1 + \frac{d_k}{T},\frac{r_kd_k}{T}\right\}} \right] \right) \notag \\
    &\;\;\; + O_p \left( d_k^{2(\alpha_{k,1}-\alpha_{k,z_k})}\frac{r_k}{T}\right), \label{eqn:MER_rate_lemma3_Lam}
\end{align}
where $d_{\text{-}k,m}$, $s_{\text{-}k,m}$ and $\hat\U_{k,m,(z_k)}$ are defined in parallel to $d_{\text{-}k}$, $s_{\text{-}k}$ and $\hat\U_{k,(z_k)}$ respectively, but for the $m$-th sample. Clearly, a smaller $\frac{d_{\text{-}k,m}}{s_{\text{-}k,m}}$ leads to a faster convergence rate, so we want to find a random sample with a smaller ratio $\frac{d_{\text{-}k,m}}{s_{\text{-}k,m}}$. Fortunately, we can do so by simply observing the eigenvalue ratios of the sample covariance matrix of each random sample, with the help of the following assumption.
\begin{itemize}
    \item[(R2)] {\em We assume $\lambda_{d_k}(\bSigma_{\epsilon,l}^{(k)})$ is uniformly bounded below from 0 for $l\in[\dmk]$, so that $\lambda_{d_k}\left(\sum_{l \in \cS_m}\bSigma_{\epsilon,l}^{(k)}\right) \geq c |\cS_m|$ holds for any random subsets $\cS_m \subseteq [d_{\text{-}k}]$ for some constant $c>0$. Let $\A_{\epsilon,T}$ be the $T \times T$ matrix with its $(t,s)$ element to be $(\A_{\epsilon,T})_{t,s} = \sum_{q\geq 0}  a_{\epsilon,q}a_{\epsilon,q+|t-s|}$. Denote $0 < y: = \lim_{d_k, T \rightarrow \infty}\frac{\min(d_k, T)}{\max(d_k, T)} \leq 1$ and $y^\ast = \min(y,1)$, then we assume there exists $c_1 \in (1 - y^\ast, 1]$ such that $\lambda_{\lfloor c_1 T \rfloor}(\A_{\epsilon,T}) > c_2 >0$ for large $T$, where $c_2$ is a positive constant.}
\end{itemize}
Together with Assumption (R1), Assumption (R2) enables us to utilize random matrix theory to bound the eigenvalues of various sample covariance matrices from below (see (\ref{eqn:smallest_eigenvalue_bound_error}), (\ref{eqn:smallest_eigenvalue_bound_error_nonzeromu}) and (\ref{eqn:lemma2.7}) in Lemma \ref{lem:1} and Lemma \ref{lem:2}).
As long as the $\bSigma_{\epsilon,\ell}^{(k)}$’s are well-conditioned, and the serial correlation of the $\epsilon_{t,\ell,j}^{(k)}$'s are not too strong, Assumption (R2) will be satisfied.
With Assumption (R2), in Lemma \ref{lem:2}, we have shown that
\begin{align}
    ER_{m,j} := \frac{\lambda_1 \left(\frac{\tilde{\X}_{k,m}^\T \left( \I_T - \frac{1}{T} \1_T \1_T^\T \right)\tilde{\X}_{k,m}}{T} \right)}{\lambda_{j} \left(\frac{\tilde{\X}_{k,m}^\T \left( \I_T - \frac{1}{T} \1_T \1_T^\T \right)\tilde{\X}_{k,m}}{T} \right)} \asymp \frac{d_k^{\alpha_{k,1}}}{\frac{d_{\text{-}k,m}}{s_{\text{-}k,m}} \left( 1+ \frac{d_k}{T} \right)}
    \label{eqn:ER}
\end{align}
for $j$ satisfying $r_k+ 1 \leq j \leq \lfloor c\min{(T,d_k)} \rfloor - r_k$. Therefore, the above eigenvalue ratios of the covariance matrix of the $m$-th sample can help measure the quantity $\frac{d_{\text{-}k,m}}{s_{\text{-}k,m}}$. The larger $ER_{m,j}$ is, the smaller $\frac{d_{\text{-}k,m}}{s_{\text{-}k,m}}$ becomes, which leads to a faster convergence rate of the factor loading space. In fact, if we set $z_k = 1$, then the convergence rate (\ref{eqn:MER_estimator_convergencerate}) can be upper bounded by
\begin{align*}
    \norm{ \hat{\Q}_{k,m,(1)} - \Q_k\ddot{\H}_{k,m} }^2 \leq O_p\left(\frac{d_{\text{-}k,m}}{s_{\text{-}k,m}} \left( 1+ \frac{d_k}{T} \right) d_k^{-\alpha_{k,1}} \right),
\end{align*}
which is exactly the inverse of the rate for $ER_{m,j}$. Hence in practice, for each random sample, we can compute $ER_{m,j}$ for some $j$ satisfying $r_k+ 1 \leq j \leq \lfloor c\min{(T,d_k)} \rfloor-r_k$. Then, we can choose the sample with the largest $ER_{m,j}$, and obtain the estimator $\hat{\Q}_{k,max,(z_k)}$ based on this sample accordingly. We call $\hat{\Q}_{k,max,(z_k)}$ the maximum eigenvalue ratio estimator. Such an estimator enjoys the best rate of convergence among all chosen samples.

Note that estimators based on partial sum of fibers can be better than summing all fibers. As a simple example, suppose $K = 2$, $r_2 = 1$ and we want to estimate $\A_1$. If $\A_2$ has half of the entries 1 and the other half -1 (perhaps with some small perturbations), then it will sum to 0 or very close to 0. Then $s_{\text{-}k} = O(1)$, so $\frac{d_{\text{-}k}}{s_{\text{-}k}} = O(d_{\text{-}k})$ for summing all fibers. For partial sum, if we choose $d_{\text{-}k,m} \asymp 1$ fibers to sum, then $s_{\text{-}k,m} = O(1)$ for maximum eigenvalue ratio sample, and $\frac{d_{\text{-}k,m}}{s_{\text{-}k,m}} = O(1)$. If we choose $d_{\text{-}k}/2$ fibers to sum, then as $M_0 \rightarrow \infty$,
we will finally find one maximum eigenvalue ratio sample to have ${s_{\text{-}k,m}} = O(d_{\text{-}k}^2)$, in which case $\frac{d_{\text{-}k,m}}{s_{\text{-}k,m}} = O(\frac{1}{d_{\text{-}k}})$. Therefore, estimators based on summing partial fibers with maximum eigenvalue ratio can perform better in theory. In practice, with $M_0$ reasonably large, say in the order of hundreds, we can already achieve $d_{\text{-}k,m}/s_{\text{-}k,m}$ being of constant order if the strongest factors in $\A_j$ for the majority of $j\in[K]$ are not too weak, and $n_l$ of the same order as $d_k$, say $n_l=d_k/2$. This helps the rates of convergence tremendously in (\ref{eqn:MER_estimator_convergencerate}) and (\ref{eqn:MER_rate_lemma3_Lam}), especially when $d_k \asymp T$.

\subsubsection{Pre-averaging estimator}\label{subsec:pre-averaging}

Theoretically, the maximum eigenvalue ratio estimator enjoys the best rate of convergence among all chosen random samples, but it can be too unstable as it depends on only one particular random sample.
To obtain a more stable estimator empirically, we can utilize more random samples and calculate an ``average estimator'' accordingly. The basic idea is to first choose a certain number of random samples with the largest eigenvalue ratios, and then obtain the factor loading estimators based on the ``average'' of these chosen samples. More specifically, suppose we have $M_0$ random samples. Among them, we choose $M$ samples with the largest eigenvalue ratios $ER_{m,j}$ as stated in the previous subsection. Then, we can calculate an aggregated covariance matrix of these $M$ chosen samples as
\begin{align*}
    \hat{\Sigma}_{\tilde{\x}_k,agg} := \frac{1}{M} \sum_{m=1}^M \frac{\tilde{\X}_{k,m}^\T  \left( \I_T - \frac{1}{T} \1_T \1_T^\T \right) \tilde{\X}_{k,m}}{T},
\end{align*}
where $\tilde{\X}_{k,m}$ is defined similar to $\tilde{\X}_{k}$, with the subscript $m$ indicating that the data comes from the $m$-th random sample (we define $\tilde{\F}_{k,m}$ and $\tilde{\E}_{k,m}$ similarly as in (\ref{eqn:X_k_double_dot})). The pre-averaging estimator $\hat{\Q}_{k,pre,(z_k)}$ is defined as the $z_k$ eigenvectors corresponding to the $z_k$ largest eigenvalues of $\hat{\Sigma}_{\tilde{\x}_k,agg}$, with the constraint $\hat{\Q}_{k,pre,(z_k)}^\T \hat{\Q}_{k,pre,(z_k)} = \I_{z_k}$.

Let $s_{\text{-}k,pre}: = \frac{1}{M}\sum_{m=1}^M s_{\text{-}k,m}$, $\tilde{\V}_{k,pre}$ be the $z_k \times z_k$ diagonal matrix of the first $z_k$ largest eigenvalues of $\hat{\Sigma}_{\tilde{\x}_{k,agg}}$ in decreasing order, and $\ddot{\V}_{k,pre} := \tilde{\V}_{k,pre} / s_{\text{-}k,pre}$. The theoretical properties of $\hat{\Q}_{k,pre,(z_k)}$ can be summarized in the following theorem.

\begin{theorem}\label{thm:preaverage_sumestimatorrate_newmethod}
Let Assumption (E1), (E2), (F1), (L1), (L2), (R1) be satisfied for all $M$ chosen random samples, and
\begin{align*}
    c_{k,pre} := \ \min{\left\{ 1 + \frac{d_k}{T},\frac{r_kd_k}{T}\right\}} \frac{\frac{1}{M}\sum_{m=1}^M d_{\text{-}k,m} s_{\text{-}k,m}}{s_{\text{-}k,pre}^2} + d_k^{\alpha_{k,1}}\left( 1+ \frac{d_k^2}{T^2} \right) \frac{\frac{1}{M} \sum_{m=1}^M{d_{\text{-}k,m}^2}}{s_{\text{-}k,pre}^2}.
\end{align*}
Then
\begin{align}\label{eqn:preaverage_sumestimatorrate}
     \norm{ \hat{\Q}_{k,pre,(z_k)} - \Q_k\ddot{\H}_{k,pre} }^2 &= O_p\left(  d_k^{-2\alpha_{k,z_k}} c_{k,pre}\right),
\end{align}
where $\ddot{\H}_{k,pre} = \frac{\D_k^{\frac{1}{2}} \frac{1}{M} \sum_{m=1}^M \left[\ddot{\F}_{k,m} \left( \I_T - \frac{1}{T} \1_T \1_T^\T \right) \ddot{\F}_{k,m}^\T \right] \A_k^\T\hat{\Q}_{k,pre,(z_k)}\ddot{\V}_{k,pre}^{-1}}{T}$ has $rank(\ddot{\H}_k) = z_k$. Moreover, further assuming (L1'), there exists $\hat\U_{k,pre,(z_k)}$ with $\hat\U_{k,pre,(z_k)}^\T \hat\U_{k,pre,(z_k)} = \I_{z_k}$ such that $\hat\Q_{k,pre,(z_k)} = \hat\U_{k,pre,(z_k)} \P_{k,pre,(z_k)}$ with $\P_{k,pre,(z_k)}$ being an orthogonal matrix, so that
\begin{align}\label{eqn:preaverage_rate_lemma3_Lam}
    \norm{\hat\U_{k,pre,(z_k)} - \U_{k,(z_k)} }^2 = O_p \left(d_k^{-2\alpha_{k,z_k}} \left[d_k^{2\alpha_{k,1}}\frac{r_k}{T} + c_{k,pre} \right]\right).
\end{align}
The matrix $\U_{k,(z_k)}$ is defined to be the matrix consisting of the first $z_k$ columns of $\U_k$.
\end{theorem}

Note that if $d_{\text{-}k,m}$ and $s_{\text{-}k,m}$ remain the same among all $m$, then the rate (\ref{eqn:preaverage_sumestimatorrate}) is the same as (\ref{eqn:MER_estimator_convergencerate}) (similarly, (\ref{eqn:preaverage_rate_lemma3_Lam}) is the same as (\ref{eqn:MER_rate_lemma3_Lam})), which is the rate of the maximum eigenvalue ratio estimator. In practice, we may have a number of samples with relatively same scale of eigenvalue ratio and cannot identify which one is the largest. Theorem \ref{thm:preaverage_sumestimatorrate_newmethod} suggests that choosing all of them and obtaining the pre-averaging estimator can perform as good as choosing the largest one. What we gain from averaging is then the stability of the estimator.

\begin{remark}\label{remark:population_parameters} (Sample parameters and population parameters)
The assumptions in Theorem \ref{thm:preaverage_sumestimatorrate_newmethod} are on each random sample, and the convergence rates in (\ref{eqn:preaverage_sumestimatorrate}) and (\ref{eqn:preaverage_rate_lemma3_Lam}) depend on the sample parameters $d_{\text{-}k,m}$ and $s_{\text{-}k,m}$. We hope to write the assumptions and convergence rates in the form of population parameters. To do this, we can easily control $d_{\text{-}k,m}$ by setting the size of random sample. For instance, if we set $d_{\text{-}k,m} \asymp d_{\text{-}k}$ for all $m$, then Assumptions (E1) -- (R1) are naturally satisfied for all samples. We do not need any extra assumptions, except that (\ref{scalecondition}) in Assumption (L2) needs to be modified to $\frac{d_{\text{-}k}}{s_{\text{-}k,m}}\left(1+\frac{d_k}{T}\right) = o\left(d_k^{\alpha_{k,z_k}}\right)$. So we need to find an expression of $s_{\text{-}k,m}$.

Suppose for each $l \neq k$, we randomly select $n_l$ rows of $\A_l$ each time with $n_l \asymp d_l$, so that $d_{\text{-}k,m} \asymp d_{\text{-}k}$ for each $m$.
Then, since we are choosing the samples with largest eigenvalue ratios, when $M_0$ is large enough, 
we have $s_{l,m} \asymp s_{l,max}$ for each chosen sample, where $s_{l,max}$ is the largest possible sum resulting from $n_l$ rows of $\A_l$, or some large order albeit smaller than the possible maximum. This way, each $s_{\text{-}k,m} \asymp s_{\text{-}k,max}$, where $s_{\text{-}k,max}: = \prod_{l \in [K] \setminus \{ k\}}  s_{l,max}$.
\end{remark}

Based on Remark \ref{remark:population_parameters}, we state the following assumption for the maximum eigenvalue ratio estimator and the pre-averaging estimator:
\begin{itemize}
    \item[(L2')] (Signal Cancellation of maximum eigenvalue ratio sample) {\em For $k\in[K]$, define \begin{align}\label{eqn:defining_slmax}
    s_{k,max} := \max_{\cS_{k,m}\in \{\cS_{k,m}\subseteq[d_k]: \; m\in[M_{k,0}], \; |\cS_{k,m}| = n_k\} } \left[\sum_{j=1}^{r_k} \left( \sum_{i \in \cS_{k,m}}(\A_k)_{ij} \right)^2 \right],
    \end{align}
    and $s_{\text{-}k,max} := \prod_{l \in [K] \setminus \{ k\}}  s_{l,max}$. Then we assume
  \begin{align*}
      \frac{d_{\text{-}k}}{s_{\text{-}k,max}}\left(1+\frac{d_k}{T}\right) = o\left(d_k^{\alpha_{k,z_k}}\right),
  \end{align*}
  for some $z_k \leq r_k$. (If $z_k = 1$, then the assumption is most relaxed.)
  }
\end{itemize}

In (\ref{eqn:defining_slmax}), $M_{k,0}$ is the number of random row sums we consider for $\A_k$, so that for $k\in[K]$, the number of random samples from the $\dmk$ fibres is $M_0 = \prod_{j\in[K]\setminus\{k\}}M_{j,0}$.
Assumption (L2') is parallel to Assumption (L2), except that it is made on the maximum eigenvalue ratio sample, which corresponds to the maximum eigenvalue ratio estimator. And for the pre-averaging estimator, we assume that the $M$ samples we choose all satisfy $s_{l,m} \asymp s_{l,max}$ as $M_0$ is large enough (by choosing the $M$ samples with the largest eigenvalue ratios as defined in (\ref{eqn:ER})). This way, we can restate Theorem \ref{thm:preaverage_sumestimatorrate_newmethod} using population parameters as:
\begin{theorem}\label{thm:preaverage_sumestimatorrate_populationparameter}
For each random sample, if the sample size is $n_l$ for each $l \neq k$, then under Assumptions (E1), (E2), (F1), (L1), (L2'), (R1), (R2), with the pre-averaging estimator or the maximum eigenvalue ratio estimator based on choosing the $M$ samples with the largest eigenvalue ratios defined in (\ref{eqn:ER}), we have
\begin{align*}
      \norm{ \hat{\Q}_{k,pre,(z_k)} - \Q_k\ddot{\H}_{k,pre} }^2 \asymp \norm{ \hat{\Q}_{k,max,(z_k)} - \Q_k\ddot{\H}_{k,max} }^2 = O_p\left(  d_k^{-2\alpha_{k,z_k}} c_{k,max} \right),
\end{align*}
where
\begin{align*}
    c_{k,max} := \min{\left\{ 1 + \frac{d_k}{T},\frac{r_kd_k}{T}\right\}} \frac{ d_{\text{-}k} }{s_{\text{-}k,max}} + d_k^{\alpha_{k,1}}\left( 1+ \frac{d_k^2}{T^2} \right) \frac{{d_{\text{-}k,}^2}}{s_{\text{-}k,max}^2}.
\end{align*}
Moreover, further assuming (L1'), for the same definition of $\hat\U_{k,pre,(z_k)}$ and $\hat\U_{k,max,(z_k)}$ as before,
\begin{align}\label{eqn:preaverage_rate_lemma3_Lam_populationassumption}
    \norm{\hat\U_{k,pre,(z_k)} - \U_{k,(z_k)} }^2  \asymp \norm{\hat\U_{k,max,(z_k)} - \U_{k,(z_k)} }^2 = O_p \left(d_k^{-2\alpha_{k,z_k}} \left[d_k^{2\alpha_{k,1}}\frac{r_k}{T} + c_{k,max} \right]\right).
\end{align}
\end{theorem}
The rates in Theorem \ref{thm:preaverage_sumestimatorrate_populationparameter}, which are based on partial sum of fibers, are usually better than the one in Theorem \ref{thm:sumestimatorrate_newmethod}, which is based on summing all fibers. This is because $s_{\text{-}k,max}$ is usually larger than $s_{\text{-}k}$. In practice, as discussed at the end of Section \ref{subsubsec:MaximumEigenvaluesRatioEstimators}, a reasonably large $M_0$ and $n_l \asymp d_k$ will usually gives very good magnitude of $s_{\text{-}k,max}$ when we choose $M$ largest eigenvalue ratios in (\ref{eqn:ER}).

The need for Assumption (R2) in Theorem \ref{thm:preaverage_sumestimatorrate_populationparameter} is because of the fact that we assumed the $M$ samples we choose from the $M_0$ samples satisfy $s_{l,m}\asymp s_{l,max}$, which implicitly assumed (R2) already (so that (\ref{eqn:ER}) is satisfied, meaning we do have the ability to observe from the eigenvalue ratios to choose those samples with the best rates to satisfy $s_{l,m} \asymp s_{l,max}$). By the same token, Theorem \ref{thm:preaverage_sumestimatorrate_newmethod} does not need Assumption (R2) since all rates in the theorem are based on individual samples.

\begin{remark}\label{remark:rate_comparisons}
Suppose in (L2'), the ratio $\dmk/s_{\text{-}k,max}$ is of order $\dmk^{-1}$, which can be achieved if, for instance, the strongest factor in $\A_j$ is pervasive and have majority of elements of the same sign for each $j\in[K]$. Suppose further that the $r_k$'s and $K$ are constants, with $d_k\asymp T$ for each $k\in[K]$. The results from Theorem \ref{thm:preaverage_sumestimatorrate_populationparameter} implies that the projection matrix
$\hat\P_{k,pre,(r_k)} := \hat\Q_{k,pre,(r_k)}\hat\Q_{k,pre,(r_k)}^\T $
has error rate
\begin{equation}\label{eqn:rate_pre}
\norm{\hat\P_{k,pre,(r_k)} - \Q_k(\Q_k^\T\Q_k)^{-1}\Q_k^\T} = \norm{\hat\P_{k,pre,(r_k)} - \U_k\U_k^\T} =
O_P(d_k^{-\alpha_{k,z_k}}( \dmk^{-1/2} + d_k^{\alpha_{k,1}/2}\dmk^{-1} ) ).
\end{equation}
This can be compared to the rates in \cite{Chenetal2022}, which needs the errors to be sub-Gaussian (compared to our Assumption (R1) where only bounded fourth moments is needed). While their $\sigma^2$ can be considered constant, their $\lambda$ is such that $\lambda \asymp \prod_{k=1}^{K}d_k^{\alpha_{k,1}}$. The TIPUP procedure has rate (in our notations, using equation (47) in \cite{Chenetal2022}, which has a faster rate of convergence than TOPUP)
\begin{equation}\label{eqn:rate_chenetal}
\norm{\hat\P_k - \U_{k}\U_k^\T} = O_P\bigg( \frac{d_k^{1/2}}{T^{1/2}\prod_{k=1}^Kd_k^{\alpha_{k,1}/2}} + \frac{d^{1/2}}{T^{1/2}\prod_{k=1}^Kd_k^{\alpha_{k,1}}} \bigg).
\end{equation}
When all factors are strong, i.e., $\alpha_{k,j}=1$, the rate in (\ref{eqn:rate_pre}) is faster than that in (\ref{eqn:rate_chenetal}). When $\alpha_{k,1}=1$ and $\alpha_{k,r_k}=0.5$, i.e., the strongest factor is pervasive but the weakest factor is quite weak, then the two rates will be the same.
Indeed,
the better performance of the pre-averaging estimator is reflected in the simulation results in Section \ref{sec:simulation} when $s_{\text{-}k,max}$ is large.

\end{remark}

\section{Re-estimation by Projection}\label{sec:re-estimation}
\setcounter{equation}{0}
Setting $z_k=1$, with the pre-averaging estimators $\hat{\Q}_{k,pre,(1)}$ for each $k\in[K]$
as described in Section \ref{subsec:pre-averaging},
we have estimated a direction for projection for the mode-$k$ unfolding matrix of $\cX_t$ for each $k\in[K]$ which is asymptotically pointing to the direction of the strongest factors (see (\ref{eqn:preaverage_rate_lemma3_Lam_populationassumption}) in Theorem \ref{thm:preaverage_sumestimatorrate_populationparameter}). This point will be explained below and in Section \ref{subsec:refiningprojection}.

From (\ref{eqn:mode-k-flattening}), using the notation $\bar{w}$ to denote the sample mean of $\{w_t\}$, we have
\begin{align*}
  \mat_k(\cX_t - \bar{\cX}) &= \A_k\mat_k(\cF_{t,\text{-}k} - \bar{\cF}_{\cdot,\text{-}k}) + \mat_k(\cE_t - \bar{\cE})\\
  &= \A_k\mat_k(\cF_{t} - \bar{\cF})\Amk^\T + \mat_k(\cE_t - \bar{\cE}),
\end{align*}
where $\Amk := \A_K\otimes\cdots\otimes\A_{k+1}\otimes\A_{k-1}\otimes\cdots\otimes\A_1$, with $\otimes$ denoting the Kronecker product. Suppose we have $\q_k = \A_k\c_k$ where $\c_k$ is a non-zero vector, $k\in[K]$. Define
\begin{align*}
\qmk :&= \q_K\otimes\cdots\otimes\q_{k+1}\otimes\q_{k-1}\otimes\cdots\otimes\q_1 = \Amk\cmk,
\end{align*}
where $\cmk := \c_K\otimes\cdots\otimes\c_{k+1}\otimes\c_{k-1}\otimes\cdots\otimes\c_1$. Then we can define the new projected data as
\begin{align}
\y_t^{(k)} :&= \mat_k(\cX_t - \bar{\cX})\qmk\label{eqn:project_data}\\
&= \A_k\mat_k(\cF_t-\bar{\cF})\Amk^\T\Amk\cmk + \mat_k(\cE_t -\bar{\cE})\qmk. \notag
\end{align}
Depending on the direction $\cmk$, we can see from above that the signals from the factors are strengthened due to the term $\Amk^\T\Amk\cmk$, while the noise level is retained or strengthened, depending on the level of cross-correlations among the noise fibres (the term $\bPsi_\ell^{(k)}\e_t^{(k)}$ in (\ref{eqn:unfoldeddecomposition})).

Note also that since we use $z_k=1$, we do not need to know the rank of the core tensor before this projection step. This is important in practice since it means that we can use the projected data to potentially improve on the accuracy of any core tensor rank estimation procedures (see Section \ref{sec:rank_estimation}) without the need to know the core tensor rank itself first. The projected data can also be used to estimate a finer projection direction, essentially iterating the projection step. Such re-projection step can be potentially advantageous since the estimated factor loading matrices for each mode $k\in[K]$ can be more accurate. See Theorem \ref{thm:reiterate_projection_direction}  below and the explanations followed. See simulation results regarding this in Section \ref{sec:simulation} as well.

\subsection{Refining the projection direction}\label{subsec:refiningprojection}
From Theorem \ref{thm:preaverage_sumestimatorrate_populationparameter}, setting $z_k=1$ there, we obtain $\hat\q_{k,pre} := \hat{\Q}_{k,pre,(1)} = \hat{\U}_{k,pre,(1)}\P_{k,pre,(1)} = \pm\hat{\U}_{k,pre,(1)}$ (WLOG we take the plus sign in the presentations hereafter), with an error rate
\begin{equation}\label{eqn:qhatkpre_rate}
\norm{\hat{\q}_{k,pre} - \U_{k,(1)}} = O_P\bigg(\sqrt{\frac{r_k}{T}}+d_k^{-\alpha_{k,1}}c_{k,max}^{1/2}\bigg).
\end{equation}
For instance, if $r_k$ is a constant and $c_{k,max}\asymp d_k^{\alpha_{k,1}}$, then the above rate is
$T^{-1/2} + d_k^{-\alpha_{k,1}/2}$, which can be slower than $T^{-1/2}$ if $d_k\asymp T$ but $\alpha_{k,1} < 1$, i.e., the strongest factor corresponding to $\A_k$ is not pervasive.

For each $k \in [K]$, we create the projected data $\y_t^{(k)}$ as in (\ref{eqn:project_data}), using
\begin{equation}\label{eqn:qmk}
\qmk = \hat\q_{\text{-}k,pre} := \hat\q_{K,pre}\otimes\cdots\otimes\hat\q_{k+1,pre}\otimes\hat\q_{k-1,pre}\otimes\cdots\otimes\hat\q_{1,pre}.
\end{equation}
Then we define $\check\q_{k}^{(1)}$ to be the eigenvector corresponding to the largest eigenvalue of the matrix
\[\wt\bSigma_y^{(k)} := T^{-1}\sum_{t=1}^T\y_t^{(k)}\y_t^{(k)\T}.\]
The superscript $(1)$ in $\check\q_{k}^{(1)}$ signals that this is the first iterated estimator for $\U_{k,(1)}$. A corresponding factor loading matrix estimator for the factor loading space spanned by the columns of $\A_k$ can be obtained as well by simply defining $\check\Q_k^{(1)}$ to contain the $r_k$ eigenvectors corresponding to the $r_k$ largest eigenvalues of $\wt\bSigma_y^{(k)}$, assuming we know the true number of factors $r_k$. We can iterate this process to obtain refinement of projection direction. More formally, we introduce the following algorithm.

\underline{Algorithm for Iterative Projection Direction Refinement}
\begin{itemize}
  \item[1.] Initialize $\check\q_k^{(0)} = \hat{\q}_{k,pre}$ for each $k\in[K]$.

  \item[2.] For $i\geq 1$, at the $i$-th step, create projected data $\y_{t,i}^{(k)} := \mat_k(\cX_t - \bar{\cX})\check\q_k^{(i-1)}$ for each $k\in[K]$.

  \item[3.] For each $k\in[K]$, define $\check\q_k^{(i)}$ the eigenvector corresponding to the largest eigenvalue of
\begin{equation}\label{eqn:Sigma_y}
  \wt\bSigma_{y,i}^{(k)} := T^{-1}\sum_{t=1}^T\y_{t,i}^{(k)}\y_{t,i}^{(k)\T}.
\end{equation}
  \item[4.] Replace $i$ by $i+1$. Go back to step 2. Stop until after the procedure has been repeated for a fixed number of times.
\end{itemize}

Before presenting the main theorems of this subsection, define
\begin{align}
\gmkhat :&= \hat\q_{\text{-}k,pre}^\T\Amk\Amk^\T\hat\q_{\text{-}k,pre}  = \hat{\U}_{\text{-}k,pre,(1)}^\T\Amk\Amk^\T\hat{\U}_{\text{-}k,pre,(1)}\label{eqn:ghat_{-k}}\\
&= \U_{\text{-}k,(1)}^\T\Amk\Amk^\T\U_{\text{-}k,(1)} +  (\Umkpre1 - \U_{\text{-}k,(1)})^\T\Amk\Amk^\T(\Umkpre1 - \U_{\text{-}k,(1)}) \notag\\
&\quad+ 2(\Umkpre1 - \U_{\text{-}k,(1)})^\T\Amk\Amk^\T\U_{\text{-}k,(1)}\notag\\
&\geq \U_{\text{-}k,(1)}\T\Amk\Amk^\T\U_{\text{-}k,(1)} + 2(\Umkpre1 - \U_{\text{-}k,(1)})^\T\Amk\Amk^\T\U_{\text{-}k,(1)} \notag\\
&= \Umkc1^\T\Umk\Gmk\Umk^\T\Umkc1 + 2(\Umkpre1 - \Umkc1)^\T\Umk\Gmk\Umk^\T\Umkc1, \notag
\end{align}
where the last line used the singular value decomposition of $\A_k$ in (\ref{eqn:SVD_Ak}), and
\begin{align*}
  \Umkpre1 &:= \U_{K,pre,(1)}\otimes\cdots\otimes\U_{k+1,pre,(1)}\otimes\U_{k-1,pre,(1)}\otimes\cdots\otimes\U_{1,pre,(1)},\\ \Umk &:=\U_K\otimes\cdots\otimes\U_{k+1}\otimes\U_{k-1}\otimes\cdots\otimes\U_1,\;\;\; \Gmk := \G_K\otimes\cdots\otimes\G_{k+1}\otimes\G_{k-1}\otimes\cdots\otimes\G_1,\\
   \Umkc1 &:= \U_{K,(1)}\otimes\cdots\U_{k+1,(1)}\otimes\U_{k-1,(1)}\otimes\cdots\otimes\U_{1,(1)}.
\end{align*}
But (\ref{eqn:lemma2.2}) in Lemma \ref{lem:2} 
proves  that $\G_k$ has the $j$-th diagonal element with order $d_k^{\alpha_{k,j}}$ for $j\in[r_k]$ under Assumption (L1), which is the same as those in $\D_k$. At the same time, induction (proof omitted) easily gives
\begin{equation}\label{eqn:qmkpre_rate}
\norm{\Umkpre1 - \Umkc1} = O_P\bigg(\sum_{j=1;j\neq k}^K\norm{\hat{\U}_{j,pre,(1)} - \U_{j,(1)}}\bigg) = O_P\bigg(K\sqrt{\frac{r_{\max}}{T}} + \sum_{j=1;j\neq k}^Kd_j^{-\alpha_{j,1}}c_{j,max}^{1/2}\bigg),
\end{equation}
where the last rate is from (\ref{eqn:qhatkpre_rate}) with $r_{\max} := \max_{k\in[K]}r_k$. Hence from the decomposition after (\ref{eqn:ghat_{-k}}), we then have, assuming the above rate is $o(1)$,
\begin{equation}\label{eqn:gmkhat_rate}
\gmkhat \asymp (1+o_P(1))(\Gmk)_{11} \asymp_P \prod_{j=1;j\neq k}^Kd_j^{\alpha_{j,1}}.
\end{equation}
We present a further assumption needed before presenting Theorem \ref{thm:reiterate_projection_direction}.

\begin{itemize}
  \item[(RE1)]   {\em For a positive integer $N$, let $\cA_{f,T} \in \mathbb{R}^{(N+1)T\times T}$  be defined as $\cA_{f,T} := (\a_{f,1},\ldots,\a_{f,T})$, where
    \[\a_{f,t} := (\0_{t-1}^\T,a_{f,NT},a_{f,NT-1},\ldots,a_{f,0},\0_{T-t}^{\T})^\T,\;\;\; t\in[T],\]
with $\0_j$ being a column vector of $j$ zeros and the $a_{f,q}$'s are from Assumption (F1). Define $\cA_{e,T}$ and $\cA_{\epsilon,T}$ similarly using coefficients from $\{a_{e,q}\}$ and $\{a_{\epsilon,q}\}$ respectively from Assumption (E2). Then we assume that (with $\cA$ can be either $\cA_{f,T},\cA_{e,T}$ or $\cA_{\epsilon,T}$) $\norm{\cA}$ is uniformly bounded above, and
\begin{align*}
  \frac{1}{T}\text{\em tr}(\cA^\T\cA) &= 1 - o(T^{-2}d^{-4}), \;\;\; \frac{1}{T}\text{\em tr}(\cA^\T\cA)^2\rightarrow a_1,\;\;\; \frac{1}{T^2}\1_T^\T(\cA^\T\cA)^2\1_T\rightarrow a_2, \;\;\; \frac{1}{T^{3/2}}\1_T^\T\cA^\T\cA\1_T \rightarrow a_3,
\end{align*}
where $\1_T$ is a column vector of $T$ ones, and the constant $a_1$ can be $a_{1,f}, a_{1,e}$ and $a_{1,\epsilon}$ for $\cA=\cA_{f,T}, \cA_{e,T}$ and $\cA_{\epsilon,T}$ respectively. Similarly for constants $a_2$ and $a_3$.}
\end{itemize}
Consider a truncated linear process $\{y_t\}_{t\in[T]}$, and the original process $\{\tilde y_t\}_{t\in[T]}$,
\[\tilde y_t = \sum_{q\geq 0}a_qz_{t-q},\;\;\; y_t = \sum_{q=0}^{NT}a_qz_{t-q},\;\text{with } \; \var(\tilde y_t) = 1,\]
where $\{z_t\}$ is a sequence of i.i.d. random variables. Construct the matrix $\cA$ using $\{a_q\} $ similar to those in Assumption (RE1). Then $\cA^\T\cA$ contains the variance of $\{y_t\}$ on the diagonal, and lag-$k$ autocovariance on the $k$-th off-diagonal. The rates in (RE1) are then controlling how fast the $a_q$'s are going to 0, and how much serial dependence between the $y_t$'s are allowed. In particular, general linear processes with absolutely summable autocovariance sequence, short range dependent processes like ARMA models, satisfy the assumption.

\begin{theorem}\label{thm:reiterate_projection_direction}
Let all the assumptions in Theorem \ref{thm:preaverage_sumestimatorrate_populationparameter} be satisfied, together with (RE1). Let $g_s := \prod_{j=1}^Kd_j^{\alpha_{j,1}}$ and $S_\psi^{(k)} := \sum_{j=1}^{\dmk}\norm{\bPsi_j^{(k)}}^2$. Assume further that for each $k\in[K]$,
\begin{align*}
  r &= O(r_e), \;\;\; d_k = O(g_s) = (r_e + \sqrt{T})S_\psi^{(k)},  \;\;\; \max_{j\in[\dmk]}\norm{\bSigma_{\epsilon,j}^{(k)}} = O\bigg(\prod_{j=1}^Kd_j^{\alpha_{j,1}}\sqrt{\frac{r}{T}}\bigg).
\end{align*}
Then
\begin{align*}
\norm{\check{\q}_k^{(1)} - \U_{k,(1)}} &= O_P\bigg\{\sqrt{\frac{r}{T}}\bigg[1 + b_k^2\frac{d}{T}\bigg]
+ g_s^{-1/2}b_k\sqrt{\frac{rd}{T}} \bigg\}, \; (\text{assumed }o_P(1)) \;\text{ where } \\
b_k &= K\sqrt{\frac{r_{\max}}{T}} + \sum_{j=1;j\neq k}^Kd_j^{-\alpha_{j,1}}c_{j,max}^{1/2} = o(1).
\end{align*}
Furthermore, if
\[K(r + \max_{j\in[\dmk]}\norm{\bSigma_{\epsilon,j}^{(k)}})\prod_{j=1}^Kd_j^{1-\alpha_{j,1}} = o(T),\]
then the Algorithm for Iterative Projection Direction Refinement will produce, after a certain number of iterations (say $m$),
\[\norm{\check\q_{k}^{(m)} - \U_{k,(1)}} = O_P\bigg(\sqrt{\frac{r}{T}}\bigg).\]

\end{theorem}
The rate $b_k$ is the rate from (\ref{eqn:qmkpre_rate}).
To put the above results into perspective, assume a very common scenario that $d_1\asymp \cdots\asymp d_K\asymp T$ (this is especially true in economic applications where $T$ is small), with K and each $r_k$ being constants for $k\in[K]$. We first note that if all $r_k$ factors in $\A_k$ are pervasive, i.e., $\alpha_{k,j}=1$ for all $j\in[r_k]$, then
since $\gmkhat \asymp_P \dmk$ (see (\ref{eqn:gmkhat_rate})),  if we also have $r_e = O(d_k)$,  $\norm{\bPsi_j^{(k)}} = O(1)$ and $\max_{j\in[\dmk]}\norm{\bSigma_{\epsilon,j}^{(k)}} = o(T)$, the first estimated projection direction will have an error $O_P(T^{-1/2})$, and any refinements will retain the same rate.
The three conditions $r_e = O(d_k)$,  $\norm{\bPsi_j^{(k)}} = O(1)$ and $\max_{j\in[\dmk]}\norm{\bSigma_{\epsilon,j}^{(k)}} = o(T)$ are quite relaxed. They allow for a general within-fibre correlation structures (as long as (E1) is satisfied; see the comments after (E1) in Section \ref{subsubsec:errorassumptions}) and not too complicated across-fibre cross-correlation structures for each pair of mode-$k$ fibres.

Even if $\alpha_{k,1} < 1$ (i.e., the strongest factor corresponding to $\A_k$ is not pervasive), $\norm{\check{\q}_k^{(1)} - \U_{k,(1)}}$ can still be $O_P(T^{-1/2})$, as long as we have strong factors from other factor loading matrices $\A_j$'s for $j\neq k$ such that the rate constraints given in Theorem \ref{thm:reiterate_projection_direction} are satisfied. One such scenario can be that $\alpha_{j,1}=1$ for $j\neq k$ and $\alpha_{k,1}=1/2$, with $\gmkhat \asymp \dmk$, $r_e = O(d_k^{1/2})$, $\norm{\bPsi_j^{(k)}} = O(1)$ and $\max_{j\in[\dmk]}\norm{\bSigma_{\epsilon,j}^{(k)}} = o(Td_k^{-1/2})$. This presents a significantly weak strongest factor corresponding to $\A_k$, and without the help of projection and strong factors from other modes' factor loading spaces, the typical rate for estimating such a weak factor would be $d_k^{-1/4}$, which is much worse than $T^{-1/2}$.

The fixed rate $O_P(\sqrt{r/T})$ in Theorem \ref{thm:reiterate_projection_direction} comes from the fact that we need to distinguish the direction of the strongest factors from all other directions of weaker factors in order to find the ``best'' projection direction. In the case of studying the whole $\U_k$, which contains the $r_k$ eigenvectors of the whole factor loading space for $\A_k$, we in fact may get a better rate of convergence even in the presence of weak factors.

\begin{theorem}\label{thm:reestimation_factor_loading_space}
  Let all the assumptions in Theorem \ref{thm:reiterate_projection_direction} be satisfied. Suppose we know the value of $r_k$, and perform an eigenanalysis on $\wt\bSigma_{y,m+1}^{(k)}$ in (\ref{eqn:Sigma_y}) which utilized the projection direction $\check{\q}_k^{(m)}$ in Theorem \ref{thm:reiterate_projection_direction}, obtaining $r_k$ eigenvectors as an estimator of the factor loading space of $\A_k$.

  Then there exists $\check{\U}_k \in \mathbb{R}^{d_k\times r_k}$ with $\check\U_k^\T\check\U_k = \I_{r_k}$ such that the $r_k$ eigenvectors obtained above is $\check\U_k$ multiplied with some orthogonal matrix, with
  \begin{align*}
  \norm{\check{\U}_k - \U_k} = O_P\Bigg\{& g_s^{-1/2}d_k^{\alpha_{k,1}-\alpha_{k,r_k}}\bigg[\sqrt{\frac{r}{T}}\bigg(\sqrt{d_k}+K\sqrt{\frac{rd}{T}} + \sqrt{r_eS_\psi^{(k)}}\bigg)\\
  &+ g_s^{-1}d_k^{\alpha_{k,1}-\alpha_{k,r_k}}\bigg(\max_{j\in[\dmk]}\norm{\bSigma_{\epsilon,j}^{(k)}}\bigg[1 + \frac{K^2rd}{T^2}\bigg] + S_\psi^{(k)}\bigg) \bigg]\Bigg\}, \; (\text{assumed } o_P(1)).
  \end{align*}
\end{theorem}

Consider $d_1\asymp \cdots \asymp d_K \asymp T$, with $K$ and $r_k$ being constants for $k\in[K]$. If all factors for $\A_k$ are pervasive, i.e., $\alpha_{k,j}=1$ for all $j\in[r_k]$, then with $r_e = O(d_k^{1/2})$, $\norm{\bPsi_j^{(k)}} = O(1)$ and $\max_{j\in[\dmk]}\norm{\bSigma_{\epsilon, j}^{(k)}} = O(d_k)$ (as long as (E1) is satisfied; see the comments after (E1) in Section \ref{subsubsec:errorassumptions}), we  have $\norm{\check\U_k - \U_k} = O_P(T^{-3/4})$. This improves to $O_P(T^{-1})$ if $r_e \asymp d_k$ but $S_\psi^{(k)} = O(\dmk/T)$, i.e., when the cross-correlations among fibres are weaker.

The above rate can be greatly improved if the terms $K\sqrt{rd/T}$ and $K^2rd/T^2$ can be removed, which are associated with the noise. They are there because the estimated projection direction is correlated with the data in general. If we have independent noise tensor $\{\cE_t\}$ (e.g., the setting in \cite{Chenetal2022}) we can split the data into half, and using only one half of it for projection direction estimation while the other half is for re-estimation only. Then the estimated projection direction will be independent of the re-estimation data, and hence the final rate indeed will be rid of these two terms. When all the factors are strong, this improved rate will be the same as the one for TIPUP in equation (47) of \cite{Chenetal2022}. We do not pursue this since our paper focuses on time series data with serial correlation in the noise, and moreover, in practice the re-estimation using projected data with projection direction estimated from the same data is performing very well already.

\subsection{Practical implementation: Setting of $M_0$ and $M$, and number of iterations}\label{subsection:Setting_M_0_and_M}
In Theorem \ref{thm:preaverage_sumestimatorrate_populationparameter}, the rate of convergence depends on $c_{k,max}$, which itself depends on $s_{\text{-}k,max}$, $k\in[K]$. We can see that in (\ref{eqn:defining_slmax}), $s_{k,max}$ depends on $M_{k,0}$, so that $s_{\text{-}k,max}$ depends on $M_0 = \prod_{j\in [K]\setminus\{k\}}M_{j,0}$, the number of random samples from the $\dmk$ mode-$k$ fibres. Ideally, we should go over all $\prod_{l\in[K]\setminus\{k\}}\binom{d_l}{n_l}$ possible samples, but this is far from practical since this number is too large for computational efficiency. In practice, we consider $M_0 = 200$ say, but with samples so that $n_l$ is fixed at $d_l/2$ for each $l\in[K]\setminus\{k\}$. Then we choose $M=5$ samples to be aggregated. These numbers are tested to strike a good balance between computational efficiency and estimation accuracy for all our simulation settings with $K=2$ or $K=3$ in Section \ref{sec:simulation}. In fact, increasing $M_0$ and $M$ by 10 folds in those settings still do not have any visible improvement of the estimators in all settings there.

As for the number of iterations used for the iterative projection estimator in Section \ref{subsec:refiningprojection}, in all our simulation settings in Section \ref{sec:simulation}, we use 30 iterations, which guaranteed convergence in all cases. In fact, 10 iterations is enough for the purpose in most cases, hence the computational time of the iterative step reported in Section \ref{sec:simulation} can in fact be made faster by lowering the number of iterations.

\section{Core Tensor Rank Estimation Using Projected Data}\label{sec:rank_estimation}
\setcounter{equation}{0}
With the projected data and the associated covariance matrix $\wt\bSigma_{y,m+1}^{(k)}$ defined in (\ref{eqn:Sigma_y}), we can, for each $k\in[K]$, form the correlation matrix
\begin{equation}\label{eqn:R_y}
\wt\R_{y,m+1}^{(k)} := \diag^{-1/2}(\wt\bSigma_{y,m+1}^{(k)})\wt\bSigma_{y,m+1}^{(k)}\diag^{-1/2}(\wt\bSigma_{y,m+1}^{(k)}).
\end{equation}
Our estimator for $r_k$ for each $k\in[K]$ is then defined to be
\begin{equation}\label{eqn:rkhat}
\hat{r}_k := \max\{j:\lambda_j(\wt\R_{y,m+1}^{(k)})>1+\eta_T, \; j\in[d_k]\},
\end{equation}
where $\eta_T \rightarrow 0$ as $T\rightarrow \infty$ and its practical choice will be discussed in Section \ref{subsec:practical_core_rank}. This estimator is inspired by the one in \cite{Fanetal2022} for independent observations from a vector factor model, where they count the number of eigenvalues larger than $1+\sqrt{p/(n-1)}$ from a correlation matrix estimator, where $p$ and $n$ are dimension and sample size respectively.

\subsection{Main results}\label{subsec:further_assumptions}
To introduce a further assumption needed in this section, similar to (\ref{eqn:ghat_{-k}}), define
\begin{align}
\gmkcheck &:= \check\q_{\text{-}k}^{(m)\T}\Amk\Amk^\T\check\q_{\text{-}k}^{(m)}\label{eqn:gmkcheck}\\
&\geq (1+o_P(1))(\Gmk)_{11} \asymp_P \prod_{j=1;j\neq k}^Kd_j^{\alpha_{j,1}},\notag
\end{align}
where $\qmkcheck^{(m)}$ is defined using a similar convention in (\ref{eqn:qmk}), and the second line is from the result of Theorem \ref{thm:reiterate_projection_direction} and a derivation similar to that in (\ref{eqn:gmkhat_rate}). Now define

\begin{equation}\label{eqn:Sigma_y_population}
\bSigma_{y,m+1}^{(k)} := \gmkcheck\A_k\A_k^\T + \sum_{j=1}^{\dmk}(\qmkcheck^{(m)})_j^2\bSigma_{\epsilon,j}^{(k)}
+ \bigg(\sum_{j=1}^{\dmk}(\qmkcheck^{(m)})_j\bPsi_j^{(k)}\bigg)^{\otimes 2},
\end{equation}
where for any matrix $A$ we define $A^{\otimes 2} := AA^\T$. Apart from $\qmkcheck^{(m)}$, the matrix $\bSigma_{y,m+1}^{(k)}$ is not random, and is in fact the expected value of $\wt\bSigma_{y,m+1}^{(k)}$, pretending that $\qmkcheck^{(m)}$ is a constant vector.

\begin{itemize}
  \item[(RE2)] (Model Parameters) {\em For each $k\in[K]$, we assume that for each $j\in[d_k]$, $\lambda_j(\diag(\A_k\A_k^\T))$ is uniformly bounded away from 0 and infinity as $T,d_k\rightarrow \infty$. Moreover,
      $r_k = o(d_k^{1-\alpha_{k,1} + \alpha_{k,r_k}})$, and
     \begin{align*}
       \max_{j\in[\dmk]}\norm{\bSigma_{\epsilon,j}^{(k)}}, \; r_eS_\psi^{(k)} = o\bigg(\prod_{j=1;j\neq k}^Kd_j^{\alpha_{j,1}}\bigg)  \text{ and } d_k = O(r_eS_\psi^{(k)}).
     \end{align*}
}
\end{itemize}
Assumption (RE2) ensures that each row of $\A_k$ has at least one non-zero value, meaning that at least one factor drivers the dynamics of the corresponding element in $\y_{t,m+1}^{(k)}$. The assumption can be weakened so that the values are vanishing, at the price of more complicated proofs and rates in Theorem \ref{thm:rkhat_consistency}. The corresponding uniform bound from infinity is only there for the simplification of rates in Theorem \ref{thm:rkhat_consistency}.  Coupled with the rates in the same assumption (which is the most relaxed when all strongest factors are pervasive), the diagonal elements in $\bSigma_{y,m+1}^{(k)}$  are dominated by those in $\gmkcheck \A_k\A_k^\T$, hence (RE2) implies that
\begin{equation}\label{eqn:Sigma_y_diagonal_dominance}
\diag(\bSigma_{y,m+1}^{(k)}) = \gmkcheck\diag(\A_k\A_k^\T)(1+o(1)).
\end{equation}
The rates assumed are reasonable. The assumption $d_k = O(r_eS_{\psi}^{(k)})$ is utilized only to simplify the final rates in Theorem \ref{thm:rkhat_consistency}, and can be dropped at the expense of more complicated rates.

\begin{theorem}\label{thm:population_eigenvalue_correlation_matrix}
Let Assumption (E1), (F1) and (RE2) hold. For each $k\in[K]$, define the correlation matrix
\[\R_{y,m+1}^{(k)} = \text{\em diag}^{-1/2}(\bSigma_{y,m+1}^{(k)})\bSigma_{y,m+1}^{(k)}\text{\em diag}^{-1/2}(\bSigma_{y,m+1}^{(k)}).\]
Then, in probability, for large enough $T,d_k$, we have $\lambda_j(\R_{y,m+1}^{(k)}) \succeq_P r_k^{-1}d_k^{1-\alpha_{k,1}+\alpha_{k,j}}>1$ for $j\in[r_k]$, whereas $\lambda_j(\R_{y,m+1}^{(k)}) \leq 1$ for $j=r_k+1,\ldots,d_k$.
\end{theorem}
This theorem is in parallel to Theorem 1 of \cite{Fanetal2022}. With this, we can write
\[r_k = \max\{j:\lambda_j(\R_{y,m+1}^{(k)})>1, \; j\in[d_k]\}.\]
In light of this, the estimator $\hat{r}_k$ in (\ref{eqn:rkhat}) makes sense. The following theorem shows further that $\hat{r}_k$ is in fact consistent for $r_k$ for a suitable choice of $\eta_T$.

\begin{theorem}\label{thm:rkhat_consistency}
Let all the assumptions in Theorem  \ref{thm:reiterate_projection_direction} hold, together with (RE2). Suppose further that
\begin{align*}
d_k^{\alpha_{k,1}-\alpha_{k,r_k}}(\max_{j\in[\dmk]}\norm{\bSigma_{\epsilon,j}^{(k)}} + S_{\psi}^{(k)}) &= o(g_s),\;\;\; d_k^{\alpha_{k,1}-\alpha_{k,r_k}}\sqrt{\frac{r}{T}}\Big[ \sqrt{r_eS_{\psi}^{(k)}} + K\sqrt{\frac{rd}{T}} \Big] = o(g_s),
\end{align*}
where $g_s$ is defined in Theorem \ref{thm:reiterate_projection_direction}. Then as $T,d_k\rightarrow \infty$, we have for each $k\in[K]$,
\begin{align*}
  \lambda_j(\wt\R_{y,m+1}^{(k)}) &= \left\{
                                     \begin{array}{ll}
                                       \lambda_j(\R_{y,m+1}^{(k)})(1+ O_P\{r_kd_k^{\alpha_{k,1}-\alpha_{k,j}-1}a_T(\alpha_{k,1}) + a_T(0)\}), & \hbox{$j\in[r_k]$;} \\
                                       \lambda_j(\R_{y,m+1}^{(k)}) + O_P\{a_T(0)\}, & \hbox{$j\in[d_k]/[r_k]$,}
                                     \end{array}
                                   \right.\\
& \left\{
          \begin{array}{ll}
            \succeq_P r_k^{-1}d_k^{1-\alpha_{k,1} + \alpha_{k,j}}(1+O_P\{r_kd_k^{\alpha_{k,1}-\alpha_{k,j}-1}a_T(\alpha_{k,1}) + a_T(0)\}), & \hbox{$j\in[r_k]$;} \\
            \leq 1 + O_P\{a_T(0)\}, & \hbox{$j\in[d_k]/[r_k]$,}
          \end{array}
        \right.
\end{align*}
where for $0\leq \delta \leq 1$,
\[a_T(\delta) := \sqrt{\frac{r}{T}}\bigg(d_k^{\delta} + Kd_k^{(\delta+\alpha_{k,1})/2}\sqrt{\frac{rd}{Tg_s}} + \frac{K^2r^{1/2}dd_k^{\alpha_{k,1}}}{T^{3/2}g_s}\max_{j\in[\dmk]}\norm{\bSigma_{\epsilon,j}^{(k)}}\bigg) = o(1).\]
Hence $\hat{r}_k$ in (\ref{eqn:rkhat}) is a consistent estimator for $r_k$ if we choose $\eta_T = Ca_T(0)$ for some constant $C>0$.
\end{theorem}
If the strongest factor for each mode-$k$ unfolded matrix is pervasive, i.e., $\alpha_{j,1} = 1$ for each $j\in[K]$, and $K$ is a constant, then the rate $a_T(\alpha_{k,1}) = a_T(1)$ and $a_T(0)$ are respectively simplified to
\[a_T(\alpha_{k,1}) = a_T(1) = O_P\bigg(d_k\sqrt{\frac{r}{T}}\bigg), \;\;\; a_T(0) = O_P\bigg(\sqrt{\frac{r}{T}}\bigg(1 + \sqrt{\frac{rd_k}{T}} + \frac{r^{1/2}d_k}{T^{3/2}}\max_{j\in[\dmk]}\norm{\bSigma_{\epsilon,j}^{(k)}}\bigg)\bigg).\]
If further $d_k \asymp T$, $r_k$ is constant for each $k\in[K]$ and $\max_{j\in[\dmk]}\norm{\bSigma_{\epsilon,j}^{(k)}} = O(T^{1/2})$,  then $a_T(0) \asymp T^{-1/2}$, showing that the best rate of convergence of $\lambda_j(\wt\R_{y,m+1}^{(k)})$  to $\lambda_j(\R_{y,m+1}^{(k)})$ for $j\in[d_k]/[r_k]$ is at best $T^{-1/2}$ when $d_k \asymp T$. It means that our search for $\eta_T$ can be in the form $CT^{-1/2}$.

The extra rate assumptions in the theorem may not be more stringent than those in Theorem \ref{thm:reiterate_projection_direction} and (RE2). For instance, if $K$ and each $r_k$ for $k\in[K]$ are constants with $d_1\asymp\cdots\asymp d_K\asymp T$ and all factors are pervasive (i.e., $\alpha_{k,j}=1$ for each $j\in[r_k]$ and $k\in[K]$), then the extra rate assumptions in Theorem \ref{thm:rkhat_consistency} become
\begin{align*}
  \max_{j\in[\dmk]}\norm{\bSigma_{\epsilon,j}^{(k)}} + S_{\psi}^{(k)} = o(g_s), \;\;\; r_eS_{\psi}^{(k)} = o(Tg_s),
\end{align*}
which are in fact far more relaxed than (RE2) and those rate assumptions in Theorem \ref{thm:reiterate_projection_direction}.

\subsection{Practical implementation for core rank estimator}\label{subsec:practical_core_rank}
Since there is only one mode-$k$ unfolding matrix from our data, we propose a method for Bootstrapping the mode-$k$ fibres given our current data, so that we can search for $\eta_T$ using the Bootstrapped data. Consider for $b=1,\ldots,B$ ($B>0$ is a pre-chosen integer), a sequence of independent and identically distributed Bernoulli random variables $\{\xi_j^{(b)}\}_{j=1,\ldots,\dmk}$, with $\{\xi_j^{(b_1)}\}$ independent of $\{\xi_j^{(b_2)}\}$ for $b_1\neq b_2$. We then create a random matrix $\W_b\in\mathbb{R}^{\dmk\times \dmk}$, where the $i$-th column is $\0$ except its $j$-th zero (with $j$ chosen uniformly from $[\dmk]$)
is replaced by $\xi_i^{(b)}$.
We then create the new projected data
\begin{equation*}\label{eqn:Bootstrapped_projected_data}
  \y_{t,m+1,b}^{(k)} := \mat_k(\cX_t - \bar{\cX})\W_b\W_b^\T\qmkcheck^{(m)}.
\end{equation*}
Essentially, we Bootstrap the mode-$k$ fibres by choosing them randomly with replacement, and augment the vector of projection $\qmkcheck^{(m)}$ accordingly by pre-multiplying it with $\W_b^\T$.

From here on, we drop the subscript $m+1$ for the ease of presentation. We then create, for each $k\in[K]$ and $b\in[B]$,
\begin{align*}
\wt\bSigma_{y,b}^{(k)} &:= T^{-1}\sum_{t=1}^T\Big(\y_{t,b}^{(k)}\Big)^{\otimes 2}, \;\;\; \wt\R_{y,b}^{(k)} := \diag^{-1/2}(\wt\bSigma_{y,b}^{(k)})\wt\bSigma_{y,b}^{(k)}\diag^{-1/2}(\wt\bSigma_{y,b}^{(k)}).
\end{align*}
For a constant $C$, we then calculate
\begin{align*}
  \hat{r}_k^{(b)}(C) := \max\{j:\lambda_j(\wt\R_{y,b}^{(k)}) > 1+CT^{-1/2}, \;j\in[d_k]\}.
\end{align*}
We propose to choose $C$ with
\[\hat{C} := \min_{C>0}\wh\var(\{\hat{r}_k^{(b)}(C)\}_{b\in[B]}),\]
where $\wh\var(\{x_t\}_{t\in\mathcal{T}})$ is the sample variance of $\{x_t\}_{t\in\mathcal{T}}$. Finally, our estimator for $r_k$ is defined to be
\begin{equation}\label{eqn:Bootstrapped_rk}
\check{r}_k := \text{Mode of } \{\hat{r}_k^{(b)}(\hat{C})\}_{b\in[B]}.
\end{equation}
The intuition of $\hat{C}$ and $\check{r}_k$ is as follows. If there are $r_k$ factors for $\A_k$, then the first $r_k$ eigenvalues of $\wt\R_{y,b}^{(k)}$ for each $b\in[B]$ should be approximately well-separated. Setting a large $C$ will create a large threshold $1+CT^{-1/2}$ that is almost always lying in between $\lambda_{j}(\wt\R_{y,b}^{(k)})$ and $\lambda_{j+1}(\wt\R_{y,b}^{(k)})$ for some fixed $j\in[r_k]$ for each $b\in[B]$, so that $\wh\var(\{\hat{r}_k^{(b)}(C)\}_{b\in[B]})$ will be small, or even equals 0.

However, if $C$ is small such that $1+CT^{-1/2}$ is now in between $\lambda_{j}(\wt\R_{y,b}^{(k)})$ and $\lambda_{j+1}(\wt\R_{y,b}^{(k)})$ for some $j\in[d_k]/[r_k]$ and some $b\in[B]$, then we expect that this particular threshold will lie in between $\lambda_{j'}(\wt\R_{y,b}^{(k)})$ and $\lambda_{j'+1}(\wt\R_{y,b}^{(k)})$ for some $j'\neq j$ and some others $b\in[B]$, since all these eigenvalues are less than or equal to 1 by Theorem \ref{thm:population_eigenvalue_correlation_matrix}, and their variability is originated from the noise series only, making them less stable compared to when $j\in[r_k]$. Hence for a small enough $C$, we expect $\wh\var(\{\hat{r}_k^{(b)}(C)\}_{b\in[B]})$ to be large. The range of values of $C$ such that $1+CT^{-1/2}$ lies in between $\lambda_{r_k}(\wt\R_{y,b}^{(k)})$ and $\lambda_{r_k+1}(\wt\R_{y,b}^{(k)})$ for the majority of $b\in[B]$ will then include $\hat{C}$. The definition of $\check{r}_k$ in (\ref{eqn:Bootstrapped_rk}) allows for variability arises from the noises and the $r_k$-th factor which can be weak and hence may not be detected in all Bootstrap samples.

Finally, in all the simulation settings in Section \ref{sec:simulation} ,we use $B=50$ Bootstrap samples. This is a safe number, since reducing it to 10 in fact hardly change the results in our simulation experiments.

\section{Simulation Experiments}\label{sec:simulation}
In this section,
we conduct simulation experiments to compare the performances of our pre-averaging estimators (PRE) and iterative projection estimators (PROJ) to other state-of-the-art competitors. We also test the performance of our proposed rank estimators (BCorTH) with bootstrapping of tensor fibres for tuning parameter selection. A real data analysis is also carried out in Section \ref{subsec:realdata}.

\subsection{Simulation settings}\label{subsec:simulation_setting}
For generating our data, we use model (\ref{eqn:tensorfactormodel}), with elements in $\mu$ being i.i.d. standard normal in each repetition of experiment. For $k\in[K]$, each factor loading matrix $\A_k$ is generated independently with $\A_k = \B_k\R_k$, where the elements in $\B_k \in\mathbb{R}^{d_k\times r_k}$ are i.i.d. $U(u_1,u_2)$, and $\R_k \in \mathbb{R}^{r_k\times r_k}$ is diagonal with the $j$th diagonal element being $d_k^{-\zeta_{k,j}}$, $0\leq \zeta_{k,j}\leq 0.5$. Pervasive (strong) factors have $\zeta_{k,j}=0$, while weak factors have $0<\zeta_{k,j}\leq 0.5$.

The elements in $\cF_t$ are independent standardized AR(5) with AR coefficients 0.7,0.3,-0.4,0.2 and -0.1, and the corresponding innovation process is i.i.d. standard normal. Same for the elements in $\e_t^{(k)}$ and $\bepsilon_{t,\ell}^{(k)}$ in (\ref{eqn:unfoldeddecomposition}), but their AR coefficients are (-0.7,-0.3,-0.4,0.2,0.1) and (0.8,0.4, -0.4,0.2,-0.1) respectively. The matrix $\bPsi^{(1)} \in \mathbb{R}^{d_1\times r_e}$ is generated with i.i.d. standard normal entries, but has an independent probability of 0.7 being set exactly to 0. Each matrix $\bSigma_{\epsilon,\ell}^{(1)}$ has randomly generated eigenvectors with eigenvalues on $U(1,3)$. Each experiment is repeated 500 times.


We consider the simulation setting (I), (II) and (III), with sub-setting (a) and (b), detailed below:
\begin{itemize}
  \item[(Ia)] All strong factors with $\zeta_{k,j}=0$ for all $k,j$. $u_1 = -2$, $u_2 = 2$ so that columns of $\A_k$ sum to normal magnitude (small $s_k$).

\item[(IIa)]  One strong factor with $\zeta_{k,1}=0$ and $\zeta_{k,2}=0.2$ for all $k$. $u_1 = -2$, $u_2 = 2$.

\item[(IIIa)] Two weak factors with $\zeta_{k,1}=0.1$ and $\zeta_{k,2}=0.2$ for all $k$. $u_1 = -2$, $u_2 = 2$.

 \item[(Ib)] Same as (Ia), except that $u_1 = 0$, $u_2 = 2$ so that column sums of $\A_k$ have large magnitude (large $s_k$).

\item[(IIb)] Same as (IIa), except that $u_1 = 0$, $u_2 = 2$.

\item[(IIIb)] Same as (IIIa), except that $u_1 = 0$, $u_2 = 2$.
\end{itemize}

For all profiles above, we  set $r_k = 2$ for all $k$, and let $r_e = 10$. Each profile will be considered for the following four settings of different dimensions:
\begin{itemize}
  \item[i.] $K=2$, $T=100$, $d_1=d_2=40$;
  \item[ii.] $K=2$, $T=200$, $d_1=d_2=40$;
  \item[iii.] $K=2$, $T=200$, $d_1=d_2=80$;
  \item[iv.] $K=3$. $T=200$, $d_1=d_2=d_3=20$.

\end{itemize}

\subsection{Results for settings with $K=2$}\label{subsec:sim_K=2}
For our pre-averaging estimation procedure, we set $M_0 = 200$, $M = 5$ and $n_l = \frac{d_l}{2}$ (refer to Section \ref{subsec:pre-averaging} and Remark \ref{remark:population_parameters} for the meaning of these parameters. See also Section \ref{subsection:Setting_M_0_and_M}). We have tried to increase $M_0$ and $M$ 10 folds in all our settings, but there are no distinguishable improvements. We first calculate the pre-averaging estimator $\wh\Q_{k,pre}$ for each $k\in[K]$, and obtain $\check\q_k^{(0)} = \wh\q_{k,pre} = \wh\Q_{k,pre,(1)}$. Then calculate $\check\q_k^{(m)}$ according to the Algorithm for Iterative Projection Direction Refinement in Section \ref{subsec:refiningprojection} for $m=1,\ldots,29$. Finally, we obtain the iterative projection estimator $\check\U_k$ by utilising $\qmkcheck^{(29)}$ as the projection direction for each $k\in[K]$. We compare the pre-averaging estimators $\wh\Q_{k,pre}$ and the iterative projection estimator $\check\U_k$ with the TOPUP/TIPUP estimators in \cite{Chenetal2022} and the iTOPUP/iTIPUP iterative estimators in \cite{Hanetal2020}. We also compare to the Higher Order Singular Value Decomposition (HOSVD) and the Higher Order Orthogonal Iterations (HOOI) (see \cite{Shenetal2016} for example), although these two methods still lack theoretical guarantees for time series tensor factor models.  For estimation accuracy, we use the measure $\|\hat\Q_{k} \hat\Q_{k}^\T - \U_k \U_k^\T\|$, i.e. the spectral norm of the difference of the estimated and true underlying projection matrices, where $\U_k$ is obtained using the singular value decomposition of $\A_k$.

\begin{figure}[!htp]
    \hspace{-0.18in}
	\includegraphics[width=17cm]{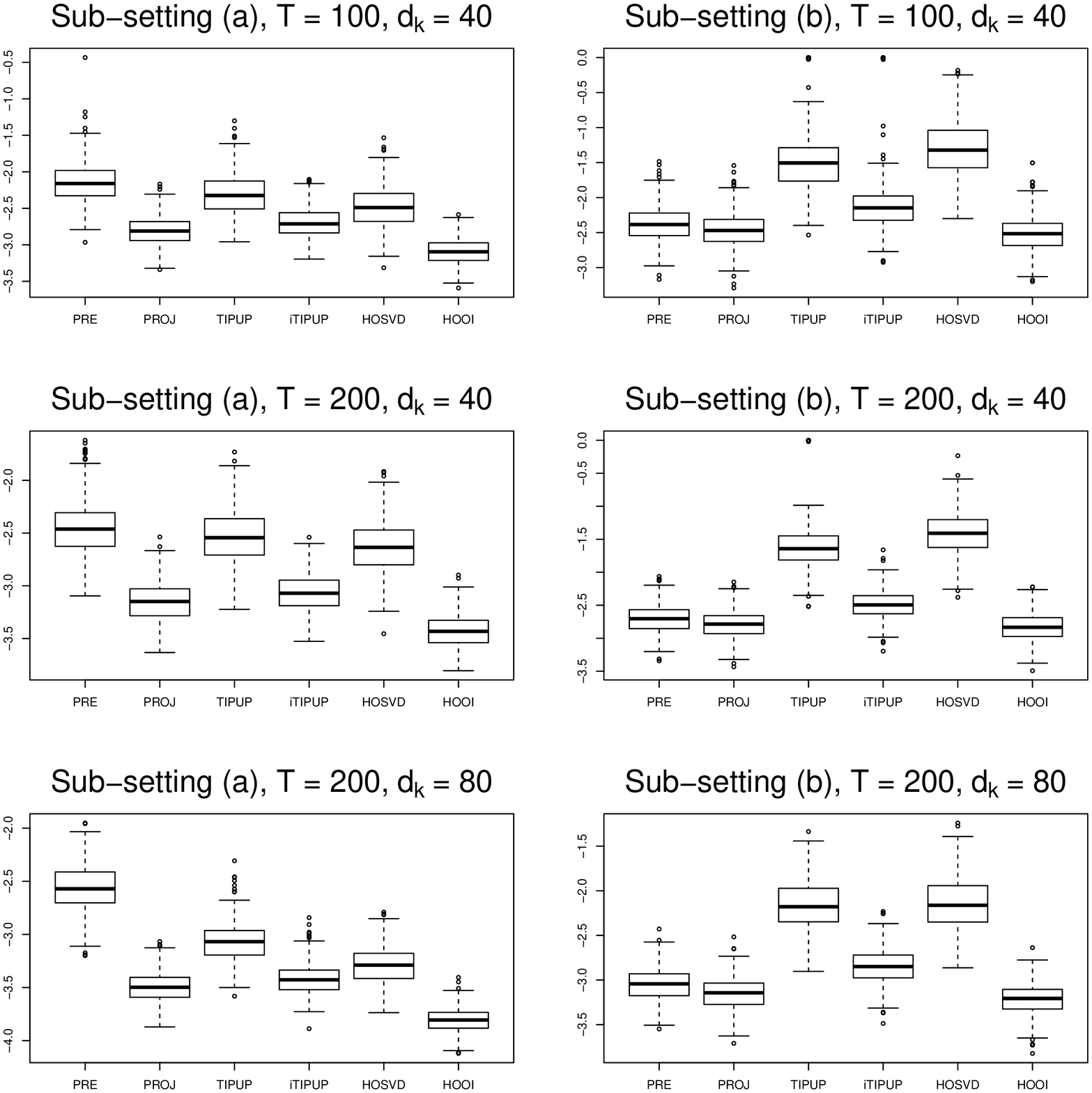}\\
	\caption{Plot of estimation error $\norm{\hat\Q_2\hat\Q_2^\T - \U_2\U_2^\T}$ (in log-scale) for Setting (I). {\em Left:} Sub-setting (a). {\em Right:} Sub-setting (b).}
	\label{fig:1}
\end{figure}

\begin{figure}[!htp]
    \hspace{-0.18in}
	\includegraphics[width=17cm]{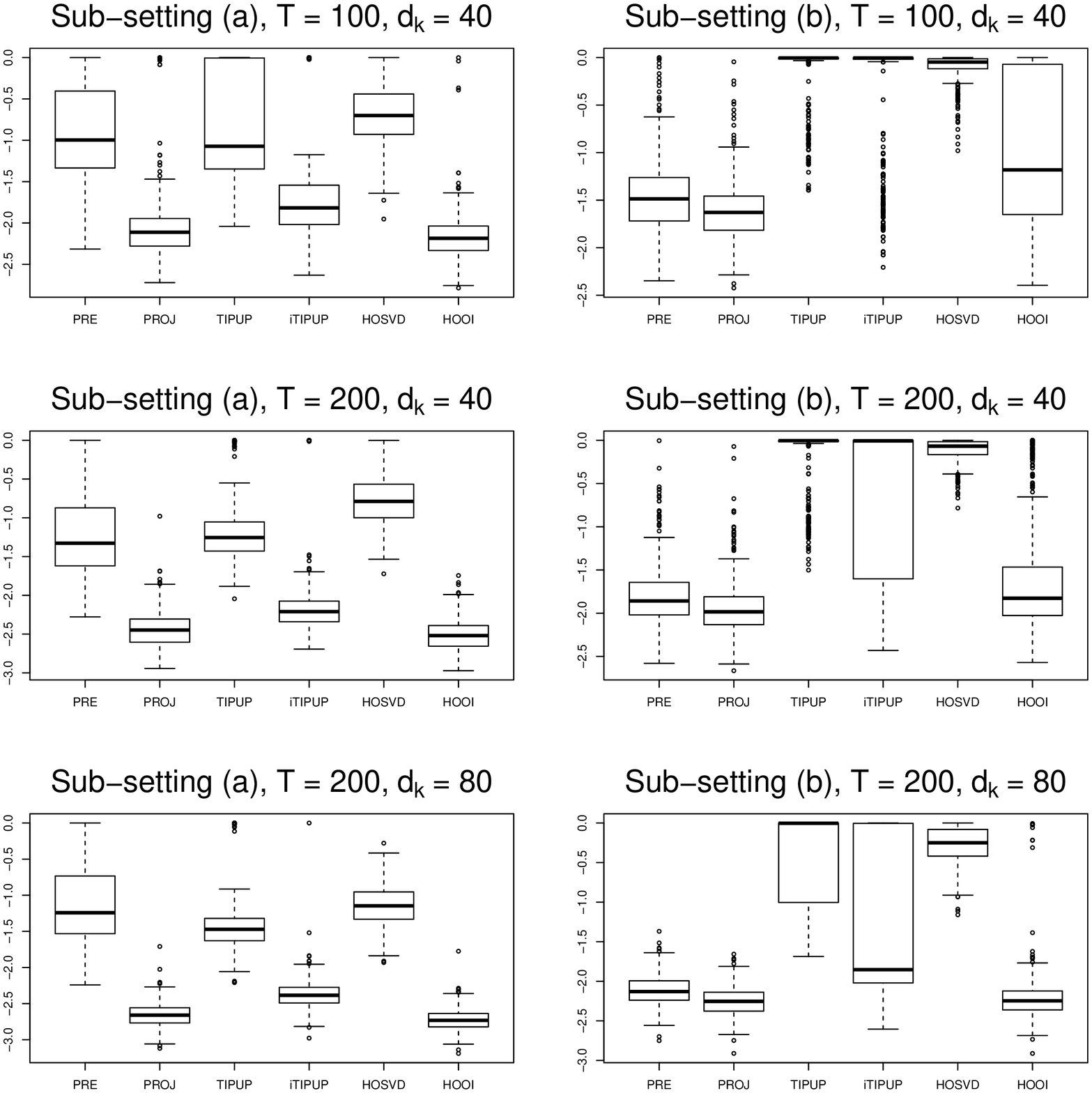}\\
	\caption{Plot of estimation error $\norm{\hat\Q_2\hat\Q_2^\T - \U_2\U_2^\T}$ (in log-scale) for Setting (II). {\em Left:} Sub-setting (a). {\em Right:} Sub-setting (b).}
	\label{fig:2}
\end{figure}

\begin{figure}[!htp]
    \hspace{-0.18in}
	\includegraphics[width=17cm]{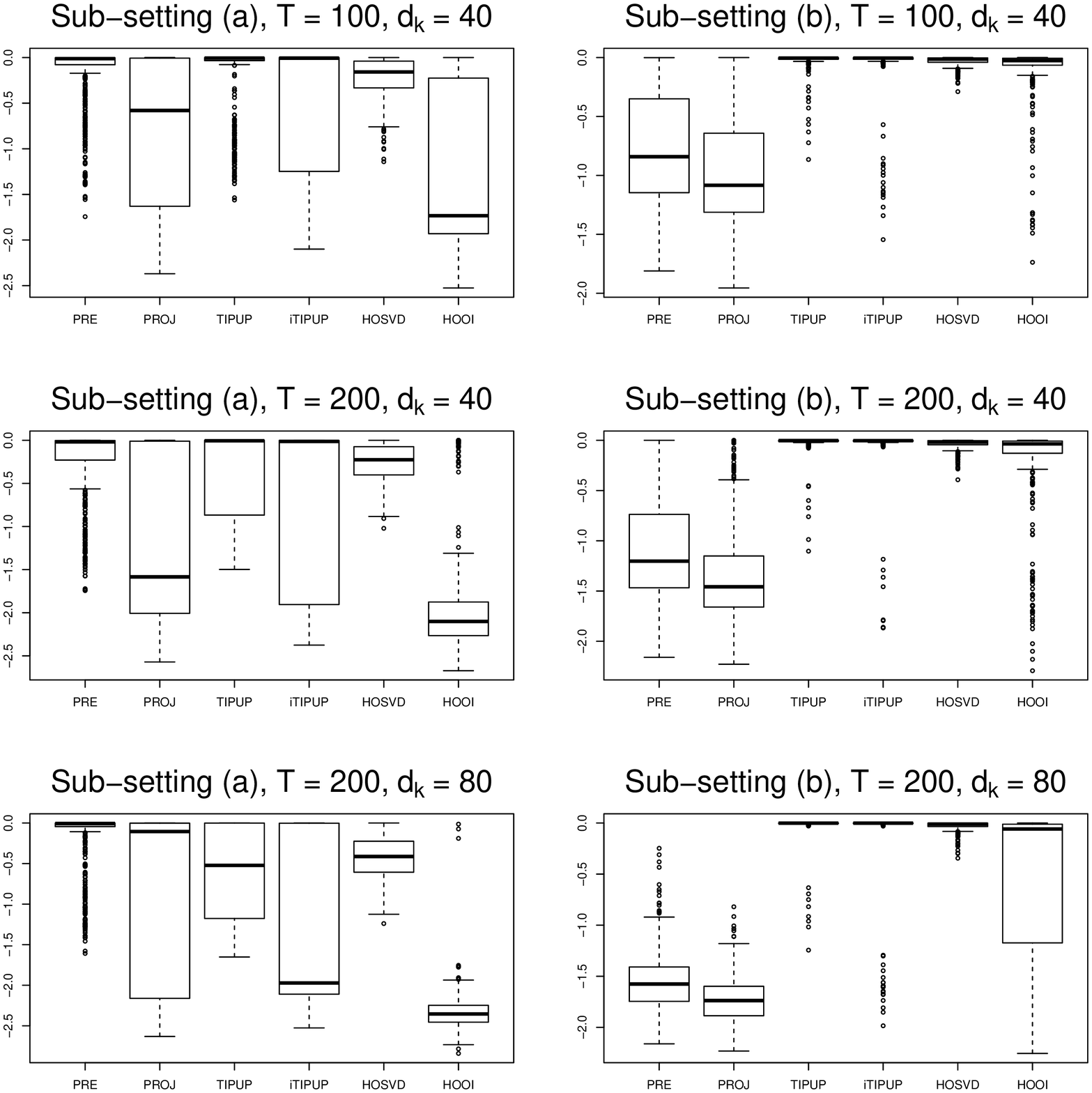}\\
	\caption{Plot of estimation error $\norm{\hat\Q_2\hat\Q_2^\T - \U_2\U_2^\T}$ (in log-scale) for Setting (III). {\em Left:} Sub-setting (a). {\em Right:} Sub-setting (b).}
	\label{fig:3}
\end{figure}

We omit the results for $\A_1$ and only display the estimation accuracy for $\A_2$ in Figures \ref{fig:1} to \ref{fig:3} since the results for $\A_1$ and $\A_2$ are very similar. We also omit comparing to TOPUP and iTOPUP in the end since their performances are far worse than TIPUP and iTIPUP in all settings.

From Figure \ref{fig:1}, when all factors are strong (pervasive), all estimators perform well, but HOOI outperforms all methods in sub-setting (a) when the elements of $\A_2$ are i.i.d. $U(-2,2)$, with our iterative projection estimator and iTIPUP follow closely. All methods improve in absolute terms also when $T$ or $d_1, d_2$ increase. The relative advantage of HOOI over the iterative projection estimator is largely removed under sub-setting (b) where the elements of $\A_2$ are i.i.d. $U(0,2)$. The main difference between sub-setting (a) and (b) is that the mean of each element of sub-setting (a) is 0 while it is non-zero in sub-setting (b). Whenever the mean of the elements are not 0, our methods, including the pre-averaging estimator, can take advantage since our methods are based in summing rows of $\A_1$ (for estimating $\A_2$) or $\A_2$ (for estimating $\A_1$).

The picture changes substantially when one of the pervasive factors are made weak factor for $\A_1$ and $\A_2$ in setting (II) (Figure \ref{fig:2}). In sub-setting (a), our iterative projection method is now as good as HOOI, which are both better than iTIPUP. In sub-setting (b), with our methods taking advantage of the non-zero mean of the elements in $\A_1$ and $\A_2$, the pre-averaging estimator and the iterative projection estimator are both performing better than the HOOI for $T=100,200$ and $d_k=40$, while our iterative projection estimator is still on par with HOOI when $d_k=80$. TIPUP and iTIPUP are now performing far worse compared to our estimators and HOOI.

The picture changes drastically again under setting (III) (Figure \ref{fig:3}), when all factors are weak. Both our iterative projection estimator and iTIPUP are now performing worse than HOOI, with large variances too, under sub-setting (a). Except for HOOI, all methods also fail to get better as $T$ or $d_k$ increase, showing the difficulty of the setting. However, under sub-setting (b), our iterative projection method now works the best and show descent performances under different $T$ and $d_k$, while all other methods are failing.

\begin{table}[ht]
\begin{center}
  \begin{tabular}{|c|c|c||c|c|c||c|c|c|}
  \hline
 \multicolumn{3}{|c||}{$ T = 100, \ \ d_1 = d_2 = 40$} & \multicolumn{3}{|c||}{$ T = 200, \ \ d_1 = d_2 = 40$} & \multicolumn{3}{|c|}{$ T = 200, \ \ d_1 = d_2 = 80$}\\
\hline
  Initial & Iterative & Total & Initial & Iterative & Total & Initial & Iterative & Total\\
\hline
 PRE & PROJ & & PRE & PROJ & & PRE & PROJ & \\
\hline
$0.71_{(2)}$ & $0.16_{(1)}$ & $0.87_{(2)}$ & $1.33_{(3)}$ & $0.29_{(2)}$ & $1.63_{(3)}$ & $4.06_{(9)}$ & $1.16_{(6)}$ & $5.22_{(11)}$\\
\hline
\hline
 TIPUP & iTIPUP & &  TIPUP & iTIPUP & & TIPUP & iTIPUP & \\
\hline
$0.08_{(1)}$ & $1.74_{(4)}$ & $1.83_{(4)}$ & $0.16_{(1)}$ & $3.27_{(5)}$ &  $3.43_{(5)}$ & $0.46_{(4)}$	& $4.89_{(11)}$	& $5.34_{(12)}$\\
\hline
\hline
HOSVD & HOOI & & HOSVD & HOOI & &HOSVD & HOOI & \\
 \hline
$0.09_{(1)}$ & $1.77_{(6)}$ & $1.86_{(6)}$ & $0.16_{(1)}$ & $3.33_{(8)}$ & $3.48_{(8)}$ & $0.41_{(2)}$ & $5.02_{(11)}$ & $5.43_{(11)}$
\\
\hline
\end{tabular}
\end{center}
\caption{Mean and SD (in bracket, presented as actual values multiplied by 100) of the run time (in seconds) for factor loading estimations under different methods and different dimensions with $K = 2$. }\label{table1}
\end{table}

Next we compare computational time for different methods. Table \ref{table1} shows the average time needed to compute the three different initial estimators, namely, our pre-averaging estimator (PRE), the TIPUP estimator, and the HOSVD. It also shows the average time needed to compute the iterative estimators corresponding to these initial estimators, namely, our iterative projection estimator (PROJ), the iTIPUP, and the HOOI. Under our different settings with $K=2$, we see that our pre-averaging estimator takes up the most computational time, while the iterative step is fast. The opposite is true for TIPUP/iTIPUP and HOSVD/HOOI, where their iterative steps are slow but take less time to calculate an initial estimator.

Finally, we compare the accuracy of our bootstrapped rank estimator in (\ref{eqn:Bootstrapped_rk}) to the iTIP-ER in \cite{Hanetal2022}. We abbreviate our method as BCorTh, which stands for Bootstrapped Correlation Thresholding. The projection vector used for constructing $\wt\bSigma_{y,m+1}^{(k)}$ is obtained from our iterative projection estimator as before, and the bootstrap is conducted 50 times. We run the simulation for 500 times for each setting, and report the mean and standard deviation of $\hat{r}_1$, $\hat{r}_2$, and the computational time for BCorTh and iTIP-ER. Table \ref{table2} shows the results. We can see that BCorTh generally outperforms iTIP-ER in most settings in terms of correct proportion. The only exception is Setting (IIIa), where iTIP-ER is better especially with large $T$ and $d_k$.
In general, iTIP-ER performs very poorly in Setting (IIb) and (IIIb), while BCorTh performs relatively poorly in Setting (IIIa).

Regarding the effect of $T$ and $d_k$, in Setting (I) and (II), both BCorTh and iTIP-ER give better performance when $T$ and $d_k$ increase. Setting (III) is different: In Setting (IIIa), increasing $T$ and $d_k$ largely enhances the performance of iTIP-ER, but not for BCorTh. In Setting (IIIb), increasing $T$ and $d_k$ largely enhances the performance of BCorTh, while it negatively affects the performance of iTIP-ER.

Also, in all settings except (IIIa), BCorTh and iTIP-ER generally tend to under-estimate the true number of factors. In Setting (IIIa), iTIP-ER tends to over-estimate $r_k$, while BCorTh also over-estimates $r_k$ when $d_k = 80$, but under-estimates $r_k$ with smaller $d_k$ and $T$.


\begin{table}[ht]
\begin{center}
  \begin{tabular}{|p{1.6cm}|c|c|c|c|c|c|c|c|c|}

\hline
 &  \multicolumn{2}{|c|}{BCorTh} & \multicolumn{2}{|c|}{iTIP-ER} & {BCorTh} & {iTIP-ER} & BCorTh & iTIP-ER\\
\hline
 Setting & $\hat{r}_1$ & $\hat{r}_2$ & $\hat{r}_1$ & $\hat{r}_2$  & \multicolumn{2}{|c|}{Correct Proportion}& \multicolumn{2}{|c|}{Run Time(s)}\\
\hline
 \multicolumn{9}{|c|}{$T=100, d_1=d_2=40$}\\
\hline
  (Ia) & $2.00_{(4)}$ & $2.01_{(8)}$ & $2.00_{(0)}$ & $2.00_{(0)}$ & $0.99$ & $1.00$  & $1.70_{(4)}$ & $3.79_{(7)}$
\\
\hline
 (Ib) & $1.99_{(10)}$ & $2.01_{(11)}$ & $1.68_{(47)}$ & $1.64_{(48)}$ & $0.98$ & $0.44$ & $1.68_{(3)}$ & $3.68_{(16)}$\\
\hline
 (IIa) & $1.98_{(16)}$ & $2.01_{(18)}$ & $1.65_{(48)}$ & $1.62_{(50)}$ & $0.96$ & $0.43$ & $1.67_{(3)}$ & $3.66_{(17)}$ \\
\hline
 (IIb) & $1.65_{(48)}$ & $1.89_{(33)}$ & $1.06_{(24)}$ & $1.03  _{(17)}$ & $0.59$ & $0.00$ & $1.66_{(3)}$ & $3.44_{(9)}$ \\
\hline
 (IIIa) & $1.55_{(53)}$ & $2.00_{(128)}$ & $3.28_{(330)}$ & $2.83_{(241)}$ & $0.30$ & $0.38$ & $1.65_{(4)}$ & $3.79_{(25)}$ \\
\hline
 (IIIb) & $1.19_{(39)}$ & $1.49_{(51)}$ & $2.10_{(269)}$ & $1.82_{(198)}$ & $0.10$ & $0.06$ & $1.66_{(3)}$ & $3.58_{(29)}$ \\

\hline
\multicolumn{9}{|c|}{$T=200, d_1=d_2=40$}\\
\hline
 (Ia) & $2.00_{(0)}$ & $2.01_{(8)}$ & $2.00_{(0)}$ & $2.00_{(0)}$ & $0.99$ & $1.00$  & $2.98_{(9)}$ & $7.20_{(10)}$
\\
\hline
 (Ib) & $2.00_{(6)}$ & $2.01_{(12)}$ & $1.86_{(35)}$ & $1.79_{(41)}$ & $0.98$ & $0.68$ & $2.96_{(6)}$ & $7.10_{(22)}$\\
\hline
 (IIa) & $1.99_{(8)}$ & $2.01_{(9)}$ & $1.86_{(34)}$ & $1.78_{(41)}$ & $0.99$ & $0.70$ & $2.96_{(5)}$ & $7.12_{(20)}$ \\
\hline
 (IIb) & $1.81_{(40)}$ & $1.97_{(21)}$ & $1.06_{(24)}$ & $1.04  _{(20)}$ & $0.77$ & $0.00$ & $2.93_{(4)}$ & $6.65_{(16)}$ \\
\hline
 (IIIa) & $1.64_{(51)}$ & $1.71_{(102)}$ & $2.61_{(239)}$ & $2.37_{(182)}$ & $0.39$ & $0.64$ & $2.95_{(6)}$ & $7.19_{(31)}$ \\
\hline
 (IIIb) & $1.25_{(43)}$ & $1.73_{(46)}$ & $1.36_{(185)}$ & $1.24_{(121)}$ & $0.19$ & $0.01$ & $2.97_{(5)}$ & $6.69_{(27)}$ \\
\hline
\multicolumn{9}{|c|}{$T=200, d_1=d_2=80$}\\
\hline
 (Ia) & $2.00_{(0)}$ & $2.01_{(1)}$ & $2.00_{(0)}$ & $2.00_{(0)}$ & $0.99$ & $1.00$  & $14.78_{(17)}$ & $9.92_{(12)}$
\\
\hline
 (Ib) & $2.00_{(4)}$ & $2.01_{(11)}$ & $2.00_{(6)}$ & $1.98_{(14)}$ & $0.99$ & $0.98$ & $14.62_{(32)}$ & $9.93_{(19)}$\\
\hline
 (IIa) & $2.00_{(0)}$ & $2.01_{(9)}$ & $1.92_{(27)}$ & $1.85_{(36)}$ & $0.99$ & $0.79$ & $14.81_{(19)}$ & $9.91_{(24)}$ \\
\hline
 (IIb) & $1.99_{(8)}$ & $2.01_{(9)}$ & $1.12_{(33)}$ & $1.07  _{(26)}$ & $0.99$ & $0.01$ & $14.80_{(63)}$ & $9.29_{(25)}$ \\
\hline
 (IIIa) & $2.72_{(107)}$ & $2.30_{(110)}$ & $2.05_{(161)}$ & $2.00_{(96)}$ & $0.38$ & $0.89$ & $14.88_{(62)}$ & $10.10_{(49)}$ \\
\hline
 (IIIb) & $1.77_{(43)}$ & $2.00_{(18)}$ & $1.05_{(21)}$ & $1.04_{(21)}$ & $0.74$ & $0.00$ & $14.89_{(62)}$ & $9.25_{(19)}$ \\
\hline

\end{tabular}
\end{center}
\caption{Rank estimation results under different settings and dimensions with $K = 2$. The columns $\hat{r}_1$ and $\hat{r}_2$ report the mean and SD (in bracket, presented as actual values multiplied by 100)
of the estimates $\hat{r}_1$ and  $\hat{r}_2$ respectively under different methods. Correct proportion counts the proportion
of $(\hat{r}_1,\hat{r}_2)$ = $(2,2)$. Run time is the mean and SD (actual values multiplied by 100) of the run time in seconds.}\label{table2}
\end{table}

\subsection{Results for settings with $K=3$}\label{subsec:results_K=3}
In addition to the previous simulation experiments, we run Settings (Ia) to (IIIb) with $K = 3$, $T = 200$, $d_1 = d_2 = d_3 = 20$. Figure \ref{fig:7} to \ref{fig:9} show the logarithm of estimation errors under different settings. From the results, HOOI performs the best in all settings. Our iterative projection estimator is better than iTIPUP in all but Setting (Ia) and (IIIa).

Table \ref{table:time_K3} records the average run time for factor loading estimators. We can see that PRE+PROJ is still the fastest method in this scenario.

\begin{table}[ht]
\begin{center}
  \begin{tabular}{|c|c|c|c|c|c|c|c|c|c|c|}

\hline
PRE & PROJ & Total & & TIPUP & iTIPUP & Total &  & HOSVD & HOOI & Total \\
\hline
 $5.37_{(19)}$ & $1.57_{(8)}$ & $6.94_{(26)}$ & & $0.57_{(7)}$ & $7.18_{(17)}$ & $7.75_{(19)}$ & & $0.47_{(3)}$ & $7.16_{(13)}$ & $7.64_{(14)}$
\\
\hline

\end{tabular}
\end{center}
\caption{Mean and SD (in bracket, presented as actual values multiplied by 100) of the run time (in seconds) for factor loading estimations under different methods with $K =3$, $T = 100, d_1=d_2=d_3 = 20$.}\label{table:time_K3}
\end{table}

Table \ref{table3} records the rank estimation results for the above settings. For $K = 3$, we set the number of bootstrap $B=10$ to save time, since $B = 50$ requires too much computational time as $\dmk$ is large. We have tested that reducing $B$ from 50 to 10 does not significantly change the results of BCorTh.

Comparing BCorTh and iTIP-ER, we can observe that in terms of correct proportions, BCorTh is more superior than iTIP-ER in Setting (IIb) and (IIIb), while iTIP-ER is better than BCorTh in Setting (IIIa). They perform similarly in all other settings.

\begin{table}[ht]
\begin{center}
  \begin{tabular}{|c|c|c|c|c|c|c|c|c|c|}

\hline
  \multicolumn{3}{|c|}{BCorTh} & \multicolumn{3}{|c|}{iTIP-ER} & {BCorTh} & {iTIP-ER} & BCorTh & iTIP-ER\\
\hline
 $\hat{r}_1$ & $\hat{r}_2$ & $\hat{r}_3$ & $\hat{r}_1$ & $\hat{r}_2$ & $\hat{r}_3$  & \multicolumn{2}{|c|}{Correct Proportion}& \multicolumn{2}{|c|}{Run Time(s)}\\
\hline
  \multicolumn{10}{|c|}{Setting (Ia) (first row) and (Ib) (second row)}\\
\hline
 $2.00_{(0)}$ & $2.00_{(0)}$ & $2.00_{(0)}$ & $2.00_{(0)}$ & $2.00_{(0)}$ & $2.00_{(0)}$ & $1.00$ & $1.00$  & $21.1_{(5)}$ & $15.4_{(3)}$
\\
\hline
 $2.00_{(6)}$ & $2.00_{(4)}$ & $2.00_{(4)}$ & $2.00_{(0)}$ & $2.00_{(0)}$ & $2.00_{(0)}$ & $0.99$ & $1.00$  & $22.1_{(5)}$ & $15.3_{(4)}$ \\
\hline
  \multicolumn{10}{|c|}{Setting (IIa) (first row) and (IIb) (second row)}\\
\hline
 $2.00_{(0)}$ & $2.00_{(0)}$ & $2.00_{(0)}$ & $2.00_{(0)}$ & $2.00_{(0)}$ & $2.00_{(0)}$ & $1.00$ & $1.00$  & $20.9_{(5)}$ & $15.4_{(3)}$ \\
\hline
 $1.99_{(9)}$ & $1.99_{(12)}$ & $1.98_{(14)}$ & $1.98_{(15)}$ & $1.95_{(21)}$ & $1.95_{(23)}$ & $0.96$ & $0.88$  & $22.1_{(5)}$ & $12.2_{(2)}$ \\
\hline
  \multicolumn{10}{|c|}{Setting (IIIa) (first row) and (IIIb) (second row)}\\
\hline
 $2.01_{(19)}$ & $2.04_{(21)}$ & $2.03_{(24)}$ & $2.00_{(0)}$ & $2.00_{(0)}$ & $2.00_{(0)}$ & $0.92$ & $1.00$  & $21.0_{(5)}$ & $15.4_{(3)}$ \\
\hline
 $1.99_{(8)}$ & $2.00_{(4)}$ & $2.00_{(10)}$ & $1.89_{(32)}$ & $1.88_{(33)}$ & $1.86_{(34)}$ & $0.98$ & $0.66$  & $21.1_{(5)}$ & $12.1_{(6)}$ \\
\hline

\end{tabular}
\end{center}
\caption{Rank estimation results under different settings with $K = 3$, $T = 200$, $d_1 = d_2 = d_3 = 20$. The columns $\hat{r}_1$, $\hat{r}_2$ and $\hat{r}_3$ report the mean and SD (in bracket, presented as actual values multiplied by 100)
of the estimates $\hat{r}_1$, $\hat{r}_2$ and $\hat{r}_3$ respectively under different methods. Correct proportion counts the proportion
of $(\hat{r}_1,\hat{r}_2, \hat{r}_3) = (2,2,2)$. Run time is the mean and SD (actual values multiplied by 10) of the run time in seconds.}\label{table3}
\end{table}


\begin{center}

\begin{figure}[!htp]
    \hspace{-0.18in}
	\includegraphics[width=17cm]{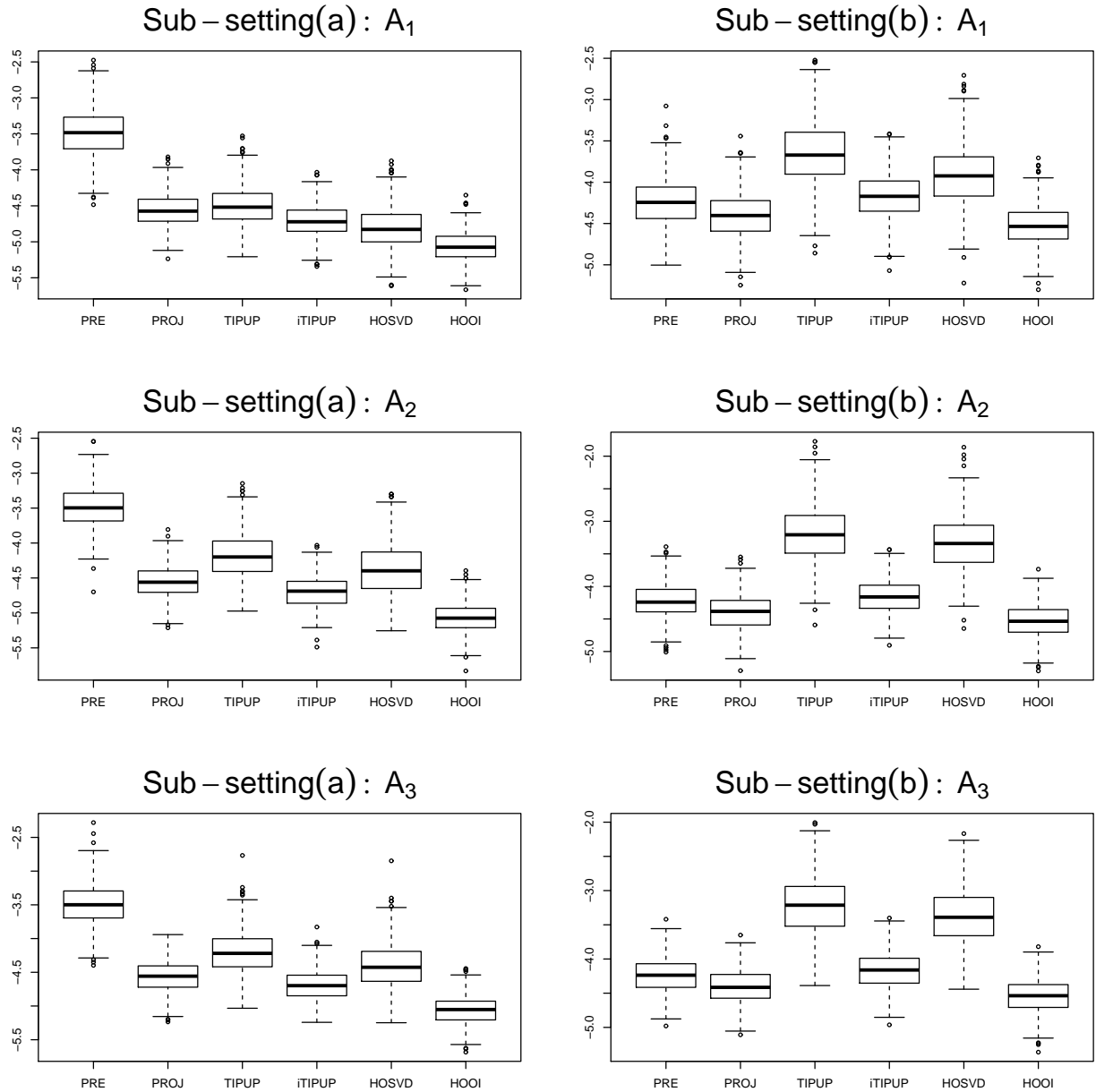}\\
	\caption{Plot of logarithm of estimation errors under Setting (I) with $K = 3$. Left column: sub-setting (a). Right column: sub-setting (b). Upper row: $k = 1$. Middle row: $k = 2$. Lower row: $k = 3$.}
	\label{fig:7}
\end{figure}
\begin{figure}[!htp]
    \hspace{-0.18in}
	\includegraphics[width=17cm]{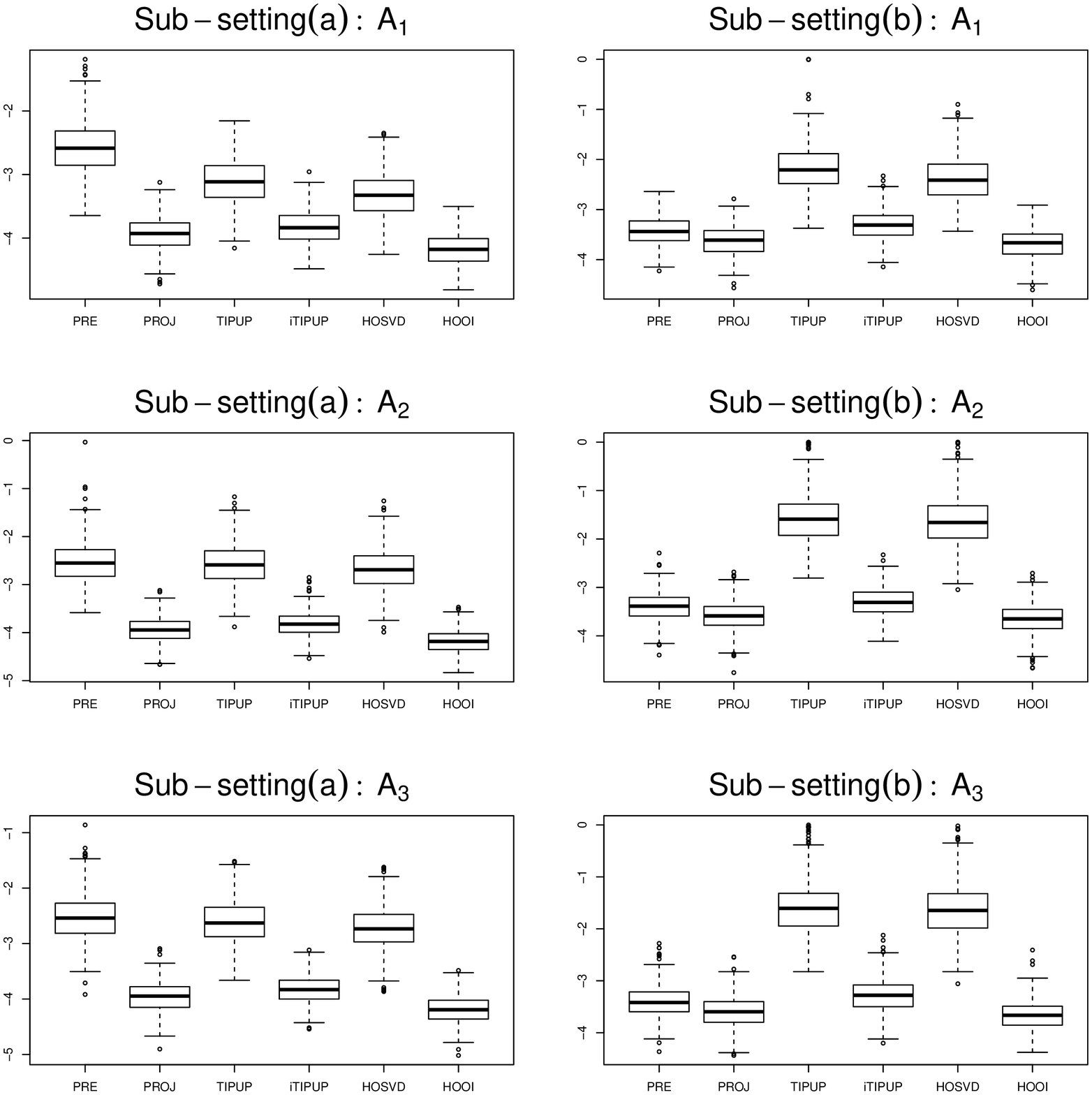}\\
	\caption{Plot of logarithm of estimation errors under Setting (II) with $K = 3$. Refer to Figure \ref{fig:7} for explanations.}
	\label{fig:8}
\end{figure}
\begin{figure}[!htp]
    \hspace{-0.18in}
	\includegraphics[width=17cm]{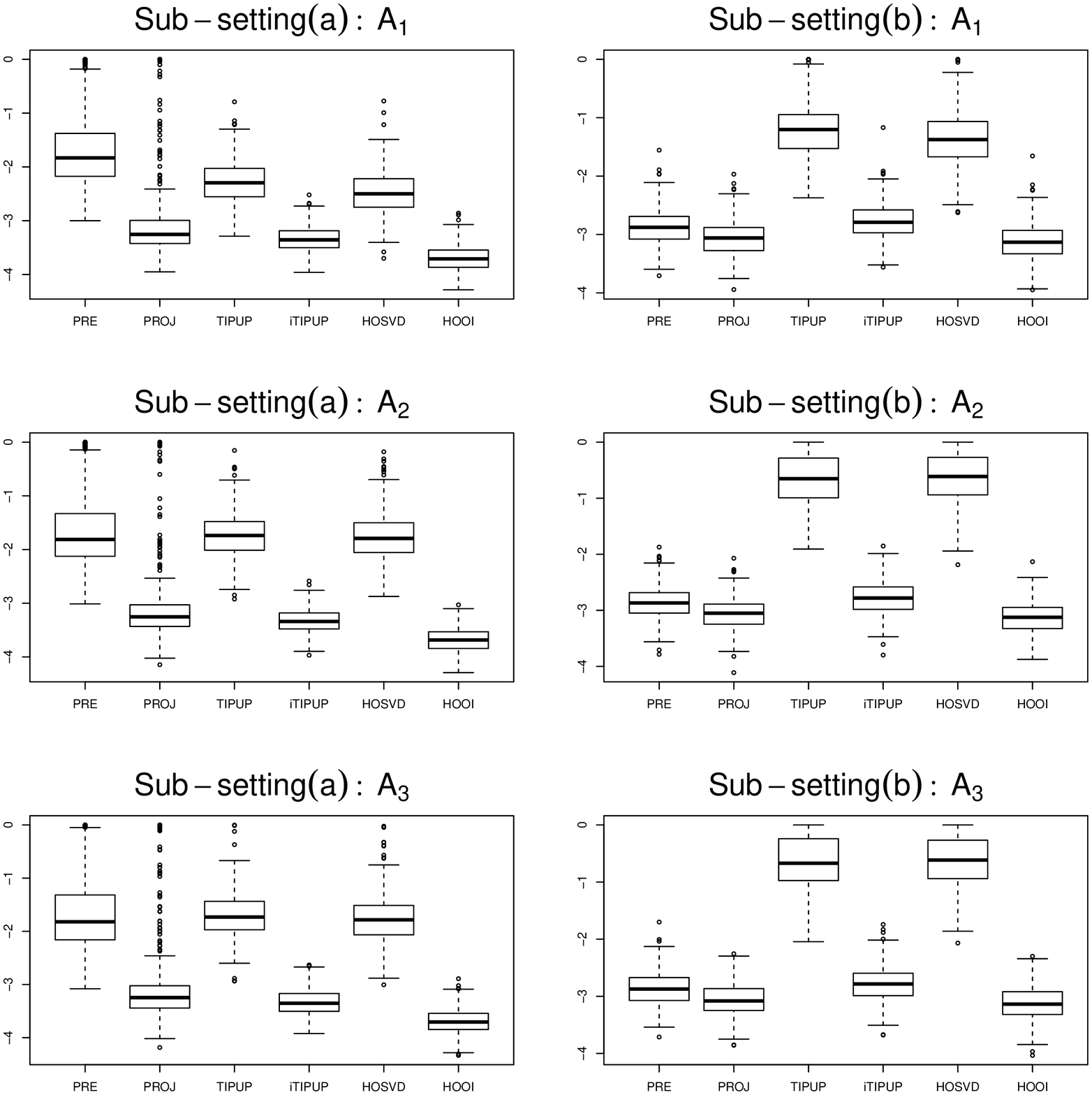}\\
	\caption{Plot of logarithm of estimation errors under Setting (III) with $K = 3$. Refer to Figure \ref{fig:7} for explanations.}
	\label{fig:9}
\end{figure}

\end{center}

\subsection{Real data analysis}\label{subsec:realdata}
We analyze a set of Fama-French portfolio returns data formed on size and operating profitability. Stocks are categorized into 10 different sizes (market equity, using NYSE market equity deciles) and 10 different operating profitability (OP) levels (using NYSE OP deciles. OP is annual revenues minus cost of goods sold, interest expense, and selling, general, and administrative expenses divided by book equity for the last fiscal year end). These levels, and hence the stocks in each category, are allocated at the end of June each year.  Moreover, the stocks in each of the $10\times 10$ categories form  exactly two portfolios, one being value weighted, and the other of equal weight.
Hence, there are two sets of $10\times 10$ portfolios with their time series of returns observed. We use monthly data from July 1973 to June 2021, so that $T=576$, and each data tensor we have thus has size $10\times 10\times 576$. For more details, please visit

\href{https://mba.tuck.dartmouth.edu/pages/faculty/ken.french/Data_Library/det_100_port_szme_op.html}
{https://mba.tuck.dartmouth.edu/pages/faculty/ken.french/Data\_Library/det\_100\_port\_szme\_op.html}

Since the market factor is certainly pervasive in financial returns, we use the following CAPM to remove its effects and facilitate detection of potentially weaker factors:
\[ \y_t = \bar{\y} + \bbeta(x_t - \bar{x}) + \e_t, \]
where $\y_t \in \mathbb{R}^{100}$ contains the returns of the 100 portfolios at time $t$, $x_t$ is the return for the NYSE composite index at time $t$, and $\bbeta = (\beta_1,\ldots,\beta_{100})^\T$ is the vector of $\beta$'s for the 100 portfolios. Least squares estimation leads us to
\[\wh\bbeta = \frac{\sum_{t=1}^T(x_t-\bar{x})(\y_t-\bar{\y})}{\sum_{t=1}^T(x_t-\bar{x})^2}.\]
Hence the data we analyze is $\{\y_t - \bar{\y} - \wh\bbeta(x_t - \bar{x})\}_{t=1,\ldots,576}$, with each observed vector reshaped into a $10\times 10$ tensor.

\begin{table}[ht]
\begin{center}
  \begin{tabular}{|c|c|c|c|c|c|c|c|c||c|c|c|c|c|c|c|c|}

\hline
& \multicolumn{8}{c||}{Value Weighted} & \multicolumn{8}{c|}{Equal Weight} \\
\hline
& \multicolumn{2}{c|}{$h_0 = 1$} & \multicolumn{2}{c|}{$h_0 = 2$}& \multicolumn{2}{c|}{$h_0 = 3$}& \multicolumn{2}{c||}{$h_0 = 4$}& \multicolumn{2}{c|}{$h_0 = 1$}& \multicolumn{2}{c|}{$h_0 = 2$}& \multicolumn{2}{c|}{$h_0 = 3$}& \multicolumn{2}{c|}{$h_0 = 4$}\\
\hline
Method & $\hat{r}_1$  & $\hat{r}_2$  & $\hat{r}_1$  & $\hat{r}_2$  & $\hat{r}_1$  & $\hat{r}_2$  & $\hat{r}_1$  & $\hat{r}_2$  & $\hat{r}_1$  & $\hat{r}_2$  & $\hat{r}_1$  & $\hat{r}_2$  & $\hat{r}_1$  & $\hat{r}_2$  & $\hat{r}_1$  & $\hat{r}_2$\\
\hline
\hline

iTIP-ER & 2& 2 & 3 & 3 & 2 &3 &2 &3 & 2 & 2 & 4 &3 &3 &3 &2 &3 \\
\hline
BCorTh  & 2& 2 & - & - & - & -& - & -& 2 & 2 & - & -& - & -& - & -\\
\hline

\end{tabular}
\end{center}
\caption{Rank estimators for Fama-French Portfolios.}\label{table4}
\end{table}

Table \ref{table4} shows that both  BCorTh and the iTIP-ER (using $h_0=1$) return $\hat r_1 = \hat r_2 = 2$ for both value weighted and equal weight portfolios, although adding more lags (larger $h_0$) for iTIP-ER results in different factors in general. Using $(\hat{r}_1, \hat{r}_2) = (2,2)$, we estimate the factor loading spaces by our iterative projection method and compared to iTIPUP and HOOI. Similar to \cite{Wangetal2019}, we show the estimated loading matrices after a varimax rotation that maximizes the variance of the squared factor loadings, scaled by 30 for a cleaner view. To save space, we only show the results for value weighted portfolios in Table \ref{table5} and \ref{table6}.

\begin{table}[ht]
\begin{center}
  \begin{tabular}{|c|c||cccccccccc|}

\hline
Method & Factor & OP1 & OP2 & OP3 & OP4 & OP5 & OP6 & OP7 & OP8 & OP9 & OP10\\
\hline
\hline
\multirow{2}{*}{PROJ} & 1 & 8 & -3 & -6 & -7 & -8 & \red{-9} & \red{-12} & \red{-11} & \red{-12} & \red{-14}\\
\cline{2 - 12}
 & 2& \red{26} & \red{10} & 7 & 5 &4 &3 &1 &2 &1 &0 \\
\hline
\hline
\multirow{2}{*}{iTIPUP} & 1 & \red{-25} & \red{-14} & \red{-9} & -1 & -2 & -1 & 0 & -3 & 1 & -2 \\
\cline{2 - 12}
 & 2& \red{12} & \red{-13} & -5 & -8 & -7 & -8 & \red{-10} & \red{-11} & \red{-9} & \red{-9} \\
\hline
\hline
\multirow{2}{*}{HOOI} & 1 & \red{28} & 8 & 4 & 2 & 1 & 0 & 0 & 0 & 0 & 2\\
\cline{2 - 12}
 & 2& 5 & -5 & -7 & \red{-9} & \red{-10} & \red{-10} & \red{-11} & \red{-11} & \red{-12} & \red{-12} \\
\hline
\hline
\end{tabular}
\end{center}
\caption{OP Factor Loading Matrices for Value Weighted Portfolios after rotation and scaling. Magnitudes larger than 8 are highlighted in red.}\label{table5}
\end{table}

\begin{table}[!ht]
\begin{center}
  \begin{tabular}{|c|c||cccccccccc|}

\hline
Method & Factor & S1 & S2 & S3 & S4 & S5 & S6 & S7 & S8 & S9 & S10\\
\hline
\hline
\multirow{2}{*}{PROJ} & 1 & \red{-11} & \red{-11} & \red{-11} & \red{-10} & \red{-11} & \red{-9} & \red{-9} & -6 & -4 & \red{10} \\
\cline{2 - 12}
 & 2& -2 & 1 & 0 & 3 &4 &6 &6 &8 &\red{10} & \red{25} \\
\hline
\hline
\multirow{2}{*}{iTIPUP} & 1 & \red{-11} & \red{-13} & \red{-13} & \red{-13} & \red{-11} & \red{-9} & -3 & -2 & 0 & \red{9} \\
\cline{2 - 12}
 & 2& 6 & 2 & -2 & -2 & -5 & \red{-9} & \red{-12} & \red{-12} & \red{-14} & \red{-16} \\
\hline
\hline
\multirow{2}{*}{HOOI} & 1 & \red{-14} & \red{-15} & \red{-12} & \red{-10} & \red{-9} & -6 & -4 & -1 & 2 & \red{9}\\
\cline{2 - 12}
 & 2& 5 & 1 & -2 & -5 & -7 & \red{-10} & \red{-11} & \red{-13} & \red{-14} & \red{-15} \\
\hline
\hline
\end{tabular}
\end{center}
\caption{Size Factor Loading Matrices for Value Weighted Portfolios after rotation and scaling. Magnitudes larger than 8 are highlighted in red.}\label{table6}
\end{table}

From Table \ref{table5}, we can see that from PROJ and HOOI, OP1 itself (possibly OP2 as well) forms one group while OP6 to OP10 form another. iTIPUP also gives a clear grouping effect but is a bit different from the other two methods. Nevertheless, we can say that OP1 and OP2 possibly represents ``low operating profitability'', while OP6 to OP10 are ``high operating profitability'', and are governed by different factors. For Size, from the three methods, S1 to S5 form one group (``small size'') while another group contains at least S9 and S10 (``large size''), possibly S6 to S8 as well.

\section{Appendix: Basic Tensor Manipulations}\label{sec:BasicTensorManipulations}
We briefly introduce the notations and review on tensor manipulations in this section just enough for the presentation of this paper. For more information, please refer to \cite{KolderBader2009}.

Let $\cX \in \mathbb{R}^{d_1\times\cdots\times d_K}$ be an order-$K$ tensor. Here $K$ represents the number of dimensions in $\cX_t$, also called the number of {\em modes}. For instance, a vector time series has $K=1$ while a matrix time series has $K=2$. If we write $\cX = (x_{i_1\cdots i_K})$, then we define a {\em mode-$k$ fibre} of $\cX$ to be a column vector (of length $d_k$)
\begin{align*}
 (x_{i_1\cdots i_{k-1},j,i_{k+1}\cdots i_{K}})_{j\in[d_k]}, \;\;\; i_\ell \in [d_\ell] \;\text{ with } \ell\in[K],
\end{align*}
where we define, for any positive integer $n$, $[n]:= \{1,\ldots,n\}$. Hence there are in total $\dmk := \prod_{\ell=1\;;\ell \neq k}^Kd_\ell$ number of mode-$k$ fibres for the tensor $\cX$. The {\em mode-$k$ unfolding matrix} $\X_{(k)} \in \mathbb{R}^{d_k\times \dmk}$ (also denoted as $\mat_k(\cX)$ in this paper) is then defined to be the matrix containing (in order) all the mode-$k$ fibres of $\cX$.

If there is a matrix $\A = (a_{ij}) \in \mathbb{R}^{I_k\times r_k}$, $k\in[K]$, and $\cF = (f_{i_1\cdots i_K}) \in \mathbb{R}^{r_1\times\cdots r_K}$ is an order-$K$ tensor,
then the {\em $k$-mode product} of $\cF$ and $\A$, denoted by $\cF \times_k \A \in \mathbb{R}^{r_1\times\cdots\times r_{k-1}\times I_k\times r_{k+1} \times \cdots \times r_K}$, is defined as
\begin{align*}
  (\cF \times_k \A)_{i_1\cdots i_{k-1},j,i_{k+1}\cdots i_K} :=  \sum_{i_k=1}^{r_k} f_{i_1\cdots i_{k} \cdots i_K}a_{ji_k}.
\end{align*}
As an example, consider $K=2$, so that $\cF$ is a matrix. Let $\A_1 \in \mathbb{R}^{I_1\times r_1}$ and $\A_2 \in \mathbb{R}^{I_2\times r_2}$.  Then the mode-1 fibres of $\cF$ are in fact the columns of $\cF$, while the mode-2 fibres of $\cF$ are the rows of $\cF$ (made column vectors). We also have
\[\cF\times_1\A_1 = \A_1\cF, \;\;\; \cF\times_2\A_2 = \cF\A_2^\T = (\A_2\cF^\T)^\T.\]
Note that the two unfolded matrices (mode-1 and mode-2 respectively) are $\F_{(1)} = \cF$ and $\F_{(2)} = \cF^\T$, hence the above also means that
\[\cF\times_1\A_1 = \A_1\F_{(1)}, \;\;\; \cF\times_2\A_2 = (\A_2\F_{(2)})^\T.\]
In general, to calculate $\cF \times_k \A$, we find the mode-$k$ unfolding matrix $\F_{(k)}$ first and then calculate $\A\F_{(k)}$, which contains all mode-$k$ fibres of $\cF$ being pre-multiplied by $\A$. Then we put the columns in $\A\F_{(k)}$ back into the original shape of the tensor $\cF$. Hence for $K=2$, to put the columns in $\A\F_{(2)}$ back into the original orientation of the tensor (rows), we take transpose of it, so that $\cF \times_2 \A = (\A\F_{(2)})^\T$.
The order of distinct mode products does not matter, in the sense that for $i\neq j$,
\[\cF \times_i \A_i \times_j \A_j = \cF \times_j \A_j \times_i \A_i.\]

Finally, if $\cC = \cF\times_1\A_1\times_2\cdots\times_K\A_K$, then we have the formula
\begin{equation*}
\C_{(k)} = \A_k\F_{(k)}\A_{\text{-}k}^\T,
\end{equation*}
where $\otimes$ is the Kronecker product, $\A_{\text{-}k} := \A_K\otimes\cdots\otimes\A_{k+1}\otimes\A_{k-1}\otimes\cdots\otimes\A_1$, and $\C_{(k)}$ is the mode-$k$ unfolding of $\cC$. Hence, denote by $\vec(\cX)$ the vectorization of $\cX$, which is to stack all model-1 fibres of $\cX$ into a column vector, we have
\[\vec(\C_{(k)}) = (\A_{\text{-}k}\otimes \A_k)\vec(\F_{(k)}).\]
In particular,
\begin{equation*}
\vec(\cC) = \vec(\C_{(1)}) = (\A_{\text{-}1}\otimes\A_1)\vec(\F_{(1)}) = (\A_{K}\otimes\cdots\otimes\A_1)\vec(\cF).
\end{equation*}

\section{Appendix: Proof}\label{sec:Proofs}
\setcounter{equation}{0}
We first present three important lemmas under our model assumptions.

\begin{lemma}\label{lem:1}
Under Assumption (E1), (E2), (L1), (L2), (R1), we have
     \begin{align}\label{eqn:largest_eigenvalue_bound_error}
         \lambda_1 \left(\frac{\ddot{\E}_k^\T \ddot{\E}_k}{T} \right) = O_p\left(\frac{d_{\text{-}k}}{s_{\text{-}k}} \left(1 + \frac{d_k}{T} \right) \right), \\
         \label{eqn:largest_eigenvalue_bound_error_nonzeromu}
         \lambda_1 \left(\frac{\ddot{\E}_k^\T \left( \I_T - \frac{1}{T} \1_T \1_T^\T \right) \ddot{\E}_k}{T} \right) = O_p\left(\frac{d_{\text{-}k}}{s_{\text{-}k}} \left(1 + \frac{d_k}{T} \right) \right).
    \end{align}
     In addition, if Assumption (R2) is satisfied, then
     \begin{align}\label{eqn:smallest_eigenvalue_bound_error}
         \mathbb{P} \left( \lambda_{\lfloor c\min{(T,d_k)} \rfloor}\left(\frac{\ddot{\E}_k^\T \ddot{\E}_k}{T} \right) \geq C\left(\frac{d_{\text{-}k}}{s_{\text{-}k}} \left(1 + \frac{d_k}{T}\right)\right)\right)= 1, \\
        \label{eqn:smallest_eigenvalue_bound_error_nonzeromu}
         \mathbb{P} \left( \lambda_{\lfloor c\min{(T,d_k)} \rfloor}\left(\frac{\ddot{\E}_k^\T \left( \I_T - \frac{1}{T} \1_T \1_T^\T \right) \ddot{\E}_k}{T} \right) \geq C\left(\frac{d_{\text{-}k}}{s_{\text{-}k}} \left(1 + \frac{d_k}{T}\right)\right)\right)= 1,
     \end{align}
     for some $c\in (0,1]$ and $C >0$.
\end{lemma}

In Lemma \ref{lem:1}, (\ref{eqn:largest_eigenvalue_bound_error}) provides an upper bound for the largest eigenvalue of $\ddot{\E}_k^\T \ddot{\E}_k$ (and  $\ddot{\E}_k^\T\left( \I_T - \frac{1}{T} \1_T \1_T^\T \right) \ddot{\E}_k$), which facilitates the proof of Theorem \ref{thm:sumestimatorrate_newmethod}.  (\ref{eqn:smallest_eigenvalue_bound_error}) suggests that at least $\lfloor c\min{(T,d_k)} \rfloor$ largest eigenvalues of $\ddot{\E}_k^\T \ddot{\E}_k$ (and  $\ddot{\E}_k^\T\left( \I_T - \frac{1}{T} \1_T \1_T^\T \right) \ddot{\E}_k$) are of the same order, which guarantees the validity of using eigenvalue ratio to detect the existence of factors, which will be further discussed in Remark \ref{remark:eigenvalue_ratio}.

\begin{lemma}\label{lem:2}
Under Assumption (E1), (E2), (F1), (L1), (L2), (R1), we have
\begin{align*}
    \frac{\ddot{\F}_k \ddot{\F}_k^\T}{T} \rightarrow \Sigma_{F,k}
\end{align*}
for some positive definite matrix $\Sigma_{F,k}$, with all eigenvalues bounded away from 0 and infinity. For $j \in [r_k]$,
\begin{align}\label{eqn:lemma2.1}
    \lambda_j\left(\frac{\ddot{\F}_k \ddot{\F}_k^\T}{T}\right) \asymp 1, \\
    \label{eqn:lemma2.2}
    \lambda_j\left(\A_k^\T\A_k\right) \asymp d_k^{\alpha_{k,j}}, \\
    \label{eqn:lemma2.3}
    \lambda_j\left(\frac{ \A_k \ddot{\F}_k \ddot{\F}_k^\T\A_k^\T}{T} \right) \asymp d_k^{\alpha_{k,j}},  \\
    \label{eqn:lemma2.4}
    \lambda_j\left(\frac{\ddot{\F}_k \left( \I_T - \frac{1}{T} \1_T \1_T^\T \right) \ddot{\F}_k^\T}{T}\right) \asymp 1, \\
    \label{eqn:lemma2.5}
    \lambda_j\left(\frac{ \A_k \ddot{\F}_k \left( \I_T - \frac{1}{T} \1_T \1_T^\T \right) \ddot{\F}_k^\T\A_k^\T}{T} \right) \asymp d_k^{\alpha_{k,j}}.
\end{align}
and for $j \in [\min(z_k,r_k)]$,
\begin{align}
    \label{eqn:lemma2.6}
    \lambda_j\left(\frac{\ddot{\X}_k^\T \left( \I_T - \frac{1}{T} \1_T \1_T^\T \right) \ddot{\X}_k}{T} \right) \asymp d_k^{\alpha_{k,j}}.
\end{align}
In addition, if Assumption (R2) is satisfied, then for $\min(z_k,r_k) + 1 \leq j \leq \lfloor c\min{(T,d_k)} \rfloor - r_k$,
\begin{align}
    \label{eqn:lemma2.7}
    \lambda_j\left(\frac{\ddot{\X}_k^\T \left( \I_T - \frac{1}{T} \1_T \1_T^\T \right) \ddot{\X}_k}{T} \right) \asymp \frac{d_{\text{-}k}}{s_{\text{-}k}} \left(1 + \frac{d_k}{T}\right).
\end{align}

\end{lemma}

The weak serial dependence of factors and errors are quantified in the following lemma.

\begin{lemma}\label{lem:3}
Define $\A_{f,T}$ and $\A_{e,T}$ similar to $\A_{\epsilon,T}$ in Assumption (R2). Define $\A_{f,T^2}$ to be the $T^2 \times T^2$ fourth moment matrix of the MA process $f_{t,l,j}^{(k)}$ for any $k,l,j$, and define $\A_{e,T^2}$ and $\A_{\epsilon,T^2}$ similarly. Then, under Assumption (E1), (E2) and (F1), we have the following results: For $\A_T$ can either be $\A_{f,T}$, $\A_{e,T}$ or $\A_{\epsilon,T}$,
  \begin{align}{\label{eqn:R3.1}}
       \1_T^\T \A_{T} \1_T = O(T), \ \ \|\A_{T}\|_F^2 = O(T).
  \end{align}
   For $\A_{T^2}$ can either be $\A_{f,T^2}$, $\A_{e,T^2}$ or $\A_{\epsilon,T^2}$,
  \begin{align}{\label{eqn:R3.2}}
      \1_{T^2}^\T \A_{T^2} \1_{T^2} = O(T^2), \ \ \|\A_{T^2}\|_F^2 = O(T^2).
  \end{align}
Moreover,
  \begin{align}{\label{eqn:R3.3}}
      \vec{(\A_{f,T})}^\T \vec{(\A_{e,T})} = O(T), \ \ \vec{(\A_{f,T})}^\T \vec{(\A_{\epsilon,T})} = O(T), \\
      {\label{eqn:R3.4}}
      \1_{T^2}^\T \A_{f,T} \otimes \A_{e,T} \1_{T^2} = O(T^2), \ \ \1_{T^2}^\T \A_{f,T} \otimes \A_{\epsilon,T} \1_{T^2} = O(T^2), \\
  {\label{eqn:R3.5}}
  \vec{(\A_{f,T} \otimes \A_{f,T})}^\T \vec{(\A_{e,T^2})} = O(T^2), \ \  \vec{(\A_{f,T} \otimes \A_{f,T})}^\T \vec{(\A_{\epsilon,T^2})} = O(T^2), \\
  {\label{eqn:R3.6}}
  \vec{(\A_{f,T^2})}^\T \vec{(\A_{e,T^2})} = O(T^2), \ \
  \vec{(\A_{f,T^2})}^\T \vec{(\A_{\epsilon,T^2})} = O(T^2).
  \end{align}

\end{lemma}

\begin{remark}\label{remark:eigenvalue_ratio}
Note that Lemma \ref{lem:2} implies the following result for eigenvalue ratio:
For $j \leq r_k - 1$,
\begin{align*}
    \frac{\lambda_j \left(\frac{\ddot{\X}_k^\T \left( \I_T - \frac{1}{T} \1_T \1_T^\T \right)\ddot{\X}_k}{T} \right)}{\lambda_{j+1} \left(\frac{\ddot{\X}_k^\T \left( \I_T - \frac{1}{T} \1_T \1_T^\T \right)\ddot{\X}_k}{T} \right)} \asymp \frac{d_k^{\alpha_{k,j}}}{d_k^{\alpha_{k,j+1}}};
\end{align*}
For $r_k + 1 \leq j \leq \lfloor c\min{(T,d_k)} \rfloor - r_k - 1$,
\begin{align*}
    \frac{\lambda_j \left(\frac{\ddot{\X}_k^\T \left( \I_T - \frac{1}{T} \1_T \1_T^\T \right)\ddot{\X}_k}{T} \right)}{\lambda_{j+1} \left(\frac{\ddot{\X}_k^\T \left( \I_T - \frac{1}{T} \1_T \1_T^\T \right)\ddot{\X}_k}{T} \right)} \asymp 1;
\end{align*}
For $j = r_k$,
\begin{align*}
    \frac{\lambda_{j} \left(\frac{\ddot{\X}_k^\T \left( \I_T - \frac{1}{T} \1_T \1_T^\T \right)\ddot{\X}_k}{T} \right)}{\lambda_{j+1} \left(\frac{\ddot{\X}_k^\T \left( \I_T - \frac{1}{T} \1_T \1_T^\T \right)\ddot{\X}_k}{T} \right)} \asymp \frac{d_k^{\alpha_{k,r_k}}}{\frac{d_{\text{-}k}}{s_{\text{-}k}}\left(1+\frac{d_k}{T}\right)} \rightarrow \infty.
\end{align*}
Therefore, \begin{align} \label{eqn:eigenratiobound}
    \frac{\lambda_1 \left(\frac{\ddot{\X}_k^\T \left( \I_T - \frac{1}{T} \1_T \1_T^\T \right)\ddot{\X}_k}{T} \right)}{\lambda_{r_k+1} \left(\frac{\ddot{\X}_k^\T \left( \I_T - \frac{1}{T} \1_T \1_T^\T \right)\ddot{\X}_k}{T} \right)} \asymp\frac{d_k^{\alpha_{k,1}}}{\frac{d_{\text{-}k}}{s_{\text{-}k}}\left(1+\frac{d_k}{T}\right)} \rightarrow \infty.
\end{align}
(\ref{eqn:eigenratiobound}) implies that at least one of ratio of subsequent eigenvalues $\left\{ \frac{\lambda_j \left(\frac{\ddot{\X}_k^\T \left( \I_T - \frac{1}{T} \1_T \1_T^\T \right)\ddot{\X}_k}{T} \right)}{\lambda_{j+1} \left(\frac{\ddot{\X}_k^\T \left( \I_T - \frac{1}{T} \1_T \1_T^\T \right)\ddot{\X}_k}{T} \right)} , j \in[r_k] \right\} $ goes to infinity. This is essential to detect the existence of factors. Note that in the special case when all factors have the same strength, we should have $\frac{\lambda_{r_k} \left(\frac{\ddot{\X}_k^\T \left( \I_T - \frac{1}{T} \1_T \1_T^\T \right)\ddot{\X}_k}{T} \right)}{\lambda_{r_k+1} \left(\frac{\ddot{\X}_k^\T \left( \I_T - \frac{1}{T} \1_T \1_T^\T \right)\ddot{\X}_k}{T} \right)} \rightarrow \infty$, which recovers the result in \cite{AhnHorenstein2013}.

In contrast, when there is no factor exist, we should have that
\begin{align*}
    \frac{\lambda_j \left(\frac{\ddot{\X}_k^\T \left( \I_T - \frac{1}{T} \1_T \1_T^\T \right)\ddot{\X}_k}{T} \right)}{\lambda_{j+1} \left(\frac{\ddot{\X}_k^\T \left( \I_T - \frac{1}{T} \1_T \1_T^\T \right)\ddot{\X}_k}{T} \right)} = \frac{\lambda_j \left(\frac{\ddot{\E}_k^\T \left( \I_T - \frac{1}{T} \1_T \1_T^\T \right)\ddot{\E}_k}{T} \right)}{\lambda_{j+1} \left(\frac{\ddot{\E}_k^\T \left( \I_T - \frac{1}{T} \1_T \1_T^\T \right)\ddot{\E}_k}{T} \right)} \asymp 1
\end{align*}
for $j = 1,..., \lfloor c\min{(T,d_k)} \rfloor$ by (\ref{eqn:smallest_eigenvalue_bound_error}) in Lemma \ref{lem:1}. Thus, every ratio (actually until $\lfloor c\min{(T,d_k)} \rfloor$) of subsequent eigenvalues remains bounded. In this way, we can use eigenvalue ratio to detect the existence of factors accordingly: when there is factor exist, at least one of the eigenvalue ratio will goes to infinity; when there is no factor, all eigenvalue ratios (until $\lfloor c\min{(T,d_k)} \rfloor$) remains bounded. Note that this is different from estimating the number of factors, since we can only know that there is factor exist, but we do not really know how many factors and what the factor strengths are. However, in the special case when all factors are of the same strength, we have $\frac{\lambda_{r_k} \left(\frac{\ddot{\X}_k^\T \left( \I_T - \frac{1}{T} \1_T \1_T^\T \right)\ddot{\X}_k}{T} \right)}{\lambda_{r_k+1} \left(\frac{\ddot{\X}_k^\T \left( \I_T - \frac{1}{T} \1_T \1_T^\T \right)\ddot{\X}_k}{T} \right)} \rightarrow \infty$ and $\frac{\lambda_{j} \left(\frac{\ddot{\X}_k^\T \left( \I_T - \frac{1}{T} \1_T \1_T^\T \right)\ddot{\X}_k}{T} \right)}{\lambda_{j+1} \left(\frac{\ddot{\X}_k^\T \left( \I_T - \frac{1}{T} \1_T \1_T^\T \right)\ddot{\X}_k}{T} \right)} \asymp 1$ for $j \neq r_k$, which is how \cite{AhnHorenstein2013} estimates the number of factors. In the same spirit, the eye-ball test of \cite{LamYao2012} also uses eigenvalue ratios in a similar manner to deduce the number of factors in a statistical factor model.
\end{remark}

{\em Proof of Lemma \ref{lem:1}.}
By Assumption (E1), (E2), (R1), we have
\begin{align*}
    \tilde{\e}_{t,k} &=  \bPsi^{(k)}\e_t^{(k)} + \sum_{\ell=1}^{\dmk}(\bSigma_{\epsilon,\ell}^{(k)})^{1/2}\bepsilon_{t,\ell}^{(k)} \notag \\
    &: = \tilde{\e}_{t,k,1} + \tilde{\e}_{t,k,2}
\end{align*}
where $\tilde{\e}_{t,k,1}:= \bPsi^{(k)}\e_t^{(k)}$ and $\tilde{\e}_{t,k,2} = \sum_{\ell=1}^{\dmk}(\bSigma_{\epsilon,\ell}^{(k)})^{1/2}\bepsilon_{t,\ell}^{(k)}$. Using such decomposition, the error matrix can be written as
\begin{align*}
    \tilde{\E}_k = \tilde{\E}_{k,1} + \tilde{\E}_{k,2},
\end{align*}
where $\tilde{\E}_{k,1} = \left( \tilde{\e}_{1,k,1}, ..., \tilde{\e}_{T,k,1} \right)^\T$ and $\tilde{\E}_{k,2} = \left( \tilde{\e}_{1,k,2}, ..., \tilde{\e}_{T,k,2} \right)^\T$. Then we can deal with $\tilde{\E}_{k,1}$ and $\tilde{\E}_{k,2}$ separately using random matrix theory. We first look at $\tilde{\E}_{k,1}$. Similar to assumptions in \cite{AhnHorenstein2013}, we further write the matrix $\tilde{\E}_{k,1}$ as
\begin{align*}
    \tilde{\E}_{k,1} = L_{e,k,1}^{\frac{1}{2}} U_{e,k,1} R_{e,k,1}^{\frac{1}{2}},
\end{align*}
where $U_{e,k,1} \in \mathbb{R}^{T \times d_k}$ are iid random variables with uniformly bounded fourth moment by Assumption (R1), and $L_{e,k,1}\in \mathbb{R}^{T \times T}$ and $R_{e,k,1} \in \mathbb{R}^{d_k \times d_k}$ are time-serial and cross-sectional covariance matrices. By Assumption (E1), $R_{e,k,1} = \bPsi^{(k)}\bPsi^{(k)\T}$, and $L_{e,k,1} = \A_{\e,T}$. Next, we need to bound the spectral norm of $L_{e,k,1}$ and $R_{e,k,1}$. Assumption (E1) implies $\norm{R_{e,k,1}} = O(d_{\text{-}k})$. For $\A_{\e,T}$, since it is symmetric, $\left\| \A_{\e,T} \right\|_1 = \left\| \A_{\e,T} \right\|_\infty$, and
\begin{align}\label{eqn:1norm_AT}
    \left\| \A_{\e,T} \right\|_1 & = \max_t \sum_{s=1}^T \left|(\A_{e,T})_{ts}\right| \notag \\
    &\leq 2 \sum_{v=0}^T \left|\sum_{q\geq 0} a_{e,q}a_{e,q+v}\right| \notag\\
    & \leq 2\left(\sum_{q \geq 0} |a_{e,q}|\right)^2 \leq C
\end{align}
by Assumption (E2). Thus, $\norm{  \A_{\e,T} } \leq \sqrt{\left\|  \A_{\e,T} \right\|_1 \left\| \A_{\e,T} \right\|_\infty} = \left\|  \A_{\e,T} \right\|_1 \leq C$. \cite{BaiYin1993} and \cite{Latala2005} show that $\lambda_1\left(U_{e,k,1}U_{e,k,1}^\T/T\right) \rightarrow (1+\sqrt{d_k/T})^2$. Thus, similar to the result obtained by \cite{AhnHorenstein2013} (see also \cite{moon_weidner_2015}), we have
\begin{align*}
    \lambda_1 \left(\frac{\tilde{\E}_{k,1}^\T \tilde{\E}_{k,1}}{T} \right) \leq \norm{L_{e,k,1} }\norm{R_{e,k,1} } \norm{U_{e,k,1}U_{e,k,1}^\T/T  } = O_p\left(d_{\text{-}k} \left(1 + \frac{d_k}{T} \right) \right).
\end{align*}
For $\tilde{\E}_{k,2}$, similarly we can write it as
\begin{align*}
    \tilde{\E}_{k,2} = L_{e,k,2}^{\frac{1}{2}} U_{e,k,2} R_{e,k,2}^{\frac{1}{2}}
\end{align*}
where  $R_{e,k,2} = \bSigma_{\epsilon}^{(k)}$, and $L_{e,k,2} = \A_{\epsilon,T}$. By Assumption (E2), $\|R_{e,k,2}\| = O(d_{\text{-}k})$, and we have $\norm{L_{e,k,2}} \leq C$ by applying the same analysis as above. Accordingly,
\begin{align*}
    \lambda_1 \left(\frac{\tilde{\E}_{k,2}^\T \tilde{\E}_{k,2}}{T} \right) \leq \norm{L_{e,k,2} }\norm{R_{e,k,2} } \norm{U_{e,k,2}U_{e,k,2}^\T/T  } = O_p\left(d_{\text{-}k} \left(1 + \frac{d_k}{T} \right) \right).
\end{align*}
Therefore,
\begin{align*}
    \lambda_1 \left(\frac{\tilde{\E}_{k}^\T \tilde{\E}_{k}}{T} \right) &\leq \bigg\|\frac{\tilde{\E}_{k,1}^\T \tilde{\E}_{k,1}}{T} \bigg\| + \bigg\|\frac{\tilde{\E}_{k,1}^\T \tilde{\E}_{k,2}}{T }\bigg\| + \bigg\|\frac{\tilde{\E}_{k,2}^\T \tilde{\E}_{k,1}}{T} \bigg\| +
    \bigg\|\frac{\tilde{\E}_{k,2}^\T \tilde{\E}_{k,2}}{T} \bigg\| \notag \\
    & = O_p\left(d_{\text{-}k} \left(1 + \frac{d_k}{T} \right) \right) + 2\bigg\|\frac{\tilde{\E}_{k,1}^\T \tilde{\E}_{k,2}}{T} \bigg\| \notag \\
    & \leq O_p\left(d_{\text{-}k} \left(1 + \frac{d_k}{T} \right) \right) + \sqrt{\bigg\|\frac{\tilde{\E}_{k,1}^\T \tilde{\E}_{k,1}}{T} \bigg\| \bigg\|\frac{\tilde{\E}_{k,2}^\T \tilde{\E}_{k,2}}{T} }\bigg\| \notag \\
    & = O_p\left(d_{\text{-}k} \left(1 + \frac{d_k}{T} \right) \right),
\end{align*}
which implies (\ref{eqn:largest_eigenvalue_bound_error}), and
\begin{align*}
    \lambda_1 \left(\frac{\tilde{\E}_{k}^\T \left( \I_T - \frac{1}{T} \1_T \1_T^\T \right) \tilde{\E}_{k}}{T} \right) & \leq \bigg\|\frac{\tilde{\E}_{k}^\T}{\sqrt{T}}\bigg\|^2 \bigg\| \I_T - \frac{1}{T} \1_T \1_T^\T \bigg\| \notag \\
    & \leq 2 \lambda_1 \left(\frac{\tilde{\E}_{k}^\T \tilde{\E}_{k}}{T}\right) \notag \\
    & =  O_p\left(d_{\text{-}k} \left(1 + \frac{d_k}{T} \right) \right),
\end{align*}
which implies (\ref{eqn:largest_eigenvalue_bound_error_nonzeromu})

 To show (\ref{eqn:smallest_eigenvalue_bound_error}), we focus on $\tilde{\E}_{k,2}$. Note that Assumption (R2) is parallel to Assumption (D) in \cite{AhnHorenstein2013}. Following Lemma A.7. in \cite{AhnHorenstein2013}, we have
 \begin{align*}
         \mathbb{P} \left( \lambda_{\lfloor c\min{(T,d_k)} \rfloor}\left(\frac{\tilde{\E}_{k,2}^\T \tilde{\E}_{k,2}}{\max{(T,d_k)}} \right) \geq Cd_{\text{-}k}\right)= 1
     \end{align*}
     for some $c\in (0,1]$ and $C >0$.
Thus,
\begin{align*}
         \mathbb{P} \left( \lambda_{\lfloor c\min{(T,d_k)} \rfloor}\left(\frac{\ddot{\E}_{k,2}^\T \ddot{\E}_{k,2}}{T} \right) \geq C\left(d_{\text{-}k} \left(1 + \frac{d_k}{T}\right)\right)\right)= 1
     \end{align*}
     for some $c\in (0,1]$ and $C >0$. Finally, by Weyl’s inequalities,
\begin{align*}
    \sqrt{ \lambda_{\lfloor c\min{(T,d_k)} \rfloor} \left(\frac{\tilde{\E}_k^\T \tilde{\E}_k}{T} \right)} &\geq \sqrt{\lambda_{\lfloor c\min{(T,d_k)} \rfloor} \left(\frac{\tilde{\E}_{k,2}^\T \tilde{\E}_{k,2}}{T} \right)} - \sqrt{\lambda_1 \left(\frac{\tilde{\E}_{k,1}^\T \tilde{\E}_{k,1} }{T}  \right)}  \notag \\
    & \asymp \sqrt{d_{\text{-}k} \left(1 + \frac{d_k}{T}\right)},
\end{align*}
which implies (\ref{eqn:smallest_eigenvalue_bound_error}), since $\lambda_1 \left(\frac{\tilde{\E}_{k,1}^\T \tilde{\E}_{k,1} }{T}  \right) = o_p\left(d_{\text{-}k} \left(1 + \frac{d_k}{T}\right) \right)$ under Assumption (R2). Similarly, following the same argument as Lemma A.7. in \cite{AhnHorenstein2013}, we can show that
\begin{align*}
         \mathbb{P} \left( \lambda_{\lfloor c\min{(T,d_k)} \rfloor}\left(\frac{\ddot{\E}_{k,2}^\T \left( \I_T - \frac{1}{T} \1_T \1_T^\T \right) \ddot{\E}_{k,2}}{T} \right) \geq C\left(d_{\text{-}k} \left(1 + \frac{d_k}{T}\right)\right)\right)= 1
     \end{align*}
     for some $c\in (0,1]$ and $C >0$, so
\begin{align*}
    \sqrt{ \lambda_{\lfloor c\min{(T,d_k)} \rfloor} \left(\frac{\tilde{\E}_k^\T \left( \I_T - \frac{1}{T} \1_T \1_T^\T \right) \tilde{\E}_k}{T} \right)} &\geq \sqrt{\lambda_{\lfloor c\min{(T,d_k)} \rfloor} \left(\frac{\tilde{\E}_{k,2}^\T \left( \I_T - \frac{1}{T} \1_T \1_T^\T \right)\tilde{\E}_{k,2}}{T} \right)}\\
    &- \sqrt{\lambda_1 \left(\frac{\tilde{\E}_{k,1}^\T \left( \I_T - \frac{1}{T} \1_T \1_T^\T \right)\tilde{\E}_{k,1} }{T}  \right)}  \notag \\
    & \asymp \sqrt{d_{\text{-}k} \left(1 + \frac{d_k}{T}\right)},
\end{align*}
which implies (\ref{eqn:smallest_eigenvalue_bound_error_nonzeromu}). This completes the proof of Lemma \ref{lem:1}. $\square$

{\em Proof of Lemma \ref{lem:2}.}
We consider the case $K = 3$ without loss of generality. By Assumption (F1), we know each element of $\cF_t$ are independent with mean 0 and variance 1. Suppose we want to estimate $\A_1$. For $i,j \in [r_1]$, we have
\begin{align*}
\mathbb{E} \left( \tilde{f}_{t,1,i} \tilde{f}_{t,1,j} \right) &=  \sum_{p=1}^{r_2}\sum_{q=1}^{r_3}\sum_{u=1}^{r_2}\sum_{v=1}^{r_3} colSum(\A_2)_p colSum(\A_3)_q  colSum(\A_2)_u colSum(\A_3)_v \mathbb{E} (f_{t,ipq} f_{t,juv}) \notag\\
& = \sum_{p=1}^{r_2}\sum_{q=1}^{r_3} colSum(\A_2)_p^2 colSum(\A_3)_q^2 \notag \\
& \asymp  s_2 s_3 = s_{\text{-}1}
\end{align*}
when $i=j$, and 0 otherwise. Thus, $\mathbb{E} \left( \ddot{f}_{t,1,i} \ddot{f}_{t,1,j} \right) \asymp 1$ for $i =j$, and 0 otherwise, which implies that $\Sigma_{F,k}$ is a diagonal matrix with all diagonal elements bounded away from 0 and infinity.

Next, we show (\ref{eqn:lemma2.1}) -- (\ref{eqn:lemma2.3}). To apply random matrix theory, similar to \cite{AhnHorenstein2013}, we can further write the matrix $\ddot{\F}_k$ as
\begin{align*}
    \ddot{\F}_k = L_{f,k}^{\frac{1}{2}} U_{f,k} R_{f,k}^{\frac{1}{2}},
\end{align*}
where $U_{f,k} \in \mathbb{R}^{r_k \times T}$ are iid random variables with uniformly bounded fourth moment by Assumption (R1), and $L_{f,k}\in \mathbb{R}^{r_k \times r_k}$ and $R_{f,k} \in \mathbb{R}^{T \times T}$ are cross-sectional and time-serial covariance matrices. By Assumption (F1), $L_{f,k} = I_{r_k}$, and $R_{f,k} = \A_{f,T}$. Following similar analysis as (\ref{eqn:1norm_AT}), we have $\|R_{f,k}\|_1 \leq C$. Thus, $\norm{ R_{f,k} } \leq \sqrt{\left\| R_{f,k} \right\|_1 \left\| R_{f,k} \right\|_\infty} = \left\| R_{f,k} \right\|_1 \leq C$.
Since we assume $r_k = o(T)$, \cite{BaiYin1993} and \cite{Latala2005} show that $\lambda_1\left(U_{f,k}U_{f,k}^\T/T\right) - (1+\sqrt{r_k/T})^2 \rightarrow 0$ and $\lambda_{r_k}\left(U_{f,k}U_{f,k}^\T/T\right) - (1-\sqrt{r_k/T})^2 \rightarrow 0$. Similar to the result obtained by \cite{AhnHorenstein2013} (see also \cite{moon_weidner_2015}), since the largest eigenvalues of $R_{f,k}$ and $L_{f,k}$ are bounded, we have
\begin{align*}
    \lambda_1 \left(\ddot{\F}_k \ddot{\F}_k^\T /T \right) \leq \norm{L_{f,k}}\norm{R_{f,k} } \norm{U_{f,k}U_{f,k}^\T/T  } = O_p(1).
\end{align*}
Similarly, since $\lambda_1(R_{f,k}) = \norm{R_{f,k} } \geq 1$, we have
\begin{align*}
    \lambda_{r_k} \left(\ddot{\F}_k \ddot{\F}_k^\T /T \right) = \lambda_{r_k} \left(U_{f,k} R_{f,k} U_{f,k}^T /T \right) = \lambda_{r_k} \left(U_{f,k}^T U_{f,k} R_{f,k}  /T \right) \geq \lambda_{r_k} \left(U_{f,k}^T U_{f,k}/T\right) \lambda_{1} (R_{f,k}) \geq 1
\end{align*}
as $r_k, T \rightarrow \infty$. Thus, we have $\lambda_j\left(\frac{\ddot{\F}_k \ddot{\F}_k^\T}{T}\right) \asymp 1$ for $j \in [r_k]$.

Next, for $j \in [r_k]$, we know
$
     \lambda_j \left(\frac{ \A_k \ddot{\F}_k \ddot{\F}_k^\T \A_k^\T}{T}  \right) = \lambda_j \left(\frac{\ddot{\F}_k\ddot{\F}_k^\T \A_k^\T \A_k}{T} \right)
$
and
 \begin{align*}
     \lambda_j \left(\A_k^\T \A_k \right) \lambda_{r_k} \left(\frac{\ddot{\F}_k\ddot{\F}_k^\T}{T}\right) \leq \lambda_j \left(\frac{\ddot{\F}_k\ddot{\F}_k^\T \A_k^\T \A_k}{T} \right) \leq \lambda_j \left(\A_k^\T \A_k \right) \lambda_1 \left(\frac{\ddot{\F}_k\ddot{\F}_k^\T}{T}\right).
 \end{align*}
Since $\lambda_j\left(\frac{\ddot{\F}_k \ddot{\F}_k^\T}{T}\right) \asymp 1$ for $j \in [r_k]$, it follows that $ \lambda_j \left(\frac{\ddot{\F}_k^\T \A_k^\T \A_k \ddot{\F}_k}{T}  \right) \asymp \lambda_j(\A_k^\T \A_k)$ for $j \in [r_k]$. Similarly,
$
    \lambda_j \left(\A_k^\T \A_k \right) = \lambda_j\left(\D_k^{\frac{1}{2}} \D_k^{-\frac{1}{2}} \A_k^\T \A_k \D_k^{-\frac{1}{2}} \D_k^{\frac{1}{2}} \right) = \lambda_j\left(\D_k \D_k^{-\frac{1}{2}} \A_k^\T \A_k \D_k^{-\frac{1}{2}} \right)
$
and
\begin{equation*}
     \lambda_j (\D_k) \lambda_{r_k}\left( \D_k^{-\frac{1}{2}} \A_k^\T \A_k \D_k^{-\frac{1}{2}} \right) \leq \lambda_j\left(\D_k \D_k^{-\frac{1}{2}} \A_k^\T \A_k \D_k^{-\frac{1}{2}} \right) \leq \lambda_j (\D_k) \lambda_1\left( \D_k^{-\frac{1}{2}} \A_k^\T \A_k \D_k^{-\frac{1}{2}} \right). \label{eqn:diagonal_eigen_equivalent}
\end{equation*}
By Assumption (L1), $\lambda_j (\D_k) \asymp d_k^{\alpha_{k,j}}$ and $\lambda_j\left( \D_k^{-\frac{1}{2}} \A_k^\T \A_k \D_k^{-\frac{1}{2}} \right) \asymp 1$ for $j \in [r_k]$ (this also holds when elements of $\A_k$ are random and independent). Thus, for $j \in [r_k]$, we have $\lambda_j \left(\frac{\A_k \ddot{\F}_k \ddot{\F}_k^\T \A_k^\T}{T}  \right) \asymp \lambda_j(\A_k^\T \A_k) \asymp d_k^{\alpha_{k,j}}$.

Next, we will show (\ref{eqn:lemma2.4}) -- (\ref{eqn:lemma2.5}) accordingly. To start with, note that
\begin{align*}
    \lambda_1\left(\frac{\ddot{\F}_k \left( \I_T - \frac{1}{T} \1_T \1_T^\T \right) \ddot{\F}_k^\T}{T}\right) &\leq \bigg\|\frac{\ddot{\F}_k \ddot{\F}_k^\T}{T}\bigg\| + \bigg\|\frac{\ddot{\F}_k  \frac{1}{T} \1_T \1_T^\T  \ddot{\F}_k^\T}{T}\bigg\| \notag \\
    &\leq \bigg\|\frac{\ddot{\F}_k \ddot{\F}_k^\T}{T}\bigg\| + \bigg\|\frac{\ddot{\F}_k}{\sqrt{T}} \bigg\|^2 \bigg\|\frac{1}{T} \1_T \1_T^\T \bigg\| = 2 \bigg\|\frac{\ddot{\F}_k \ddot{\F}_k^\T}{T}\bigg\|
    \asymp 1.
\end{align*}
And
\begin{align}\label{eqn:psi_rk_FZF}
    \lambda_{r_k}\left(\frac{\ddot{\F}_k \left( \I_T - \frac{1}{T} \1_T \1_T^\T \right) \ddot{\F}_k^\T}{T}\right) &= \lambda_{r_k}\left(\frac{\ddot{\F}_k \ddot{\F}_k^\T}{T} - \frac{\ddot{\F}_k \frac{1}{T} \1_T \1_T^\T \ddot{\F}_k^\T}{T}\right) \notag \\
    & \geq \lambda_{r_k}\left(\frac{\ddot{\F}_k \ddot{\F}_k^\T}{T}\right) - \lambda_1\left(\frac{\ddot{\F}_k \1_T \1_T^\T \ddot{\F}_k^\T}{T^2}\right).
\end{align}
We next obtain an upper bound for $\lambda_1\left(\frac{\ddot{\F}_k \1_T \1_T^\T \ddot{\F}_k^\T}{T^2}\right)$. Note that $\ddot{\F}_k \1_T \in \mathbb{R}^{r_k \times 1}$ is a random vector, with its $i$-th element to be $\sum_{t=1}^T \ddot{f}_{t,i}$. For its moment bounds, $\mathbb{E} \left(\sum_{t=1}^T \ddot{f}_{t,i}\right)^2 = \1_T^\T \A_{f,T}\1_T = O(T)$ and $\mathbb{E} \left(\sum_{t=1}^T \ddot{f}_{t,i}\right)^4 = \1_{T^2}^\T \A_{f,T^2}\1_{T^2} = O(T^2)$ by Lemma \ref{lem:3}. Therefore, following similar argument in previous analysis, we can decompose $\ddot{\F}_k \1_T = \sqrt{T} \I_{r_k} U$, where $U \in \mathbb{R}^{r_k \times 1}$ has iid random entries with bounded second and fourth moments. Thus,
\begin{align*}
    \lambda_1\left(\frac{\ddot{\F}_k \1_T \1_T^\T \ddot{\F}_k^\T}{T^2}\right) = \frac{1}{T^2}\lambda_1\left(\ddot{\F}_k \1_T \1_T^\T \ddot{\F}_k^\T\right) \leq \frac{1}{T^2}O_p(r_kT) = O_p\left(\frac{r_k}{T}\right) = o_p(1),
\end{align*}
which implies $\lambda_{r_k}\left(\frac{\ddot{\F}_k \left( \I_T - \frac{1}{T} \1_T \1_T^\T \right) \ddot{\F}_k^\T}{T}\right) \succeq 1$ by  (\ref{eqn:psi_rk_FZF}). Therefore, $\lambda_{j}\left(\frac{\ddot{\F}_k \left( \I_T - \frac{1}{T} \1_T \1_T^\T \right) \ddot{\F}_k^\T}{T}\right)$ for $j \in [r_k]$.

Next, for $j \in [r_k]$, it is easy to see
\begin{align*}
   \lambda_j\left(\A_k^\T\A_k\right) \lambda_{r_k}\left(\frac{ \ddot{\F}_k \left( \I_T - \frac{1}{T} \1_T \1_T^\T \right) \ddot{\F}_k^\T}{T} \right) &\leq \lambda_j\left(\frac{ \A_k \ddot{\F}_k \left( \I_T - \frac{1}{T} \1_T \1_T^\T \right) \ddot{\F}_k^\T\A_k^\T}{T} \right)\\
   &\leq \lambda_j\left(\A_k^\T\A_k\right) \lambda_1\left(\frac{ \ddot{\F}_k \left( \I_T - \frac{1}{T} \1_T \1_T^\T \right) \ddot{\F}_k^\T}{T} \right),
\end{align*}
which implies $ \lambda_j\left(\frac{ \A_k \ddot{\F}_k \left( \I_T - \frac{1}{T} \1_T \1_T^\T \right) \ddot{\F}_k^\T\A_k^\T}{T} \right) \asymp d_k^{\alpha_{k,j}}$ for $j \in [r_k]$.

Observe that $\left(\I_T - \frac{1}{T} \1_T \1_T^\T\right)^2 = \I_T - \frac{1}{T} \1_T \1_T^\T$, and $\ddot{\X}_k^\T \left(\I_T - \frac{1}{T} \1_T \1_T^\T\right) =  \left(\A_k \ddot{\F}_k + \ddot{\E}_k^\T\right) \left(\I_T - \frac{1}{T} \1_T \1_T^\T\right)$. Hence to show (\ref{eqn:lemma2.6}), following the similar argument as Theorem 1 in \cite{Freyaldenhoven2022}, for $j \in [r_k]$,
\begin{align}\label{eqn:upper_bound_X'X}
    \sqrt{\lambda_j \left(\frac{\ddot{\X}_k^\T \left(\I_T - \frac{1}{T} \1_T \1_T^\T\right) \ddot{\X}_k}{T} \right)} &\geq \sqrt{\lambda_j \left(\frac{\A_k \ddot{\F}_k \left(\I_T - \frac{1}{T} \1_T \1_T^\T\right) \ddot{\F}_k^\T \A_k^\T}{T}  \right)} - \sqrt{\lambda_1 \left(\frac{\ddot{\E}_k^\T \left(\I_T - \frac{1}{T} \1_T \1_T^\T\right) \ddot{\E}_k }{T}  \right)}  \notag \\
    & \succeq \sqrt{d_k^{\alpha_{k,j}}} - \sqrt{O_p\left(\frac{d_{\text{-}k}}{s_{\text{-}k}} \left(1 + \frac{d_k}{T} \right) \right)} \asymp \sqrt{d_k^{\alpha_{k,j}}}.
\end{align}
where the second line follows from Lemma \ref{lem:1} and the last line follows from Assumption (L2). Similarly, for $j \in [r_k]$,
\begin{align}\label{eqn:lower_bound_X'X}
     \lambda_j \left(\frac{\ddot{\X}_k^\T \left(\I_T - \frac{1}{T} \1_T \1_T^\T\right) \ddot{\X}_k}{T} \right) &\leq \lambda_j \left(\frac{\A_k \ddot{\F}_k \left(\I_T - \frac{1}{T} \1_T \1_T^\T\right) \ddot{\F}_k^\T \A_k^\T}{T}  \right) + \lambda_1 \left(\frac{\ddot{\E}_k^\T \left(\I_T - \frac{1}{T} \1_T \1_T^\T\right) \ddot{\E}_k }{T}  \right) \notag \\
     & \ \ \ \ + 2 \sqrt{\lambda_j \left(\frac{\A_k \ddot{\F}_k \left(\I_T - \frac{1}{T} \1_T \1_T^\T\right) \ddot{\F}_k^\T \A_k^\T}{T}  \right)} \sqrt{\lambda_1 \left(\frac{\ddot{\E}_k^\T \left(\I_T - \frac{1}{T} \1_T \1_T^\T\right) \ddot{\E}_k }{T}  \right)}  \notag \\
     & \asymp d_k^{\alpha_{k,j}} + O_p\left(\frac{d_{\text{-}k}}{s_{\text{-}k}} \left(1 + \frac{d_k}{T} \right) \right) \asymp d_k^{\alpha_{k,j}}.
\end{align}
Therefore,  $\lambda_j\left(\frac{\ddot{\X}_k^\T \left(\I_T - \frac{1}{T} \1_T \1_T^\T\right) \ddot{\X}_k}{T} \right) \asymp d_k^{\alpha_{k,j}}$ for $j \in [r_k]$.

Finally, to show (\ref{eqn:lemma2.7}), we apply similar technique as Lemma A.8 in \cite{AhnHorenstein2013}. For simplicity of expression, in the following proof, we write $\M_T = \I_T - \frac{1}{T} \1_T \1_T^\T$ and $\M = (\ddot{\F}_k\M_T)^\T (\ddot{\F}_k\M_T\ddot{\F}_k^\T)^{-1} \ddot{\F}_k\M_T$. Then $rank(\M) \leq r_k$, and
\begin{align*}
    \ddot{\X}_k^\T \M_T \ddot{\X}_k = \left(\A_k \ddot{\F}_k + \ddot{\E}_k^\T \M \right) \M_T \left(\A_k \ddot{\F}_k + \ddot{\E}_k^\T \M \right)^\T + \ddot{\E}_k^\T \left(\M_T - \M\M_T\M^\T \right) \ddot{\E}_k.
\end{align*}
Hence, for $r_k + 1 \leq j \leq \lfloor c\min{(T,d_k)} \rfloor - r_k$,
\begin{align*}
    \lambda_{j}\left(\ddot{\E}_k^\T \left(\M_T - \M\M_T\M^\T \right) \ddot{\E}_k \right) &\leq  \lambda_{j}\left(\ddot{\X}_k^\T \M_T \ddot{\X}_k \right) \\ &\leq  \lambda_{j - r_k} \left(\ddot{\E}_k^\T \left(\M_T - \M\M_T\M^\T \right) \ddot{\E}_k \right)\\ &+ \lambda_{r_k + 1}\left[\left(\A_k \ddot{\F}_k + \ddot{\E}_k^\T \M \right) \M_T \left(\A_k \ddot{\F}_k + \ddot{\E}_k^\T \M \right)^\T \right] \notag \\
     &\leq  \lambda_{j - r_k} \left(\ddot{\E}_k^\T \left(\M_T - \M\M_T\M^\T \right) \ddot{\E}_k \right),
\end{align*}
since $\left(\A_k \ddot{\F}_k + \ddot{\E}_k^\T \M \right) \M_T \left(\A_k \ddot{\F}_k + \ddot{\E}_k^\T \M \right)^\T$ is positive semi-definite and has rank at most $r_k$. Similarly, we can show that
\begin{align*}
    \lambda_{j - r_k} \left(\ddot{\E}_k^\T \left(\M_T - \M\M_T\M^\T \right) \ddot{\E}_k \right) &\leq \lambda_{j - r_k} \left(\ddot{\E}_k^\T \left(\M_T - \M\M_T\M^\T \right) \ddot{\E}_k + \ddot{\E}_k^\T  \M\M_T\M^\T  \ddot{\E}_k\right)\\ &= \lambda_{j - r_k} \left(\ddot{\E}_k^\T \M_T  \ddot{\E}_k\right),
\end{align*}
\begin{align*}
    \lambda_{j+r_k}\left(\ddot{\E}_k^\T \M_T \ddot{\E}_k \right) &\leq
    \lambda_{j}\left(\ddot{\E}_k^\T \left(\M_T - \M\M_T\M^\T \right) \ddot{\E}_k \right)  + \lambda_{r_k+1}\left(\ddot{\E}_k^\T  \M\M_T\M^\T  \ddot{\E}_k \right)\\ &= \lambda_{j}\left(\ddot{\E}_k^\T \left(\M_T - \M\M_T\M^\T \right) \ddot{\E}_k \right).
\end{align*}
Therefore, for $r_k + 1 \leq j \leq \lfloor c\min{(T,d_k)} \rfloor - r_k$,
\begin{align*}
    \lambda_{j+r_k}\left(\ddot{\E}_k^\T \M_T \ddot{\E}_k \right) \leq \lambda_{j}\left(\ddot{\X}_k^\T \M_T \ddot{\X}_k \right) \leq \lambda_{j - r_k} \left(\ddot{\E}_k^\T \M_T  \ddot{\E}_k\right),
\end{align*}
which implies (\ref{eqn:lemma2.7}) with the result of Lemma \ref{lem:1}. This completes the proof of Lemma \ref{lem:2}. $\square$

{\em Proof of Lemma \ref{lem:3}}.
For $\A_T$ can either be $\A_{f,T}$, $\A_{e,T}$ or $\A_{\epsilon,T}$, we have already shown that $\|\A_{T}\|_1 = O(1)$ in the Proof of Lemma \ref{lem:1} and Lemma \ref{lem:2} (see (\ref{eqn:1norm_AT}) for example). Hence
\begin{align*}
    \1_T^\T \A_{T} \1_T \leq \sum_{t,s=1}^T |(\A_T)_{ts}| \leq T \|\A_{T}\|_1 = O(T).
\end{align*}
In addition, it is not difficult to see every entry of $\A_T$ has absolute value bounded above by 1, hence,
\begin{align*}
    \|\A_{T}\|_F^2 = \sum_{t,s=1}^T |(\A_T)_{ts}|^2 \leq \sum_{t,s=1}^T |(\A_T)_{ts}| \leq T \|\A_{T}\|_1 = O(T).
\end{align*}
Next, for $\A_{T^2}$ can either be $\A_{f,T^2}$, $\A_{e,T^2}$ or $\A_{\epsilon,T^2}$, we have
\begin{align*}
    \left\| \A_{T^2} \right\|_1 &=  \max_{t_1,s_1} \sum_{t_2,s_2 = 1}^T \left|(\A_{T^2})_{t_1s_1,t_2s_2}\right|  \leq 2 \max_{q_1q_2} \sum_{q_3,q_4} \left|a_{q_1}a_{q_2}a_{q_3}a_{q_4} \right| \leq 2\left(\sum_{q \geq 0} |a_{q}|\right)^4 \leq C.
\end{align*}
Hence,
\begin{align*}
    \1_T^\T \A_{T^2} \1_T \leq \sum_{t,s=1}^T |(\A_{T^2})_{ts}| \leq T^2 \|\A_{T^2}\|_1 = O(T^2).
\end{align*}
Similarly, we can also observe that every entry of $\A_{T^2}$ has absolute value bounded above by 1. To see this, take $f_{t,l,j}^{(k)}$ for example, note that for the MA process $f_t:=f_{t,l,j}^{(k)}$ for any $(l,j,k)$, $\left|\mathbb{E} (f_{t_1}f_{t_2}f_{t_3}f_{t_4})\right| \leq 0.5 \left[\mathbb{E} (f_{t_1}^2f_{t_2}^2) + \mathbb{E} (f_{t_3}^2f_{t_4}^2)\right] \leq 0.25 \left[\mathbb{E} (f_{t_1}^4) +\mathbb{E} (f_{t_2}^4) + \mathbb{E} (f_{t_3}^4) + \mathbb{E} (f_{t_4}^4)  \right] \leq 1$, because $\mathbb{E} (f_{t_i}^4) = \sum_{q\geq 0} a_q^4 \leq \left(\sum_{q\geq 0} a_q^2\right)^2 = 1$ for any $t_i$. Therefore,
\begin{align*}
    \|\A_{T^2}\|_F^2 = \sum_{t,s=1}^T |(\A_{T^2})_{ts}|^2 \leq \sum_{t,s=1}^T |(\A_{T^2})_{ts}| \leq T^2 \|\A_{T^2}\|_1 = O(T^2).
\end{align*}
(\ref{eqn:R3.3}) -- (\ref{eqn:R3.6}) can be easily implied by (\ref{eqn:R3.1}) and (\ref{eqn:R3.2}). To see this, first note that for any matrices $A$ and $B$, $\|A \otimes B\|_F = \|A\|_F  \|B\|_F$, and  $|\vec{(A)}^\T \vec{(B)}| \leq \|A\|_F  \|B\|_F$ by Cauchy-Schwarz, which implies (\ref{eqn:R3.3}), (\ref{eqn:R3.5}) and (\ref{eqn:R3.6}). Finally, $\1_{T^2}^\T \A_{f,T} \otimes \A_{e,T} \1_{T^2} \leq T^2 \|\A_{f,T} \otimes \A_{e,T}\|_1 = T^2 \|\A_{f,T}\|_1 \|\A_{e,T}\|_1 = O(T^2)$, which gives (\ref{eqn:R3.4}). This completes the proof of Lemma \ref{lem:3}. $\square$

{\em Proof of Theorem \ref{thm:sumestimatorrate_newmethod}}. Note that we have
\begin{align}{\label{eqn:covariance_decomposition}}
    \frac{\ddot{\X}_{k}^\T \left( \I_T - \frac{1}{T} \1_T \1_T^\T \right) \ddot{\X}_{k}}{T} &= \frac{\left(\A_k \ddot{\F}_k + \ddot{\E}_k^\T \right) \left( \I_T - \frac{1}{T} \1_T \1_T^\T \right) \left(\A_k \ddot{\F}_k + \ddot{\E}_k^\T \right)^\T}{T} \notag \\
    &= \A_k \frac{\ddot{\F}_k \left( \I_T - \frac{1}{T} \1_T \1_T^\T \right)\ddot{\F}_k^\T}{T} \A_k^\T + R_k,
\end{align}
where
\begin{align}{\label{eqn:decomposition_R}}
    R_k = \frac{\A_k\ddot{\F}_k \left( \I_T - \frac{1}{T} \1_T \1_T^\T \right) \ddot{\E}_k}{T} + \frac{\ddot{\E}_k^\T \left( \I_T - \frac{1}{T} \1_T \1_T^\T \right)\ddot{\F}_k^\T \A_k^\T}{T} + \frac{\ddot{\E}_k^\T \left( \I_T - \frac{1}{T} \1_T \1_T^\T \right) \ddot{\E}_k}{T}.
\end{align}
We estimate $\hat\Q_{k,(z_k)}$ as the first $z_k$ eigenvectors of (\ref{eqn:covariance_decomposition}). Let's define $\U_{k,(z_k)}$ to be the matrix consisting of the first $z_k$ columns of $\U_k$, and $\B$ be its orthogonal complement. Then $\U_{k,(z_k)}$ is an invariant subspace for $\A_k \A_k^\T$ and
\begin{align*}
    \left[
\begin{array}{c}
  \U_{k,(z_k)}^\T  \\
  \B^\T
\end{array}
\right] \A_k \A_k^\T \left[\U_{k,(z_k)} \ \ \B \right] =  \left[\begin{array}{cc}
  \G_{k,(z_k)} & 0  \\
  0 & \Lambda_{k,(z_k)}
\end{array}
\right],
\end{align*}
where $\G_{k,(z_k)}$ is a $z_k \times z_k$ diagonal matrix consisting the largest $z_k$ eigenvalues of $\A_k^\T \A_k$, and $\Lambda_{k,(z_k)}$ is a $(d_k - z_k) \times (d_k - z_k)$ diagonal matrix where the first $r_k - z_k$ entries are the $z_k + 1$ to $r_k$ eigenvalues of $\A_k^\T \A_k$, and the remaining entries are all 0's. In this way, we can apply Lemma 3 of \cite{Lametal2011} and know that there exists $\hat\U_{k,(z_k)}$ (with $\hat\U_{k,(z_k)}^\T \hat\U_{k,(z_k)} = \I_{z_k}$) such that $\hat\Q_{k,(z_k)} = \hat\U_{k,(z_k)} \P_{k,(z_k)}$ where $\P_{k,(z_k)}$ is orthogonal matrix, so that
\begin{align}{\label{eqn:lemme3_Lam_decomposition}}
    \norm{\hat\U_{k,(z_k)} - \U_{k,(z_k)} } &\leq  \frac{8\bigg\|\A_k \left[ \frac{\ddot{\F}_k \left( \I_T - \frac{1}{T} \1_T \1_T^\T \right)\ddot{\F}_k^\T}{T} - \I_{r_k} \right] \A_k^\T + R_k \bigg\|}{\text{sep}(\G_{k,z_k}, \Lambda_{k,z_k})}  \notag \\
    & \preceq \frac{\bigg\| \A_k \left[ \frac{\ddot{\F}_k \left( \I_T - \frac{1}{T} \1_T \1_T^\T \right)\ddot{\F}_k^\T}{T} - \I_{r_k} \right] \A_k^\T\bigg\| + \norm{ R_k } }{d_k^{\alpha_{k,z_k}}}.
\end{align}
Next, we bound the norms on the numerator of (\ref{eqn:lemme3_Lam_decomposition}). For the first term,
\begin{align*}
    \bigg\| \A_k \left[ \frac{\ddot{\F}_k \left( \I_T - \frac{1}{T} \1_T \1_T^\T \right)\ddot{\F}_k^\T}{T} - \I_{r_k} \right] \A_k^\T\bigg\| &\leq \norm{\A_k }^2 \bigg\|\frac{\ddot{\F}_k \left( \I_T - \frac{1}{T} \1_T \1_T^\T \right)\ddot{\F}_k^\T}{T} - \I_{r_k} \bigg\| \notag \\
    & \asymp d_k^{\alpha_{k,1}}\bigg\|\frac{\ddot{\F}_k \left( \I_T - \frac{1}{T} \1_T \1_T^\T \right)\ddot{\F}_k^\T}{T} - \I_{r_k} \bigg\|.
\end{align*}
By Assumption (F1), $\{f_{t,l,j}^{(k)}\}$ is a linear process with absolutely summable autocovariance sequence. In Lemma \ref{lem:2}, we have shown that elements of $\ddot{\F}_k$ retain the covariance structure of $\{f_{t,l,j}^{(k)}\}$, so each row of $\ddot{\F}_k$ is a linear process with absolutely summable autocovariance sequence, which satisfies Assumption (R2) in \cite{Lam2021}. Thus, applying the result from Lemma 3 (or more specifically, equation (8.26)) of \cite{Lam2021} implies
\begin{align}\label{eqn:bound_FF_minus_I}
    \bigg\|\frac{\ddot{\F}_k \left( \I_T - \frac{1}{T} \1_T \1_T^\T \right)\ddot{\F}_k^\T}{T} - \I_{r_k} \bigg\| = O_p\left(\sqrt{\frac{r_k}{T}} \right),
\end{align}
which further gives
\begin{align*}
    \bigg\| \A_k \left[ \frac{\ddot{\F}_k \left( \I_T - \frac{1}{T} \1_T \1_T^\T \right)\ddot{\F}_k^\T}{T} - \I_{r_k} \right] \A_k^\T\bigg\| = O_p\left(d_k^{\alpha_{k,1}} \sqrt{\frac{r_k}{T}}\right).
\end{align*}
Next, we bound the norm of $R_k$. First, note that $\norm{ \I_T - \frac{1}{T} \1_T \1_T^\T } \leq \norm{ \I_T} + \norm{\frac{1}{T} \1_T \1_T^\T } = 2$. Bounding the squared norm of each term on the right hand side of  (\ref{eqn:decomposition_R}), we have
\begin{align}\label{eqn:original_rate_FE}
    \bigg\|\frac{\A_k\ddot{\F}_k \left( \I_T - \frac{1}{T} \1_T \1_T^\T \right) \ddot{\E}_k}{T}\bigg\|^2 &\leq \bigg\| \I_T - \frac{1}{T} \1_T \1_T^\T \bigg\|^2 \bigg\|\frac{\A_k\ddot{\F}_k }{T^{\frac{1}{2}}}\bigg\|^2 \bigg\| \frac{\ddot{\E}_k}{T^{\frac{1}{2}}}\bigg\|^2  \notag \\
    & \leq 4 \lambda_1\left(\frac{\A_k\ddot{\F}_k\ddot{\F}_k^\T \A_k^\T}{T}\right)\lambda_1\left(\frac{\ddot{\E}_k^\T \ddot{\E}_k}{T}\right) \notag \\
    & = O_p\left(d_k^{\alpha_{k,1}}\right) O_p\left(\frac{d_{\text{-}k}}{s_{\text{-}k}} \left(1 + \frac{d_k}{T} \right) \right),
\end{align}
where the last line follows from Lemma \ref{lem:1} and Lemma \ref{lem:2}. Similarly,
\begin{align*}
    \bigg\|\frac{\ddot{\E}_k^\T \left( \I_T - \frac{1}{T} \1_T \1_T^\T \right) \ddot{\F}_k^\T \A_k^\T}{T}\bigg\|^2 =  O_p\left(d_k^{\alpha_{k,1}}\right) O_p\left(\frac{d_{\text{-}k}}{s_{\text{-}k}} \left(1 + \frac{d_k}{T} \right) \right).
\end{align*}
As for the last term,
\begin{align}\label{eqn:rate_EE}
    \bigg\|\frac{\ddot{\E}_k^\T \left( \I_T - \frac{1}{T} \1_T \1_T^\T \right) \ddot{\E}_k}{T}  \bigg\|^2 & \leq 4 \lambda_1\left(\frac{\ddot{\E}_k^\T \ddot{\E}_k}{T}\right)^2 = O_p\left(\frac{d_{\text{-}k}^2}{s_{\text{-}k}^2} \left(1 + \frac{d_k^2}{T^2} \right) \right)
\end{align}
by Lemma \ref{lem:1} and Lemma \ref{lem:2}.

For the first two terms on the right hand side of (\ref{eqn:decomposition_R}), there may exist potentially better bounds. To see this, note that we can equivalently write
\begin{align*}
    \bigg\|\frac{\A_k\ddot{\F}_k \left( \I_T - \frac{1}{T} \1_T \1_T^\T \right)  \ddot{\E}_k}{T}\bigg\|^2 &\leq  \norm{\A_k}^2 \bigg\| \frac{\ddot{\F}_k\ddot{\E}_k}{T}\bigg\|^2   + \norm{\A_k}^2 \bigg\| \frac{\ddot{\F}_k\frac{1}{T} \1_T \1_T^\T\ddot{\E}_k}{T}\bigg\|^2  \notag \\
    & = O_p\left(d_k^{\alpha_{k,1}}\right) \left(\bigg\|\frac{\ddot{\F}_k\ddot{\E}_k}{T}\bigg\|^2  + \bigg\| \frac{\ddot{\F}_k\frac{1}{T} \1_T \1_T^\T\ddot{\E}_k}{T}\bigg\|^2 \right).
\end{align*}
Since $\frac{\ddot{\F}_k\ddot{\E}_k}{T}$ and $\frac{\ddot{\F}_k\frac{1}{T} \1_T \1_T^\T\ddot{\E}_k}{T}$ are $r_k \times d_k$ matrices with each element having a certain rate of convergence, we actually can have a better rate by simply counting the numbers of elements of them. More specifically, the $(i,j)$ element of $\ddot{\F}_k\ddot{\E}_k$ is
\begin{align*}
    \left(\ddot{\F}_k\ddot{\E}_k\right)_{ij} = \sum_{t=1}^T \ddot{f}_{t,k,i} \ddot{e}_{t,k,j},
\end{align*}
where $\ddot{f}_{t,k,i}$ is the $i$-th entry of $\ddot{\f}_{t,k}$, and $\ddot{e}_{t,k,j}$ are defined similarly. We now focus on bounding the Frobenius norm of $\ddot{\F}_k\ddot{\E}_k$. Note that $\|\ddot{\F}_k\ddot{\E}_k \|_F^2 = \sum_{(i,j) \in (r_k, d_k)} \left(\ddot{\F}_k\ddot{\E}_k\right)_{ij}^2 $ and
\begin{align}{\label{eqn:FE_rate_decomposition}}
    \|\ddot{\F}_k\ddot{\E}_k \|_F^2 &= \mathbb{E} \|\ddot{\F}_k\ddot{\E}_k \|_F^2 + O_p\left(\sqrt{Var(\|\ddot{\F}_k\ddot{\E}_k \|_F^2)} \right) \notag \\
    &= \mathbb{E} \|\ddot{\F}_k\ddot{\E}_k \|_F^2 + O_p\left(\sqrt{\mathbb{E}\left(\|\ddot{\F}_k\ddot{\E}_k \|_F^4\right) - \left(\mathbb{E} \|\ddot{\F}_k\ddot{\E}_k \|_F^2 \right)^2} \right).
\end{align}
Thus, we need to obtain bounds for $\mathbb{E} \|\ddot{\F}_k\ddot{\E}_k \|_F^2$ and $\mathbb{E} \|\ddot{\F}_k\ddot{\E}_k \|_F^4$. We start with $\mathbb{E} \|\ddot{\F}_k\ddot{\E}_k \|_F^2$. For each entry of $\ddot{\F}_k\ddot{\E}_k$, we have
\begin{align*}
    \mathbb{E} \left(\ddot{\F}_k\ddot{\E}_k\right)_{ij}^2 &= \mathbb{E} \left(\sum_{t=1}^T \ddot{f}_{t,k,i} \ddot{e}_{t,k,j} \right)^2 =\sum_{t=1}^T \sum_{s=1}^T \mathbb{E}\left(\ddot{f}_{t,k,i}\ddot{f}_{s,k,i} \right) \mathbb{E}\left(\ddot{e}_{t,k,i}\ddot{e}_{s,k,i} \right).
\end{align*}
By Assumption (F1),
\begin{align}{\label{eqn:ftsi_expectation}}
    \mathbb{E}\left(\ddot{f}_{t,k,i}\ddot{f}_{s,k,i} \right) = \sum_{q \geq 0} a_{f,q} a_{f,q-|t-s|} = \left(\A_{f,T}\right)_{ts}.
\end{align}
Similarly, by Assumption (E1) and (E2),
\begin{align}{\label{eqn:etsi_expectation}}
    \mathbb{E}\left(\ddot{e}_{t,k,j}\ddot{e}_{s,k,j} \right) = \frac{1}{s_{\text{-}k}} \left[ \left(\bPsi^{(k)} \bPsi^{(k)\T} \right)_{jj} \left(\A_{e,T}\right)_{ts} + \left(\bSigma_{\epsilon}^{(k)}\right)_{jj} \left(\A_{\epsilon,T}\right)_{ts} \right].
\end{align}
Hence,
\begin{align*}
    \mathbb{E} \left(\ddot{\F}_k\ddot{\E}_k\right)_{ij}^2 &= \frac{1}{s_{\text{-}k}}\sum_{t=1}^T \sum_{s=1}^T  \left(\A_{f,T}\right)_{ts} \left[\left(\bPsi^{(k)} \bPsi^{(k)\T} \right)_{jj} \left(\A_{e,T}\right)_{ts} + \left(\bSigma_{\epsilon}^{(k)}\right)_{jj} \left(\A_{\epsilon,T}\right)_{ts}\right]  \notag \\
    & = \frac{1}{s_{\text{-}k}} \left[\left(\bPsi^{(k)} \bPsi^{(k)\T} \right)_{jj} \vec{(\A_{f,T})}^\T \vec{(\A_{e,T})} + \left(\bSigma_{\epsilon}^{(k)}\right)_{jj} \vec{(\A_{f,T})}^\T \vec{(\A_{\epsilon,T})}\right],
\end{align*}
and
\begin{align}\label{eqn:second_moment_fe}
    \mathbb{E} \left\|\ddot{\F}_k\ddot{\E}_k \right\|_F^2 &= \sum_{i = 1}^{r_k} \sum_{j=1}^{d_k} \mathbb{E} \left(\ddot{\F}_k\ddot{\E}_k\right)_{ij}^2 \notag \\
    & = \frac{r_k}{s_{\text{-}k}} \left[\tr(\bPsi^{(k)} \bPsi^{(k)\T}) \vec{(\A_{f,T})}^\T \vec{(\A_{e,T})} + \tr(\bSigma_{\epsilon}^{(k)}) \vec{(\A_{f,T})}^\T \vec{(\A_{\epsilon,T})} \right] \notag \\
    & = O_p\left(r_k d_k T\frac{d_{\text{-}k}}{s_{\text{-}k}}\right),
\end{align}
where the last line follows from Assumption (E1) and Lemma \ref{lem:3}. Next, for $\mathbb{E} \left\|\ddot{\F}_k\ddot{\E}_k \right\|_F^4$, note that
\begin{align*}
     \mathbb{E} \left\|\ddot{\F}_k\ddot{\E}_k \right\|_F^4 & = \sum_{i_1=1}^{r_k} \sum_{j_1=1}^{d_k} \sum_{i_2=1}^{r_k} \sum_{j_2=1}^{d_k} \mathbb{E} \left[(\ddot{\F}_k\ddot{\E}_k)_{i_1j_1}^2 (\ddot{\F}_k\ddot{\E}_k)_{i_2j_2}^2\right],
\end{align*}
and
\begin{align*}
    \mathbb{E} \left[(\ddot{\F}_k\ddot{\E}_k)_{i_1j_1}^2 (\ddot{\F}_k\ddot{\E}_k)_{i_2j_2}^2\right] & = \mathbb{E} \left[ \left(\sum_{t=1}^T \ddot{f}_{t,k,i_1} \ddot{e}_{t,k,j_1} \right)^2 \left(\sum_{t=1}^T \ddot{f}_{t,k,i_2} \ddot{e}_{t,k,j_2} \right)^2   \right] \notag \\
    & = \sum_{t_1=1}^T \sum_{s_1=1}^T \sum_{t_2=1}^T \sum_{s_2=1}^T \mathbb{E} \left(\ddot{f}_{t_1,k,i_1} \ddot{f}_{s_1,k,i_1} \ddot{f}_{t_2,k,i_2} \ddot{f}_{s_2,k,i_2} \right)\\  &\cdot\mathbb{E} \left(\ddot{e}_{t_1,k,j_1} \ddot{e}_{s_1,k,j_1} \ddot{e}_{t_2,k,j_2} \ddot{e}_{s_2,k,j_2} \right).
\end{align*}
We deal with the terms separately. For the term related to $\ddot{f}$, we separate the cases for $i_1 = i_2$ and $i_1 \neq i_2$. When $i_1 = i_2$, by Assumption (F1), we have
\begin{align*}
    \mathbb{E} \left(\ddot{f}_{t_1,k,i_1} \ddot{f}_{s_1,k,i_1} \ddot{f}_{t_2,k,i_2} \ddot{f}_{s_2,k,i_2} \right) = \mathbb{E} \left(\ddot{f}_{t_1,k,i_1} \ddot{f}_{s_1,k,i_1} \ddot{f}_{t_2,k,i_1} \ddot{f}_{s_2,k,i_1} \right) = \left( \A_{f,T^2} \right)_{t_1s_1,t_2s_2},
\end{align*}
where $\left( \A_{f,T^2} \right)_{t_1s_1,t_2s_2}$ is the $(t_1s_1,t_2s_2)$ entry of $\A_{f,T^2}$. When $i_1 \neq i_2$,
\begin{align*}
    \mathbb{E} \left(\ddot{f}_{t_1,k,i_1} \ddot{f}_{s_1,k,i_1} \ddot{f}_{t_2,k,i_2} \ddot{f}_{s_2,k,i_2} \right) = \mathbb{E} \left(\ddot{f}_{t_1,k,i_1} \ddot{f}_{s_1,k,i_1}\right) \mathbb{E}\left( \ddot{f}_{t_2,k,i_2} \ddot{f}_{s_2,k,i_2} \right) = \left( \A_{f,T} \right)_{t_1s_1} \left( \A_{f,T} \right)_{t_2s_2}.
\end{align*}
Next, for the term related to $\ddot{e}$, by Assumption (E1) and (E2),
\begin{align}{\label{eqn:four_products_of_e}}
     &\mathbb{E} \left(\ddot{e}_{t_1,k,j_1} \ddot{e}_{s_1,k,j_1} \ddot{e}_{t_2,k,j_2} \ddot{e}_{s_2,k,j_2} \right) \notag \\
     \leq & \frac{1}{s_{\text{-}k}^2} \left[ \left(\bPsi^{(k)} \bPsi^{(k)\T} \right)_{j_1j_1} \left(\bPsi^{(k)} \bPsi^{(k)\T} \right)_{j_2j_2} \left( \A_{e,T^2} \right)_{t_1s_1,t_2s_2} + \sum_{l=1}^{d_{\text{-}k}}(\bSigma_{\epsilon,l}^{(k)})_{j_1j_1}(\bSigma_{\epsilon,l}^{(k)})_{j_2j_2} \left( \A_{\epsilon,T^2} \right)_{t_1s_1,t_2s_2} \right]\notag \\
     \leq & \frac{1}{s_{\text{-}k}^2}  \left[ \left(\bPsi^{(k)} \bPsi^{(k)\T} \right)_{j_1j_1} \left(\bPsi^{(k)} \bPsi^{(k)\T} \right)_{j_2j_2} \left( \A_{e,T^2} \right)_{t_1s_1,t_2s_2} + (\bSigma_{\epsilon}^{(k)})_{j_1j_1}(\bSigma_{\epsilon}^{(k)})_{j_2j_2} \left( \A_{\epsilon,T^2} \right)_{t_1s_1,t_2s_2} \right].
\end{align}
Hence, when $i_1 = i_2$,
\begin{align*}
    &\mathbb{E} \left[(\ddot{\F}_k\ddot{\E}_k)_{i_1j_1}^2 (\ddot{\F}_k\ddot{\E}_k)_{i_2j_2}^2\right] \notag \\
    \leq & \frac{1}{s_{\text{-}k}^2} \bigg[\left(\bPsi^{(k)} \bPsi^{(k)\T} \right)_{j_1j_1} \left(\bPsi^{(k)} \bPsi^{(k)\T} \right)_{j_2j_2} \vec{(\A_{f,T^2})}^\T \vec{(\A_{e,T^2})}\\ &+ (\bSigma_{\epsilon}^{(k)})_{j_1j_1}(\bSigma_{\epsilon}^{(k)})_{j_2j_2} \vec{(\A_{f,T^2})}^\T \vec{(\A_{\epsilon,T^2})}\bigg],
\end{align*}
and when $i_1 \neq i_2$,
\begin{align*}
    &\mathbb{E} \left[(\ddot{\F}_k\ddot{\E}_k)_{i_1j_1}^2 (\ddot{\F}_k\ddot{\E}_k)_{i_2j_2}^2\right] \notag \\
    \leq & \frac{1}{s_{\text{-}k}^2} \bigg[\left(\bPsi^{(k)} \bPsi^{(k)\T} \right)_{j_1j_1} \left(\bPsi^{(k)} \bPsi^{(k)\T} \right)_{j_2j_2} \vec{(\A_{f,T} \otimes \A_{f,T})}^\T \vec{(\A_{e,T^2})}\\ &+ (\bSigma_{\epsilon}^{(k)})_{j_1j_1}(\bSigma_{\epsilon}^{(k)})_{j_2j_2} \vec{(\A_{f,T} \otimes \A_{f,T})}^\T \vec{(\A_{\epsilon,T^2})}\bigg].
\end{align*}
Thus, by Lemma \ref{lem:3}, we have for any $(i_1,j_1,i_2,j_2)$,
\begin{align*}
    &\mathbb{E} \left[(\ddot{\F}_k\ddot{\E}_k)_{i_1j_1}^2 (\ddot{\F}_k\ddot{\E}_k)_{i_2j_2}^2\right] \notag \\
    \leq & O\left(T^2 \frac{1}{s_{\text{-}k}^2}\right)\left[\left(\bPsi^{(k)} \bPsi^{(k)\T} \right)_{j_1j_1} \left(\bPsi^{(k)} \bPsi^{(k)\T} \right)_{j_2j_2} + (\bSigma_{\epsilon}^{(k)})_{j_1j_1}(\bSigma_{\epsilon}^{(k)})_{j_2j_2} \right] .
\end{align*}
Therefore,
\begin{align}\label{eqn:fourth_moment_fe}
    \mathbb{E} \left\|\ddot{\F}_k\ddot{\E}_k \right\|_F^4  & = \sum_{i_1=1}^{r_k} \sum_{j_1=1}^{d_k} \sum_{i_2=1}^{r_k} \sum_{j_2=1}^{d_k} \mathbb{E} \left[(\ddot{\F}_k\ddot{\E}_k)_{i_1j_1}^2 (\ddot{\F}_k\ddot{\E}_k)_{i_2j_2}^2\right] \notag \\
    & \leq r_k^2 O\left(T^2 \frac{1}{s_{\text{-}k}^2}\right) \left\{\left[\tr{(\bPsi^{(k)} \bPsi^{(k)\T})}\right]^2 + \left[\tr{(\bSigma_{\epsilon}^{(k)})}\right]^2 \right\}  = O\left(r_k^2 T^2 d_k^2 \frac{d_{\text{-}k}^2}{s_{\text{-}k}^2}\right),
\end{align}
by Assumption (E1). (\ref{eqn:fourth_moment_fe}) together with (\ref{eqn:second_moment_fe}) and ({\ref{eqn:FE_rate_decomposition}}) imply $\left\|\ddot{\F}_k\ddot{\E}_k\right\|_F^2 = O_p\left(r_k d_k T\frac{d_{\text{-}k}}{s_{\text{-}k}}\right)$.

Next, we will show $\left\|\ddot{\F}_k\frac{1}{T} \1_T \1_T^\T\ddot{\E}_k\right\|_F^2 = O_p\left(r_k d_k \frac{d_{\text{-}k}}{s_{\text{-}k}}\right)$ in a similar manner. First, note that
\begin{align*}
    \left(\ddot{\F}_k\frac{1}{T} \1_T \1_T^\T\ddot{\E}_k\right)_{ij} = \frac{1}{T} \left(\sum_{t=1}^T \ddot{f}_{t,k,i}\right) \left(\sum_{t=1}^T \ddot{e}_{t,k,j}\right),
\end{align*}
and
\begin{align*}
    \mathbb{E} \left(\ddot{\F}_k\frac{1}{T} \1_T \1_T^\T\ddot{\E}_k\right)_{ij}^2 &= \frac{1}{T^2} \sum_{t_1=1}^T \sum_{s_1=1}^T \sum_{t_2=1}^T \sum_{s_2=1}^T \mathbb{E} \left(\ddot{f}_{t_1,k,i} \ddot{f}_{s_1,k,i} \right) \mathbb{E} \left(\ddot{e}_{t_2,k,j} \ddot{e}_{s_2,k,j} \right) \notag \\
    &= \frac{1}{T^2} \sum_{t_1,t_2,s_1,s_2=1}^T  (\A_{f,T})_{t_1s_1} \frac{1}{s_{\text{-}k}} \left[ \left(\bPsi^{(k)} \bPsi^{(k)\T} \right)_{jj} \left(\A_{e,T}\right)_{t_2s_2} + \left(\bSigma_{\epsilon}^{(k)}\right)_{jj} \left(\A_{\epsilon,T}\right)_{t_2s_2} \right] \notag \\
    & = \frac{1}{T^2s_{\text{-}k}} \left[ \left(\bPsi^{(k)} \bPsi^{(k)\T}\right)_{jj} \1_{T^2}^\T \A_{f,T} \otimes \A_{e,T} \1_{T^2} + \left(\bSigma_{\epsilon}^{(k)}\right)_{jj} \1_{T^2}^\T \A_{f,T} \otimes \A_{\epsilon,T} \1_{T^2} \right] \notag \\
    &= O\left(\frac{1}{s_{\text{-}k}}\right) \left[ \left(\bPsi^{(k)} \bPsi^{(k)\T}\right)_{jj} + \left(\bSigma_{\epsilon}^{(k)}\right)_{jj} \right]
\end{align*}
where the second line follows from (\ref{eqn:ftsi_expectation}) and (\ref{eqn:etsi_expectation}), and the last line follows from Lemma \ref{lem:3}. Hence,
\begin{align}\label{eqn:second_moment_f11e}
    \mathbb{E} \left\|\ddot{\F}_k\frac{1}{T} \1_T \1_T^\T\ddot{\E}_k\right\|_F^2 &= \sum_{i = 1}^{r_k} \sum_{j = 1}^{d_k} O\left(\frac{1}{s_{\text{-}k}}\right) \left[ \left(\bPsi^{(k)} \bPsi^{(k)\T}\right)_{jj} + \left(\bSigma_{\epsilon}^{(k)}\right)_{jj} \right]  \notag \\
    & = O\left(r_k\frac{1}{s_{\text{-}k}}\right) \left[ \tr{(\bPsi^{(k)} \bPsi^{(k)\T})} + \tr{(\bSigma_{\epsilon}^{(k)})}  \right] = O\left(r_k d_k \frac{d_{\text{-}k}}{s_{\text{-}k}}\right).
\end{align}
Next, we know
\begin{align*}
    \mathbb{E} \left\|\ddot{\F}_k\frac{1}{T} \1_T \1_T^\T \ddot{\E}_k \right\|_F^4 & = \frac{1}{T^4} \sum_{i_1=1}^{r_k} \sum_{j_1=1}^{d_k} \sum_{i_2=1}^{r_k} \sum_{j_2=1}^{d_k} \mathbb{E} \left[(\ddot{\F}_k\1_T \1_T^\T\ddot{\E}_k)_{i_1j_1}^2 (\ddot{\F}_k\1_T \1_T^\T\ddot{\E}_k)_{i_2j_2}^2\right],
\end{align*}
where
\begin{align*}
    \mathbb{E} \left[(\ddot{\F}_k\1_T \1_T^\T\ddot{\E}_k)_{i_1j_1}^2 (\ddot{\F}_k\1_T \1_T^\T\ddot{\E}_k)_{i_2j_2}^2\right] & = \mathbb{E} \left[ \left(\sum_{t=1}^T \ddot{f}_{t,k,i_1}\right)^2 \left(\sum_{t=1}^T \ddot{f}_{t,k,i_2}\right)^2 \right] \mathbb{E}\left[ \left(\sum_{t=1}^T \ddot{e}_{t,k,j_1}\right)^2 \left(\sum_{t=1}^T \ddot{e}_{t,k,j_2}\right)^2  \right].
\end{align*}
For the terms related to $\ddot{f}$, when $i_1 = i_2$,
\begin{align*}
    \mathbb{E} \left[ \left(\sum_{t=1}^T \ddot{f}_{t,k,i_1}\right)^2 \left(\sum_{t=1}^T \ddot{f}_{t,k,i_2}\right)^2 \right] & = \mathbb{E} \left[ \left(\sum_{t=1}^T \ddot{f}_{t,k,i_1}\right)^4 \right] = \sum_{t_1,t_2,s_1,s_2 = 1}^T  \mathbb{E} \left[\ddot{f}_{t_1,k,i_1} \ddot{f}_{s_1,k,i_1} \ddot{f}_{t_2,k,i_1} \ddot{f}_{s_2,k,i_1} \right] \notag \\
    & =  \sum_{t_1,t_2,s_1,s_2 = 1}^T \left(\A_{f,T^2}\right)_{t_1s_1,t_2s_2} = \1_{T^2}^\T \A_{f,T^2} \1_{T^2} = O(T^2)
\end{align*}
by Lemma \ref{lem:3}. When $i_1 \neq i_2$,
\begin{align*}
    \mathbb{E} \left[ \left(\sum_{t=1}^T \ddot{f}_{t,k,i_1}\right)^2 \left(\sum_{t=1}^T \ddot{f}_{t,k,i_2}\right)^2 \right] & = \mathbb{E} \left[ \left(\sum_{t=1}^T \ddot{f}_{t,k,i_1}\right)^2 \right] \mathbb{E} \left[\left(\sum_{t=1}^T \ddot{f}_{t,k,i_2}\right)^2 \right] \notag \\
    &= \left[\sum_{t=1}^T \sum_{s=1}^T \mathbb{E} \left(\ddot{f}_{t,k,i_1} \ddot{f}_{s,k,i_1} \right) \right]^2  = \left(\1_T^\T \A_{f,T} \1_T\right)^2  = O(T^2)
\end{align*}
by Lemma \ref{lem:3}. Therefore,  $\mathbb{E} \left[ \left(\sum_{t=1}^T \ddot{f}_{t,k,i_1}\right)^2 \left(\sum_{t=1}^T \ddot{f}_{t,k,i_2}\right)^2 \right] = O(T^2)$ for any $(i_1,i_2)$. Finally, for terms with $\ddot{e}$,
\begin{align*}
    &\mathbb{E}\left[ \left(\sum_{t=1}^T \ddot{e}_{t,k,j_1}\right)^2 \left(\sum_{t=1}^T \ddot{e}_{t,k,j_2}\right)^2  \right]
    = \sum_{t_1 = 1}^T \sum_{s_1 = 1}^T \sum_{t_2 = 1}^T \sum_{s_2 = 1}^T \mathbb{E} \left[ \ddot{e}_{t_1,k,j_1} \ddot{e}_{s_1,k,j_1} \ddot{e}_{t_2,k,j_2} \ddot{e}_{s_2,k,j_2}  \right] \notag \\
    \leq & \sum_{t_1,t_2,s_1,s_2 = 1}^T \frac{1}{s_{\text{-}k}^2} \left[\left(\bPsi^{(k)} \bPsi^{(k)\T} \right)_{j_1j_1} \left(\bPsi^{(k)} \bPsi^{(k)\T} \right)_{j_2j_2} \left( \A_{e,T^2} \right)_{t_1s_1,t_2s_2} + (\bSigma_{\epsilon}^{(k)})_{j_1j_1}(\bSigma_{\epsilon}^{(k)})_{j_2j_2} \left( \A_{\epsilon,T^2} \right)_{t_1s_1,t_2s_2}\right] \notag \\
    = & \frac{1}{s_{\text{-}k}^2}\left[\left(\bPsi^{(k)} \bPsi^{(k)\T} \right)_{j_1j_1} \left(\bPsi^{(k)} \bPsi^{(k)\T} \right)_{j_2j_2} \1_{T^2}^\T \A_{e,T^2} \1_{T^2} + (\bSigma_{\epsilon}^{(k)})_{j_1j_1}(\bSigma_{\epsilon}^{(k)})_{j_2j_2} \1_{T^2}^\T \A_{\epsilon,T^2} \1_{T^2} \right] \notag \\
    =& O\left(\frac{T^2}{s_{\text{-}k}^2}\right) \left[\left(\bPsi^{(k)} \bPsi^{(k)\T} \right)_{j_1j_1} \left(\bPsi^{(k)} \bPsi^{(k)\T} \right)_{j_2j_2} + (\bSigma_{\epsilon}^{(k)})_{j_1j_1}(\bSigma_{\epsilon}^{(k)})_{j_2j_2} \right],
\end{align*}
where the third line follows from ({\ref{eqn:four_products_of_e}}), and the last line from Lemma \ref{lem:3}. Therefore,
\begin{align*}
    \mathbb{E} \left[(\ddot{\F}_k\1_T \1_T^\T\ddot{\E}_k)_{i_1j_1}^2 (\ddot{\F}_k\1_T \1_T^\T\ddot{\E}_k)_{i_2j_2}^2\right] & = O\left(\frac{T^4}{s_{\text{-}k}^2}\right) \left[\left(\bPsi^{(k)} \bPsi^{(k)\T} \right)_{j_1j_1} \left(\bPsi^{(k)} \bPsi^{(k)\T} \right)_{j_2j_2} + (\bSigma_{\epsilon}^{(k)})_{j_1j_1}(\bSigma_{\epsilon}^{(k)})_{j_2j_2} \right].
\end{align*}
Hence,
\begin{align}\label{eqn:fourth_moment_of_f11e}
     \mathbb{E} \left\|\ddot{\F}_k\frac{1}{T} \1_T \1_T^\T \ddot{\E}_k \right\|_F^4 & = O\left(\frac{1}{s_{\text{-}k}^2}\right) \sum_{i_1=1}^{r_k} \sum_{j_1=1}^{d_k} \sum_{i_2=1}^{r_k} \sum_{j_2=1}^{d_k}  \left[\left(\bPsi^{(k)} \bPsi^{(k)\T} \right)_{j_1j_1} \left(\bPsi^{(k)} \bPsi^{(k)\T} \right)_{j_2j_2} + (\bSigma_{\epsilon}^{(k)})_{j_1j_1}(\bSigma_{\epsilon}^{(k)})_{j_2j_2} \right]  \notag \\
     &= r_k^2 O\left(\frac{1}{s_{\text{-}k}^2}\right)  \left\{\left[\tr{(\bPsi^{(k)} \bPsi^{(k)\T})}\right]^2 + \left[\tr{(\bSigma_{\epsilon}^{(k)})}\right]^2 \right\} = O\left(r_k^2 d_k^2 \frac{d_{\text{-}k}^2}{s_{\text{-}k}^2}\right).
\end{align}
(\ref{eqn:fourth_moment_of_f11e}) together with (\ref{eqn:second_moment_f11e}) imply $\left\|\ddot{\F}_k\frac{1}{T} \1_T \1_T^\T\ddot{\E}_k\right\|_F^2 = O_p\left(r_k d_k \frac{d_{\text{-}k}}{s_{\text{-}k}}\right)$. Therefore, we can conclude that
\begin{align}\label{eqn:new_rate_FE}
    \bigg\|\frac{\A_k\ddot{\F}_k \left( \I_T - \frac{1}{T} \1_T \1_T^\T \right)  \ddot{\E}_k}{T}\bigg\|^2
    & = O_p\left(d_k^{\alpha_{k,1}}\right) \left(\bigg\|\frac{\ddot{\F}_k\ddot{\E}_k}{T}\bigg\|^2  + \bigg\| \frac{\ddot{\F}_k\frac{1}{T} \1_T \1_T^\T\ddot{\E}_k}{T}\bigg\|^2 \right) = O_p\left(\frac{d_{\text{-}k}}{s_{\text{-}k}} \frac{r_kd_k}{T}  d_k^{\alpha_{k,1}}\right).
\end{align}
Compared with the original rate (\ref{eqn:original_rate_FE}), we can obtain a potentially better rate (\ref{eqn:new_rate_FE}) by directly bounding and counting the elements in a large matrix. In other words, the rate for $\bigg\|\frac{\A_k\ddot{\F}_k \left( \I_T - \frac{1}{T} \1_T \1_T^\T \right)  \ddot{\E}_k}{T}\bigg\|^2$ will be the minimum between the two, which is
\begin{align*}
    \bigg\|\frac{\A_k\ddot{\F}_k \left( \I_T - \frac{1}{T} \1_T \1_T^\T \right)  \ddot{\E}_k}{T}\bigg\|^2
    & = O_p\left(\frac{d_{\text{-}k}}{s_{\text{-}k}}  d_k^{\alpha_{k,1}} \min{\left\{1+\frac{d_k}{T},\frac{r_kd_k}{T}\right\}}\right).
\end{align*}
Combining the rates (\ref{eqn:original_rate_FE}), (\ref{eqn:new_rate_FE}) and  (\ref{eqn:rate_EE}) yields that
\begin{align*}
    \norm{ R_k }^2 = \frac{d_{\text{-}k}^2}{s_{\text{-}k}^2}\left(1+\frac{d_k^2}{T^2}\right) + \frac{d_{\text{-}k}}{s_{\text{-}k}}  d_k^{\alpha_{k,1}} \min{\left\{1+\frac{d_k}{T},\frac{r_kd_k}{T}\right\}} := c_k.
\end{align*}
Hence,
\begin{align*}
    \norm{\hat\U_{k,(z_k)} - \U_{k,(z_k)} }^2 = O_p \left(d_k^{-2\alpha_{k,z_k}} \left[d_k^{2\alpha_{k,1}}\frac{r_k}{T} + c_k \right]\right).
\end{align*}

Finally, to show (\ref{eqn:sumestimatorrate_newmethod_newrate}), let $\tilde{\V}_k$ be the $z_k \times z_k$ diagonal matrix of the first $z_k$ largest eigenvalues of $\hat{\Sigma}_{\tilde{\x}_{k}}$ in decreasing order, and $\ddot{\V}_k := \tilde{\V}_k / s_{\text{-}k}$. Then it follows from (\ref{eqn:covariance_decomposition}) that
\begin{align*}
    \hat{\Q}_{k,(z_k)} - \Q_k\ddot{\H}_k &= R_k\hat{\Q}_{k,(z_k)}\ddot{\V}_k^{-1}, \text{ implying }\\
    \norm{\hat{\Q}_{k,(z_k)} - \Q_k\ddot{\H}_k }^2 &\leq \norm{R_k }^2 \norm{\ddot{\V}_k^{-1} }^2 O_p\left(d_k^{-2\alpha_{k,z_k}} c_k \right).
\end{align*}
This completes the proof of Theorem \ref{thm:sumestimatorrate_newmethod}.

%

{\em Proof of Theorem \ref{thm:preaverage_sumestimatorrate_newmethod}}.
First, by the definition of $s_{\text{-}k,m}$, we have $\lambda_j\left(\frac{\tilde{\F}_{k,m}\tilde{\F}_{k,m}^\T}{T} \right) \asymp s_{\text{-}k,m}$ for $j \in [r_k]$ and for each $m$. Then,
\begin{align*}
\lambda_1\left(\frac{1}{M}\sum_{m=1}^M\frac{\tilde{\F}_{k,m}\tilde{\F}_{k,m}^\T}{T} \right) &\leq \frac{1}{M}\sum_{m=1}^M\lambda_1\left(\frac{\tilde{\F}_{k,m}\tilde{\F}_{k,m}^\T}{T} \right) \asymp s_{\text{-}k,pre},\\
\lambda_{r_k}\left(\frac{1}{M}\sum_{m=1}^M\frac{\tilde{\F}_{k,m}\tilde{\F}_{k,m}^\T}{T} \right) &\geq \frac{1}{M}\sum_{m=1}^M\lambda_{r_k}\left(\frac{\tilde{\F}_{k,m}\tilde{\F}_{k,m}^\T}{T} \right) \asymp s_{\text{-}k,pre}.
\end{align*}
Therefore, for $j \in [r_k]$,
\begin{align*}
    \lambda_j\left(\frac{1}{M}\sum_{m=1}^M\frac{\tilde{\F}_{k,m}\tilde{\F}_{k,m}^\T}{T} \right) \asymp s_{\text{-}k,pre}.
\end{align*}
Next, following the same analysis as in the proof of Lemma \ref{lem:2}, it is easy to see that
\begin{align*}
    \lambda_j\left(\frac{1}{M}\sum_{m=1}^M\frac{\A_k\tilde{\F}_{k,m}\tilde{\F}_{k,m}^\T\A_k^\T}{T} \right) \asymp s_{\text{-}k,pre}
    \asymp \lambda_j\left(\frac{1}{M}\sum_{m=1}^M\frac{\A_k\tilde{\F}_{k,m}\left( \I_T - \frac{1}{T} \1_T \1_T^\T \right)\tilde{\F}_{k,m}^\T\A_k^\T}{T} \right),
\end{align*}
for $j \in [r_k]$. On the other hand, by Lemma \ref{lem:1},
\begin{align*}
    \lambda_1\left(\frac{1}{M}\sum_{m=1}^M\frac{\tilde{\E}_{k,m}^\T\tilde{\E}_{k,m}}{T} \right) \leq \frac{1}{M}\sum_{m=1}^M \lambda_1\left(\frac{\tilde{\E}_{k,m}^\T\tilde{\E}_{k,m}}{T} \right) = O_p\left( d_{\text{-}k,pre} \left(1 + \frac{d_k}{T}\right)\right),
\end{align*}
where $d_{\text{-}k,pre} = \frac{1}{M}\sum_{m=1}^M d_{\text{-}k,m}$. Similarly,
\begin{align*}
    \lambda_1\left(\frac{1}{M}\sum_{m=1}^M\frac{\tilde{\E}_{k,m}^\T\left( \I_T - \frac{1}{T} \1_T \1_T^\T \right)\tilde{\E}_{k,m}}{T} \right) \leq O_p\left( d_{\text{-}k,pre} \left(1 + \frac{d_k}{T}\right)\right).
\end{align*}
If Assumptions (E1) -- (R1) are satisfied for all $M$ chosen samples, then $\frac{d_{\text{-}k,m}}{s_{\text{-}k,m}} \left(1 + \frac{d_k}{T}\right) = o\left(d_k^{\alpha_{k,z_k}}\right)$ for all $m$, which implies $\frac{d_{\text{-}k,pre}}{s_{\text{-}k,pre}} \left(1 + \frac{d_k}{T}\right)= o\left(d_k^{\alpha_{k,z_k}}\right)$. Therefore, if we define the new normalized version $\ddot{\X}_{k,m} = \tilde{\X}_{k,m} / s_{\text{-}k,pre}^{\frac{1}{2}}$, and $\ddot{\F}_{k,m}$ and $\ddot{\E}_{k,m}$ similarly, then following (\ref{eqn:upper_bound_X'X}) and (\ref{eqn:lower_bound_X'X}),  (we can write $\frac{1}{M}\sum_{m=1}^M\frac{\ddot{\X}_{k,m}^\T\left( \I_T - \frac{1}{T} \1_T \1_T^\T \right)\ddot{\X}_{k,m}}{T} = \X^\T \X$ by Cholesky decomposition, and similarly define $\F$ and $\E$, so that $\X = \F + \E$.) we have
\begin{align*}
      \lambda_j\left(\frac{1}{M}\sum_{m=1}^M\frac{\ddot{\X}_{k,m}^\T\left( \I_T - \frac{1}{T} \1_T \1_T^\T \right)\ddot{\X}_{k,m}}{T} \right) \asymp 1
\end{align*}
for $j \in [r_k]$. Then, we can make use of the decomposition
\begin{align}\label{eqn:preaverage_covariance_decomposition}
    \frac{1}{M} \sum_{m=1}^M \frac{\ddot{\X}_{k,m}^\T  \left( \I_T - \frac{1}{T} \1_T \1_T^\T \right) \ddot{\X}_{k,m}}{T} &= \frac{1}{M} \sum_{m=1}^M \frac{\left(\A_k \ddot{\F}_{k,m} + \ddot{\E}_{k,m}^\T \right) \left( \I_T - \frac{1}{T} \1_T \1_T^\T \right) \left(\A_k \ddot{\F}_{k,m} + \ddot{\E}_{k,m}^\T \right)^\T}{T} \notag \\
    & = \A_k \left[ \frac{1}{M}\sum_{m=1}^M \frac{\ddot{\F}_{k,m} \left( \I_T - \frac{1}{T} \1_T \1_T^\T \right)\ddot{\F}_{k,m}^\T}{T} \right] \A_k^\T + R_{k,pre},
\end{align}
where
\begin{align}{\label{eqn:decomposition_R_pre}}
    R_{k,pre} = \frac{1}{M}\sum_{m=1}^M\bigg[ \frac{\A_k\ddot{\F}_{k,m} \left( \I_T - \frac{1}{T} \1_T \1_T^\T \right) \ddot{\E}_{k,m}}{T} &+ \frac{\ddot{\E}_{k,m}^\T \left( \I_T - \frac{1}{T} \1_T \1_T^\T \right)\ddot{\F}_{k,m}^\T \A_k^\T}{T}\notag\\  &+ \frac{\ddot{\E}_{k,m}^\T \left( \I_T - \frac{1}{T} \1_T \1_T^\T \right) \ddot{\E}_{k,m}}{T} \bigg].
\end{align}
Then, following similar argument as in the proof of Theorem \ref{thm:sumestimatorrate_newmethod}, we can apply Lemma 3 of \cite{Lametal2011} and know that there exists $\hat\U_{k,pre,(z_k)}$ (with $\hat\U_{k,pre,(z_k)}^\T \hat\U_{k,pre,(z_k)} = \I_{z_k}$) such that $\hat\Q_{k,pre,(z_k)} = \hat\U_{k,pre,(z_k)} \P_{k,pre,(z_k)}$ where $\P_{k,pre,(z_k)}$ is an orthogonal matrix, so that
\begin{align*}
    \norm{\hat\U_{k,pre,(z_k)} - \U_{k,(z_k)} } &\preceq \frac{d_k^{\alpha_{k,1}}\bigg\|   \frac{1}{M}\sum_{m=1}^M \frac{\ddot{\F}_{k,m} \left( \I_T - \frac{1}{T} \1_T \1_T^\T \right)\ddot{\F}_{k,m}^\T}{T} - \I_{r_k}  \bigg\| + \norm{ R_{k,pre} } }{d_k^{\alpha_{k,z_k}}}.
\end{align*}
To proceed, note that by (\ref{eqn:bound_FF_minus_I}), we have $\bigg\| \frac{\tilde{\F}_{k,m} \left( \I_T - \frac{1}{T} \1_T \1_T^\T \right)\tilde{\F}_{k,m}^\T}{T s_{\text{-}k,m}} - \I_{r_k}  \bigg\| = O_p\left(\sqrt{r_k/T}\right)$ for each $m$, so $\bigg\| \frac{\ddot{\F}_{k,m} \left( \I_T - \frac{1}{T} \1_T \1_T^\T \right)\ddot{\F}_{k,m}^\T}{T} - \frac{s_{\text{-}k,m}}{s_{\text{-}k,pre}}\I_{r_k}  \bigg\| = O_p\left(\frac{s_{\text{-}k,m}}{s_{\text{-}k,pre}}\sqrt{\frac{r_k}{T}}\right)$, and thus,
\begin{align*}
     \bigg\|\frac{1}{M}\sum_{m=1}^M \frac{\ddot{\F}_k \left( \I_T - \frac{1}{T} \1_T \1_T^\T \right)\ddot{\F}_k^\T}{T} - \I_{r_k}  \bigg\| &=
    \frac{1}{M}\bigg\|\sum_{m=1}^M \frac{\ddot{\F}_k \left( \I_T - \frac{1}{T} \1_T \1_T^\T \right)\ddot{\F}_k^\T}{T} - M\I_{r_k}  \bigg\| \notag \\
    &= \frac{1}{M}\bigg\|\sum_{m=1}^M \left[\frac{\ddot{\F}_{k,m} \left( \I_T - \frac{1}{T} \1_T \1_T^\T \right)\ddot{\F}_{k,m}^\T}{T} - \frac{s_{\text{-}k,m}}{s_{\text{-}k,pre}}\I_{r_k}\right] \bigg\| \notag \\
    & \leq \frac{1}{M}\sum_{m=1}^M \bigg\| \frac{\ddot{\F}_{k,m} \left( \I_T - \frac{1}{T} \1_T \1_T^\T \right)\ddot{\F}_{k,m}^\T}{T} - \frac{s_{\text{-}k,m}}{s_{\text{-}k,pre}}\I_{r_k}  \bigg\| \notag \\
    & = \frac{1}{M}\sum_{m=1}^M O_p\left(\frac{s_{\text{-}k,m}}{s_{\text{-}k,pre}}\sqrt{\frac{r_k}{T}} \right) \notag  = O_p\left(\sqrt{\frac{r_k}{T}} \right).
\end{align*}

To bound $\norm{ R_{k,pre}}^2$, we can bound each term on the right hand side of (\ref{eqn:decomposition_R_pre}) by using the similar technique as in the proof of Theorem \ref{thm:sumestimatorrate_newmethod}. We have that
\begin{align*}
    \bigg\|\frac{1}{M} \sum_{m=1}^M \frac{\A_k\ddot{\F}_{k,m} \left( \I_T - \frac{1}{T} \1_T \1_T^\T \right) \ddot{\E}_{k,m}}{T} \bigg\|^2 &\leq \frac{1}{M} \sum_{m=1}^M \bigg\| \frac{\A_k\ddot{\F}_{k,m} \left( \I_T - \frac{1}{T} \1_T \1_T^\T \right)    \ddot{\E}_{k,m}}{T}\bigg\|^2 \notag \\
    &= O_p\left(  d_k^{\alpha_{k,1}} \min{\left\{ 1 + \frac{d_k}{T},\frac{r_kd_k}{T}\right\}} \frac{\frac{1}{M}\sum_{m=1}^M d_{\text{-}k,m} s_{\text{-}k,m}}{s_{\text{-}k,pre}^2} \right),
\end{align*}
and similarly,
\begin{align*}
    \bigg\|\frac{1}{M} \sum_{m=1}^M \frac{\ddot{\E}_{k,m}^\T \left( \I_T - \frac{1}{T} \1_T \1_T^\T \right)\ddot{\F}_{k,m}^\T \A_k^\T}{T} \bigg\|^2 = O_p\left(  d_k^{\alpha_{k,1}} \min{\left\{ 1 + \frac{d_k}{T},\frac{r_kd_k}{T}\right\}} \frac{\frac{1}{M}\sum_{m=1}^M d_{\text{-}k,m} s_{\text{-}k,m}}{s_{\text{-}k,pre}^2} \right),
\end{align*}
and
\begin{align*}
   \bigg\| \frac{1}{M} \sum_{m=1}^M \frac{\ddot{\E}_{k,m}^\T \left( \I_T - \frac{1}{T} \1_T \1_T^\T \right) \ddot{\E}_{k,m}}{T}  \bigg\|^2 & \leq \frac{1}{M} \sum_{m=1}^M \bigg\|  \frac{\ddot{\E}_{k,m}^\T \left( \I_T - \frac{1}{T} \1_T \1_T^\T \right) \ddot{\E}_{k,m}}{T}  \bigg\|^2 \notag \\
   &  = O_p\left(\left( 1+ \frac{d_k^2}{T^2} \right) \frac{\frac{1}{M} \sum_{m=1}^M{d_{\text{-}k,m}^2}}{s_{\text{-}k,pre}^2}   \right).
\end{align*}
Combining the above results give
\begin{align*}
    c_{k,pre} := \norm{ R_{k,pre} }^2 = O_p\bigg(  \min{\left\{ 1 + \frac{d_k}{T},\frac{r_kd_k}{T}\right\}} \frac{\frac{1}{M}\sum_{m=1}^M d_{\text{-}k,m} s_{\text{-}k,m}}{s_{\text{-}k,pre}^2} + d_k^{\alpha_{k,1}}\left( 1+ \frac{d_k^2}{T^2} \right) \frac{\frac{1}{M} \sum_{m=1}^M{d_{\text{-}k,m}^2}}{s_{\text{-}k,pre}^2}  \bigg),
\end{align*}
and thus,
\begin{align*}
    \bigg\|\hat\U_{k,pre,(z_k)} - \U_{k,(z_k)} \bigg\|^2 & = O_p \left(d_k^{-2\alpha_{k,z_k}} \left[d_k^{2\alpha_{k,1}}\frac{r_k}{T} + c_{k,pre} \right]\right).
\end{align*}

Finally, to show (\ref{eqn:preaverage_sumestimatorrate}), let $\tilde{\V}_{k,pre}$ be the $z_k \times z_k$ diagonal matrix of the first $z_k$ largest eigenvalues of $\hat{\Sigma}_{\tilde{\x}_{k,agg}}$ in decreasing order, and $\ddot{\V}_{k,pre} := \tilde{\V}_{k,pre} / s_{\text{-}k,pre}$. Then it follows from (\ref{eqn:preaverage_covariance_decomposition}) that
\begin{align*}
    \hat{\Q}_{k,pre,(z_k)} - \Q_k\ddot{\H}_{k,pre} &= R_{k,pre}\hat{\Q}_{k,pre,(z_k)}\ddot{\V}_{k,pre}^{-1}, \text{ implying }\\
        \norm{\hat{\Q}_{k,pre,(z_k)} - \Q_k\ddot{\H}_{k,pre} }^2 &\leq \norm{R_{k,pre} }^2 \norm{\ddot{\V}_{k,pre}^{-1} }^2 = O_p\left(d_k^{-2\alpha_{k,z_k}} c_{k,pre} \right).
\end{align*}
This completes the proof of Theorem \ref{thm:preaverage_sumestimatorrate_newmethod}. $\square$

Before the proof of Theorem \ref{thm:reiterate_projection_direction}, we decompose
\begin{align*}
  \wt\bSigma_{y,i}^{(k)} &= T^{-1}\sum_{t=1}^T\y_{t,i}^{(k)}\y_{t,i}^{(k)\T}
  = T^{-1}\sum_{t=1}^T\mat_k(\cX_t - \bar{\cX})\qmkcheck^{(i-1)}\qmkcheck^{(i-1)\T}\mat_k^\T(\cX_t - \bar{\cX})\\
  &= T^{-1}\sum_{t=1}^T\big(\A_k\mat_k(\cF_t - \bar{\cF})\Amk^\T\qmkcheck^{(i-1)} + \mat_k(\cE_t - \bar{\cE})\qmkcheck^{(i-1)}\big)^{\otimes 2}\\
  &= T^{-1}\sum_{t=1}^T\bigg\{ \A_k\mat_k(\cF_t - \bar{\cF})\Amk^\T\qmkcheck^{(i-1)} +
  \big[\bPsi_{1}^{(k)}(\e_t^{(k)} - \bar{\e}^{(k)}),\ldots,\bPsi_{\dmk}^{(k)}(\e_t^{(k)} - \bar{\e}^{(k)})\big]\qmkcheck^{(i-1)} \\
  &\quad\quad\quad\quad + \Big[\big(\bSigma_{\epsilon,1}^{(k)}\big)^{1/2}(\bepsilon_{t,1}^{(k)} - \bar{\bepsilon}_{\cdot,1}^{(k)}),\ldots,\big(\bSigma_{\epsilon,\dmk}^{(k)}\big)^{1/2}
  (\bepsilon_{t,\dmk}^{(k)} - \bar{\bepsilon}_{\cdot,\dmk}^{(k)})\Big]\qmkcheck^{(i-1)} \bigg\}^{\otimes 2}\\
  &= \sum_{i=1}^3\wt{\S}_{ii} + \sum_{i,j=1;i< j}^3(\wt{\S}_{ij} + \wt{\S}_{ij}^\T), \text{ where }\\
  \wt{\S}_{11} &:= \A_k(\I_{r_k}\otimes \qmkcheck^{(i-1)\T}\Amk)T^{-1}\wt\F^{(k)}\M_T\wt\F^{(k)\T}(\I_{r_k}\otimes \qmkcheck^{(i-1)\T}\Amk)^\T\A_k^\T, \\
  \wt{\S}_{12} &:= \A_k(\I_{r_k}\otimes \qmkcheck^{(i-1)\T}\Amk)T^{-1}\wt\F^{(k)}\M_T\wt{\bTheta}^{(k)\T}
  \diag^{1/2}(\bSigma_{\epsilon,1}^{(k)},\ldots,\bSigma_{\epsilon,\dmk}^{(k)})(\qmkcheck^{(i-1)\T}\otimes \I_{d_k})^\T,\\
  \wt{\S}_{13} &:= (\I_{r_k}\otimes \qmkcheck^{(i-1)\T}\Amk)T^{-1}\wt\F^{(k)}\M_T\wt{\E}^{(k)\T}\bPsi^{(k)\T}(\qmkcheck^{(i-1)\T}\otimes \I_{d_k})^\T,\\
  \wt\S_{22} &:= (\qmkcheck^{(i-1)\T}\otimes \I_{d_k})\diag^{1/2}(\bSigma_{\epsilon,1}^{(k)},\ldots,\bSigma_{\epsilon,\dmk}^{(k)})T^{-1}
  \wt\bTheta^{(k)}\M_T\wt\bTheta^{(k)}\diag^{1/2}(\bSigma_{\epsilon,1}^{(k)},\ldots,\bSigma_{\epsilon,\dmk}^{(k)})(\qmkcheck^{(i-1)\T}\otimes \I_{d_k})^\T,\\
  \wt\S_{23} &:= (\qmkcheck^{(i-1)\T}\otimes \I_{d_k})\diag^{1/2}(\bSigma_{\epsilon,1}^{(k)},\ldots,\bSigma_{\epsilon,\dmk}^{(k)})T^{-1}
  \wt\bTheta^{(k)}\M_T\wt{\E}^{(k)\T}\bPsi^{(k)\T}(\qmkcheck^{(i-1)\T}\otimes \I_{d_k})^\T,\\
  \wt\S_{33} &:= (\qmkcheck^{(i-1)\T}\otimes \I_{d_k})\bPsi^{(k)}T^{-1}\wt\E^{(k)}\M_T\wt{\E}^{(k)\T}\bPsi^{(k)\T}(\qmkcheck^{(i-1)\T}\otimes \I_{d_k})^\T,
\end{align*}
with $\bPsi^{(k)} := (\bPsi_1^{(k)\T},\ldots,\bPsi_{\dmk}^{(k)\T})^\T$, $\M_T = \I_T - T^{-1}\1_T\1_T^\T$,
\begin{align*}
  \wt\F^{(k)} &:= [\vec(\mat_k^\T(\cF_1)),\ldots,\vec(\mat_k^\T(\cF_T))] = [\f_1^{(k)},\ldots,\f_T^{(k)}] \text{ with } \\
  \f_t^{(k)} &:= (f_{t,1,1}^{(k)},\ldots,f_{t,\rmk,1}^{(k)},\ldots,f_{t,1,r_k}^{(k)},\ldots,f_{t,\rmk,r_k}^{(k)})^\T,\\
  \wt\bTheta^{(k)} &:= [\bepsilon_1^{(k)}, \ldots, \bepsilon_{T}^{(k)}] \text{ with } \bepsilon_t^{(k)} := (\bepsilon_{t,1}^{(k)\T},\ldots,\bepsilon_{t,\dmk}^{(k)\T})^\T, \text{ and }\\
  \wt\E^{(k)} &:= [\e_1^{(k)},\ldots,\e_T^{(k)}].
\end{align*}
With the notations in Assumption (E2) and (F1), for $t\in[T]$, define
\begin{align*}
  \z_{e,t}^{(k)} &:= (z_{e,t,1}^{(k)},\ldots,z_{e,t,r_e}^{(k)})^\T,\\
  \z_{\epsilon, t}^{(k)} &:= (z_{\epsilon,t,1,1}^{(k)},\ldots,z_{\epsilon,t,1,d_k}^{(k)},\ldots,z_{\epsilon,t,\dmk,1}^{(k)},\ldots,z_{\epsilon,t,\dmk,d_k}^{(k)})^\T,\\
\z_{f,t}^{(k)} &:= (z_{f,t,1,1}^{(k)}, \ldots, z_{f,t,\rmk,1}^{(k)},\ldots,z_{f,t,1,r_k}^{(k)},\ldots,z_{f,t,\rmk,r_k}^{(k)})^\T.
\end{align*}
We further split, for a fixed integer $N\geq 1$,
\begin{align*}
  \wt\F^{(k)} &:= \F^{(k)} + \check\F^{(k)}, \;\;\; \wt\bTheta^{(k)} := \bTheta^{(k)} + \check\bTheta^{(k)}, \;\;\; \wt\E^{(k)} := \E^{(k)} + \check\E^{(k)}, \; \text{ where }\\
  \F^{(k)} &:= \bigg(\sum_{q=0}^{NT}a_{f,q}\z_{f,1-q}^{(k)},\ldots, \sum_{q=0}^{NT}a_{f,q}\z_{f,T-q}^{(k)}\bigg) = (\z_{f,1-NT}^{(k)},\ldots,\z_{f,T}^{(k)})\cA_{f,T} =: \Z_f^{(k)}\cA_{f,T},\\
  \wt\bTheta^{(k)} &:= \bigg(\sum_{q=0}^{NT}a_{\epsilon,q}\z_{\epsilon,1-q}^{(k)},\ldots,\sum_{q=0}^{NT}a_{\epsilon,q}\z_{\epsilon,T-q}^{(k)}\bigg) =  (\z_{\epsilon,1-NT}^{(k)},\ldots,\z_{\epsilon,T}^{(k)})\cA_{\epsilon,T} =: \Z_\epsilon^{(k)}\cA_{\epsilon,T},\\
  \wt\E^{(k)} &:= \bigg(\sum_{q=0}^{NT}a_{e,q}\z_{e,1-q}^{(k)},\ldots,\sum_{q=0}^{NT}a_{e,q}\z_{e,T-q}^{(k)}\bigg) =  (\z_{e,1-NT}^{(k)},\ldots,\z_{e,T}^{(k)})\cA_{e,T} =: \Z_e^{(k)}\cA_{e,T},
\end{align*}
with $\cA_{f,T}$, $\cA_{\epsilon,T}$ and $\cA_{e,T}$ defined in Assumption (RE1), and $\check\F^{(k)}$, $\check\bTheta^{(k)}$ and $\check\E^{(k)}$ the remainders of the truncations. Then we define
\begin{align}
  \wt\S_{ij} &= \S_{ij} + \check\S_{ij}, \;\text{ where }\notag\\
  \S_{11} &:= \A_k(\I_{r_k}\otimes \qmkcheck^{(i-1)\T}\Amk)T^{-1}\F^{(k)}\M_T\F^{(k)\T}(\I_{r_k}\otimes \qmkcheck^{(i-1)\T}\Amk)^\T\A_k^\T, \label{eqn:S_11}
\end{align}
and similarly for other $\S_{ij}$'s. We first prove a lemma on how small each remainder term $\check\S_{ij}$ is.

\begin{lemma}\label{lem:4}
  Let all the assumptions in Theorem \ref{thm:reiterate_projection_direction} be satisfied.  Then for each $k\in[K]$,
  \[\sum_{i,j=1;i\leq j}^3\norm{\check\S_{ij}} = o_P(T^{-1}).\]
\end{lemma}

{\em Proof of Lemma \ref{lem:4}.}
Define
  \begin{equation}\label{eqn:Q}
  \begin{split}
    \Q_1^{(k)} &:= \A_k(\I_{r_k}\otimes \qmkcheck^{(i-1)\T}\Amk),\\
    \Q_2^{(k)} &:= (\qmkcheck^{(i-1)\T}\otimes \I_{d_k})\diag^{1/2}(\bSigma_{\epsilon,1}^{(k)},\ldots,\bSigma_{\epsilon,\dmk}^{(k)}),\\
    \Q_3^{(k)} &:= (\qmkcheck^{(i-1)\T}\otimes \I_{d_k})\bPsi^{(k)}.
    \end{split}
  \end{equation}
Then by Assumption (L1) and the rates for $\max_{j\in[\dmk]}\norm{\bSigma_{\epsilon,j}^{(k)}}$ and $S_{\psi}^{(k)}$ in Assumption (RE2),
\begin{align*}
  \norm{\Q_1^{(k)}}^2 &= (\qmkcheck^{(i-1)\T}\Amk\Amk^\T\qmkcheck^{(i-1)})\norm{\A_k}^2  = O\bigg(\prod_{j=1;j\neq k}^K d_j^{\alpha_{j,1}} \cdot d_k^{\alpha_{k,1}}\bigg) = o(d),\\
  \norm{\Q_2^{(k)}}^2 &= O\Big(\max_{j\in[\dmk]}\norm{\bSigma_{\epsilon,j}^{(k)}}\Big) = O(\dmk) = o(d),\;\;\; \norm{\Q_3^{(k)}}^2 = O(S_{\psi}^{(k)}) = O(d),
\end{align*}
where the first line used (\ref{eqn:gmkcheck}). We also have, by Assumption (RE1),
\begin{align*}
  E\norm{T^{-1/2}\F^{(k)}}_F^2 &= T^{-1}E\tr(\cA_{f,T}^\T\Z_f^{(k)\T}\Z_f^{(k)}\cA_{f,T}) = rT^{-1}\tr(\cA_{f,T}\cA_{f,T}^\T) = r(1 - o(T^{-2}d^{-4})),\\
  E\norm{T^{-1/2}\check\F^{(k)}}_F^2 &= O(rT\cdot T^{-1}\cdot o(T^{-2}d^{-4}) = o(rT^{-2}d^{-4})),\\
  E\norm{T^{-1/2}\bTheta^{(k)}}_F^2 &= d(1-o(T^{-2}d^{-4})), \;\;\;
  E\norm{T^{-1/2}\check\bTheta^{(k)}}_F^2 = o(dT^{-2}d^{-4}) = o(T^{-2}d^{-3}),\\
  E\norm{T^{-1/2}\E^{(k)}}_F^2 &=  r_e(1-o(T^{-2}d^{-4})), \;\;\;
  E\norm{T^{-1/2}\check\E^{(k)}}_F^2 = o(r_eT^{-2}d^{-4}).
\end{align*}

Then writing $\check\S_{11} = I_{1} + I_{1}^\T + I_{2}$, where
\begin{align*}
  I_{1} &:= T^{-1}\Q_1^{(k)}\check\F^{(k)}\M_T\F^{(k)}\Q_1^{(k)\T}, \;\;\;
  I_2 := T^{-1}\Q_1^{(k)}\check\F^{(k)}\M_T\check\F^{(k)}\Q_1^{(k)\T},
\end{align*}
we have for $a>0$,
\begin{align*}
  P(\norm{I_1} \geq a) &\leq \norm{\Q_1^{(k)}}^2 E^{1/2}\norm{T^{-1/2}\F^{(k)}}_F^2\cdot E^{1/2}\norm{T^{-1/2}\check\F^{(k)}}_F^2/a\\
  &=  o(d\cdot r^{1/2}\cdot r^{1/2}T^{-2/2}d^{-4/2}/a) = o(rT^{-1}d^{-1}/a),
\end{align*}
showing that $\norm{I_1} = o_P(rT^{-1}d^{-1})$. Similarly (details omitted),
$\norm{I_2} = O_P(rT^{-2}d^{-3})$, so that $\norm{\check\S_{11}} = o_P(rT^{-1}d^{-1})$. Similar to the above arguments, we can show that (details omitted)
\begin{align*}
  \norm{\check\S_{12}} &= o_P(r^{1/2}T^{-1}d^{-1/2}), \;\;\; \norm{\check\S_{13}} = o_P(r^{1/2}r_e^{1/2}T^{-1}d^{-1}), \;\;\; \norm{\check\S_{22}} = o_P(T^{-1}),\\
  \norm{\check\S_{23}} &= o_P(r_e^{1/2}T^{-1/2}d^{-1/2}), \;\;\; \norm{\check\S_{33}} = o_P(r_eT^{-1}d^{-1}).
\end{align*}
The proof completes by noting that $r_e = O(\dmk) = o(d)$ by (RE2), and $r=O(r_e)$ by the assumption in Theorem \ref{thm:reiterate_projection_direction}. $\square$

Before presenting the next lemma, define, for $j\in[\dmk]$ and $k\in[K]$, and $m$ a non-negative integer,
\begin{align*}
  \S_{11,m}' &= (\qmkcheck^{(m)\T}\Amk\Amk^\T\qmkcheck^{(m)})\A_k\A_k^\T, \\
  \S_{11,m}'' &= \A_k(\I_{r_k}\otimes \qmkcheck^{(m)\T}\Amk)(T^{-1}\F^{(k)}\M_T\F^{(k)\T} - \I_{r})(\I_{r_k}\otimes \qmkcheck^{(m)\T}\Amk)^\T\A_k^\T,
\end{align*}
so that $\S_{11} = \S_{11,i-1}' + \S_{11,i-1}''$, where $\S_{11}$ is defined in (\ref{eqn:S_11}). Define $\S_{ij,m} := \S_{ij}$ in the spirit of (\ref{eqn:S_11}) for $i,j=1,2,3$, with $\qmkcheck^{(i-1)}$ there replaced by $\qmkcheck^{(m)}$. Then we have $\S_{11,m} = \S_{11,m}' + \S_{11,m}''$, and
\begin{align}
\wt\bSigma_{y,m+1}^{(k)} &= T^{-1}\sum_{t=1}^T\y_{t,m+1}^{(k)}\y_{t,m+1}^{(k)\T} = \S_{11,m}' + \S_{11,m}'' + \S_{22,m} + \S_{33,m} + \sum_{i,j=1;i<j}^3(\S_{ij,m} + \S_{ij,m}^\T) + \sum_{i,j=1}^3\check\S_{ij}\notag\\
&=: \S_{11,m}' + \E. \label{eqn:remainder_E}
\end{align}

\begin{lemma}\label{lem:5}
Let all the assumptions in Theorem \ref{thm:reiterate_projection_direction} be satisfied. Then for $j\in[\dmk]$ and $k\in[K]$, we have
\begin{align*}
  \norm{\S_{11,m}''} &=  O_P\bigg(\gmkcheck d_k^{\alpha_{k,1}}\sqrt{\frac{r}{T}}\bigg), \;\;\;
  \norm{\S_{12,m}} = O_P\bigg\{\gmkcheck^{1/2}d_k^{\alpha_{k,1}/2}\bigg(\sqrt{\frac{rd_k}{T}} + \norm{\qmkcheck^{(m)} - \U_{\text{-}k,(1)}}\sqrt{\frac{rd}{T}}\bigg)\bigg\},\\
  \norm{\S_{13,m}} &= O_P\bigg(\gmkcheck^{1/2}d_k^{\alpha_{k,1}/2}\sqrt{\frac{rr_eS_{\psi}^{(k)}}{T}}\bigg),\\
  \norm{\S_{22,m}} &= O_P\bigg(\max_{j\in[\dmk]}\norm{\bSigma_{\epsilon,j}^{(k)}}\Big[1 + \norm{\qmkcheck^{(m)} - \U_{\text{-}k,(1)}}^2\frac{d}{T}\Big] + \frac{d_k}{\sqrt{T}} + \norm{\qmkcheck^{(m)} - \U_{\text{-}k,(1)}}\sqrt{\frac{d_kd}{T}}\bigg),\\
  \norm{\S_{23,m}} &= O_P\bigg\{(S_{\psi}^{(k)})^{1/2}\bigg(\sqrt{\frac{r_ed_k}{T}} + \norm{\qmkcheck^{(m)} - \U_{\text{-}k,(1)}}\sqrt{\frac{r_ed}{T}}\bigg)\bigg\},\;\;\; \norm{\S_{33,m}} = O_P(S_{\psi}^{(k)}),
\end{align*}
where $\U_{\text{-}k,(1)} := \U_{K,(1)}\otimes\cdots\otimes \U_{k+1,(1)}\otimes\U_{k-1,(1)}\otimes\cdots\otimes \U_{1,(1)}$.
\end{lemma}

{\em Proof of Lemma \ref{lem:5}}.
Using (\ref{eqn:gmkcheck}), (\ref{eqn:SVD_Ak}) and that (\ref{eqn:lemma2.2})  proves that $\lambda_j(\G_k) \asymp d_k^{\alpha_{k,j}}$ for $j\in[r_k]$, we have
\begin{align*}
  \norm{\S_{11,m}''} &\leq (\qmkcheck^{(m)\T}\Amk\Amk^\T\qmkcheck^{(m)})\norm{\A_k}^2\cdot
  \norm{T^{-1}\Z_f^{(k)}\cA_{f,T}\M_T\cA_{f,T}^\T\Z_f^{(k)\T} - \I_r}\\
  &= \gmkcheck\norm{\G_k}\cdot
  \norm{T^{-1}\Z_f^{(k)}\cA_{f,T}\M_T\cA_{f,T}^\T\Z_f^{(k)\T} - \I_r}= O_P\bigg(\gmkcheck d_k^{\alpha_{k,1}} \sqrt{\frac{r}{T}}\bigg),
\end{align*}
where the last equality used Theorem 2.8 of \cite{WangPaul2014}, which can be applied since we have $r_k = o(T^{1/3})$ and with fourth order moments exist for the elements in $\Z_f^{(k)}$ from Assumption (L1) and (R1) respectively, and that by Assumption (RE1),
\begin{align*}
  \norm{\cA_{f,T}\M_T\cA_{f,T}^\T} &\leq  \norm{\A_{f,T}}^2 < \infty,\\
  \frac{1}{(N+1)T}\tr(\cA_{f,T}\M_T\cA_{f,T}^\T) &=  \frac{1}{(N+1)T}(\tr(\cA_{f,T}^\T\cA_{f,T}) - T^{-1}\1_T^\T\cA_{f,T}^\T\cA_{f,T}\1_T) \rightarrow \frac{1}{N+1},\\
  \frac{1}{(N+1)T}\tr(\cA_{f,T}\M_T\cA_{f,T}^\T)^2 &= \frac{1}{(N+1)T}\big\{\tr(\cA_{f,T}^\T\cA_{f,T})^2 - 2T^{-1}\1_T^\T(\cA_{f,T}^\T\cA_{f,T})^2\1_T\\
  &\hspace{0.8in} + T^{-2}(\1_T^\T\cA_{f,T}^\T\cA_{f,T}\1_T)^2\big\} \rightarrow \frac{a_1-2a_2+a_3^2}{N+1}.
\end{align*}

For $\norm{\S_{12,m}}$, define the notation $\U_S$ to be a sub-matrix of $\U$ restricted to the rows indexed by $S$. Let $\Z_j := (\Z_{\epsilon}^{(k)})_{B_j} \in \mathbb{R}^{d_k\times (M+1)T}$,
where $B_j := \{{(j-1)d_k+1,\ldots,jd_k}\}$, $j=1,\ldots,\dmk$. Let also $\a_{\ell,j} := (\bSigma_{\epsilon,j}^{(k)})_{\ell}^\T$, $\ell\in[d_k]$. Using the notation in (\ref{eqn:Q}), for $h\in[r]$ and $\ell\in[d_k]$, the $(h,\ell)$-th element of $T^{-1}\Z_f^{(k)}\cA_{f,T}\M_T\cA_{\epsilon,T}^\T\Z_{\epsilon}^{(k)\T}\Q_2^{(k)\T}$ is $J_1 + J_2$, where
\begin{align*}
  J_1 &:= T^{-1}\sum_{j=1}^{\dmk}(\Umkc1)_j\a_{\ell,j}^\T\Z_j\cA_{f,T}\M_T\cA_{\epsilon,T}^\T(\Z_f^{(k)})_h,\\
  J_2 &:= T^{-1}\sum_{j=1}^{\dmk}(\qmkcheck^{(m)} - \Umkc1)_j\a_{\ell,j}^\T\Z_j\cA_{f,T}\M_T\cA_{\epsilon,T}^\T(\Z_f^{(k)})_h.
\end{align*}
We have $EJ_1=0$, and by Assumption (E1),
\begin{align*}
  E|J_1|^2 &= T^{-2}\sum_{j=1}^{\dmk}(\Umkc1)_j^2\norm{\a_{\ell,j}}^2 \norm{\cA_{f,T}\M_T\cA_{\epsilon,T}^\T}_F^2\leq O(T^{-2})\cdot \norm{\cA_{f,T}}\norm{\M_T}\cdot\norm{\cA_{\epsilon,T}}_F^2
  = O(T^{-1}),
\end{align*}
where the last equality is from Assumption (RE1). It means that $J_1 = O_P(T^{-1/2})$. For $J_2$, using the Cauchy-Schwarz inequality and that $\a_{\ell,j}^\T\Z_j\cA_{f,T}\M_T\cA_{\epsilon,T}^\T(\Z_f^{(k)})_h = O_P(T^{1/2})$ from the analysis of $J_2$ above,
\begin{align*}
  J_2 &\leq \norm{\qmkcheck^{(m)} - \Umkc1}\cdot T^{-1}\bigg( \sum_{j=1}^{\dmk}(\a_{\ell,j}^\T\Z_j\cA_{f,T}\M_T\cA_{\epsilon,T}^\T(\Z_f^{(k)})_h)^2 \bigg)^{1/2}\\
  &= \norm{\qmkcheck^{(m)} - \Umkc1}\cdot T^{-1} O_P(\dmk T)^{1/2} = O_P\bigg(\norm{\qmkcheck^{(m)} - \Umkc1}\sqrt{\frac{\dmk}{T}}\bigg).
\end{align*}

With the above, we then have
\begin{align*}
  \norm{\S_{12,m}} &= \norm{\Q_1^{(k)}T^{-1}\Z_f^{(k)}\cA_{f,T}\M_T\cA_{\epsilon,T}^\T\Z_{\epsilon}^{(k)\T}\Q_2^{(k)\T}}\leq \norm{\Q_1^{(k)}}\cdot \sqrt{rd_k}\cdot O_P(J_1+J_2)\\
  &= O_P\bigg\{\gmkcheck^{1/2}d_k^{\alpha_{k,1}/2}\bigg(\sqrt{\frac{rd_k}{T}} + \norm{\qmkcheck^{(m)} - \U_{\text{-}k,(1)}}\sqrt{\frac{rd}{T}}\bigg)\bigg\}.
\end{align*}

For $\S_{13,m}$, using the definitions in (\ref{eqn:Q}) and the bounds for $\norm{\Q_3^{(k)}}$ in Lemma \ref{lem:4},
\begin{align*}
  \norm{\S_{13,m}} &\leq \norm{\Q_1^{(k)}}\cdot\norm{\Q_3^{(k)}}\cdot\norm{T^{-1}\Z_f^{(k)}\cA_{f,T}\M_T\cA_{e,T}^\T\Z_{e}^{(k)\T}}\\
  &\leq \gmkcheck^{1/2}d_k^{\alpha_{k,1}/2}\cdot (S_{\psi}^{(k)})^{1/2}\cdot O_P(T^{-1}\sqrt{rr_e}\norm{\cA_{f,T}\M_T\cA_{e,T}^\T}_F) = O_P\bigg(\gmkcheck^{1/2}d_k^{\alpha_{k,1}/2}(S_{\psi}^{(k)})^{1/2}\sqrt{\frac{rr_e}{T}}\bigg),
\end{align*}
where the last line used the fact that an element in $\Z_f^{(k)}\cA_{f,T}\M_T\cA_{e,T}^\T\Z_e^{(k)\T}$ is $O_P(\norm{\cA_{f,T}\M_T\cA_{e,T}^\T}_F) = O_P(T^{1/2})$ by a similar calculation for treating $J_1$ above.

For $\S_{22,m}$, decompose $\Q_2^{(k)} = \Q_{2,0}^{(k)} + \Q_{2,e}^{(k)}$, where
\[\Q_{2,0}^{(k)} := (\Umkc1^\T\otimes I_{d_k})\diag^{1/2}(\bSigma_{\epsilon,1}^{(k)}, \ldots, \bSigma_{\epsilon,\dmk}^{(k)}), \;\;\; \Q_{2,e}^{(k)} := ((\qmkcheck^{(k)} - \Umkc1)^\T\otimes I_{d_k})\diag^{1/2}(\bSigma_{\epsilon,1}^{(k)}, \ldots, \bSigma_{\epsilon,\dmk}^{(k)}).\]
Then $\S_{22,m} = I_{0} + I_1 + I_2 + I_2^\T + I_3$, where
\begin{align*}
  I_0 &:= \Q_2^{(k)}\Q_2^{(k)\T},\\
  I_1 &:= \Q_{2,0}^{(k)}(T^{-1}\Z_{\epsilon}^{(k)}\cA_{\epsilon,T}\M_T\cA_{\epsilon,T}^\T\Z_{\epsilon}^{(k)\T} - \I_d)\Q_{2,0}^{(k)\T},\\
  I_2 &:= \Q_{2,e}^{(k)}(T^{-1}\Z_{\epsilon}^{(k)}\cA_{\epsilon,T}\M_T\cA_{\epsilon,T}^\T\Z_{\epsilon}^{(k)\T} - \I_d)\Q_{2,0}^{(k)\T},\\
  I_3 &:= \Q_{2,e}^{(k)}(T^{-1}\Z_{\epsilon}^{(k)}\cA_{\epsilon,T}\M_T\cA_{\epsilon,T}^\T\Z_{\epsilon}^{(k)\T} - \I_d)\Q_{2,e}^{(k)\T}.
\end{align*}
Firstly,
\[\norm{I_0} = \norm{\Q_2^{(k)}}^2 \leq \max_{j\in[\dmk]}\norm{\bSigma_{\epsilon,j}^{(k)}}.\]
Using the same notations as in the treatment of $\S_{12,m}$ before within this proof, for $\ell,h\in[d_k]$, the $(\ell,h)$ entry of $I_1$ is given by
\begin{align*}
(I_1)_{\ell,h} =   T^{-1}\bigg(\sum_{j=1}^{\dmk}(\Umkc1)_j\a_{\ell,j}^\T\Z_j\bigg)\cA_{\epsilon,T}\M_T\cA_{\epsilon,T}^\T
  \bigg(\sum_{j=1}^{\dmk}(\Umkc1)_j\a_{h,j}^\T\Z_j\bigg)^\T - \sum_{j=1}^{\dmk}(\Umkc1)_j^2\a_{\ell,j}^\T\a_{h,j}.
\end{align*}
Hence by Assumption (RE1) and (E1), writing $\G := \cA_{\epsilon,T}\M_T\cA_{\epsilon,T}^\T$,
\begin{align*}
  E(I_1)_{\ell,h} &= \sum_{j=1}^{\dmk}(\Umkc1)_j^2\a_{\ell,j}^\T\a_{h,j}\{T^{-1}\tr(\G) -1\} = O(T^{-1/2}).
\end{align*}
Also,
\begin{align}
  E(I_1)_{\ell,h}^2 &= T^{-2}\sum_{i,j=1}^{\dmk}(\Umkc1)_i^2(\Umkc1)_j^2
  E\{(\a_{\ell,i}^\T\Z_i\G\Z_i^\T\a_{h,i})
  (\a_{\ell,j}^\T\Z_j\G\Z_j^\T\a_{h,j})\}\notag\\
  &+ T^{-2}\sum_{i\neq j}(\Umkc1)_i^2(\Umkc1)_j^2E\{(\a_{\ell,i}^\T\Z_i\G\Z_j^\T\a_{h,j})^2 + E\{\a_{\ell,i}^\T\Z_i\G\Z_j^\T\a_{h,j}\a_{\ell,j}^\T\Z_j\G\Z_i^\T\a_{h,i}\}\}\notag\\
  &-2T^{-1}\tr(\G)\bigg(\sum_{j=1}^{\dmk}(\Umkc1)_j^2\a_{\ell,j}^\T\a_{h,j}\bigg)^2 +  \bigg(\sum_{j=1}^{\dmk}(\Umkc1)_j^2\a_{\ell,j}^\T\a_{h,j}\bigg)^2\notag\\
  &=T^{-2}\sum_{i=1}^{\dmk}(\Umkc1)_i^4E(\a_{\ell,i}^\T\Z_i\G\Z_i^\T\a_{h,i})^2
  +T^{-2}\sum_{i\neq j} (\Umkc1)_i^2(\Umkc1)_j^2(\a_{\ell,i}^\T\a_{h,i})(\a_{\ell,j}^\T\a_{h,j})\tr^2(\G)\notag\\
  &+T^{-2}\sum_{i\neq j}(\Umkc1)_i^2(\Umkc1)_j^2(\norm{\a_{\ell,i}}^2\norm{\a_{h,j}}^2
  +(\a_{\ell,i}^\T\a_{h,i})(\a_{\ell,j}^\T\a_{h,j}))\tr(\G^2)\notag\\
  &-(2T^{-1}\tr(\G) - 1)\bigg(\sum_{j=1}^{\dmk}(\Umkc1)_j^2\a_{\ell,j}^\T\a_{h,j}\bigg)^2. \label{eqn:I_1squared}
  \end{align}
Now define $\w_{\ell} := \Z_i^\T\a_{\ell,i}$. Then $E(\w_{\ell}) = \0$ and $\cov(\w_{\ell},\w_{h}) = \a_{\ell,i}^\T\a_{h,i}\I_{(M+1)T}$, with elements in $\w_{\ell}$ independent of each other for each $\ell \in [d_k]$ and $i \in [\dmk]$. Hence
\begin{align}
  E(\a_{\ell,i}^\T\Z_i\G\Z_i^\T\a_{h,i})^2 &= E(\w_{\ell}^\T\G\w_h) = E\bigg(\sum_{j=1}^{(N+1)T}(\G)_{jj}(\w_\ell)_j(\w_h)_j + \sum_{j_1\neq j_2}(\G)_{j_1j_2}(\w_{\ell})_{j_1}(\w_{h})_{j_2}\bigg)^2\notag\\
  &= \sum_{j=1}^{(N+1)T}(\G)_{jj}^2E[(\w_\ell)_j^2(\w_h)_j^2] + \sum_{j_1\neq j_2}(\G)_{j_1j_1}(\G)_{j_2j_2} (\a_{\ell,i}^\T\a_{h,i})^2\notag\\
  &+ \sum_{j_1\neq j_2} (\G)_{j_1j_2}^2(\norm{\a_{\ell,i}}^2\norm{\a_{h,i}}^2 + (\a_{\ell,i}^\T\a_{h,i})^2)\notag\\
  &= \sum_{j=1}^{(N+1)T}(\G)_{jj}^2\Big\{ \norm{\a_{h,i}}^2\norm{\a_{\ell,i}}^2 + (\nu_4-3)\a_{h,i}^\T\diag(\a_{\ell,i}\a_{\ell,i}^\T)\a_{h,i} + 2(\a_{\ell,i}^\T\a_{h,i})^2 \Big\}\notag\\
  &+\sum_{j_1\neq j_2}(\G)_{j_1j_1}(\G)_{j_2j_2} (\a_{\ell,i}^\T\a_{h,i})^2
  + \sum_{j_1\neq j_2} (\G)_{j_1j_2}^2(\norm{\a_{\ell,i}}^2\norm{\a_{h,i}}^2 + (\a_{\ell,i}^\T\a_{h,i})^2)\notag\\
  &= \norm{\a_{h,i}}^2\norm{\a_{\ell,i}}^2 \tr(\G^2) + (\a_{\ell,i}^\T\a_{h,i})^2(\tr^2(\G) + \tr(\G^2))
  \notag\\
  &+  (\nu_4-3)\tr(\diag^2(\G))\a_{h,i}^\T\diag(\a_{\ell,i}\a_{\ell,i}^\T)\a_{h,i}, \label{eqn:quadraticformsquared}
\end{align}
where $\nu_4 := E(\Z_i)_{11}^4 < \infty$ by Assumption (R1), and we used Lemma (A.2) of \cite{Lietal2019}.
Substitute this back to (\ref{eqn:I_1squared}), we have $ E(I_1)_{\ell,h}^2 = H_1 + H_2 + H_3 + H_4$, where
\begin{align*}
 H_1 &:= (T^{-2}\tr^2(\G) - 2T^{-1}\tr(\G) + 1)\bigg(\sum_{j=1}^{\dmk}(\Umkc1)_j^2\a_{\ell,j}^\T\a_{h,j}\bigg)^2,\\
 H_2 &:= T^{-2}\tr(\G^2)\bigg(\sum_{j=1}^{\dmk}(\Umkc1)_j^2\a_{\ell,j}^\T\a_{h,j}\bigg)^2,\\
 H_3 &:= T^{-2}\tr(\G^2)\bigg(\sum_{j=1}^{\dmk}(\Umkc1)_j^2\a_{\ell,j}^\T\a_{\ell,j}\bigg)
 \bigg(\sum_{j=1}^{\dmk}(\Umkc1)_j^2\a_{h,j}^\T\a_{h,j}\bigg),\\
 H_4 &:= T^{-2}\tr(\diag^2(\G))\sum_{j=1}^{\dmk}(\Umkc1)_j^4(\nu_4-3)\a_{h,j}^\T\diag(\a_{\ell,j}\a_{\ell,j}^\T)\a_{h,j}.
\end{align*}
By Assumption (RE1), we have
\begin{align*}
  |H_1|, |H_2|, |H_3| &= O(T^{-1})\cdot O(\max_{j\in[\dmk]}\norm{\bSigma_{\epsilon,j}^{(k)}}_{\max}^2) = O(T^{-1}),\\
  |H_4| &\leq T^{-2}\tr(\G^2)(\nu_4-3)\max_{j\in[\dmk]}\norm{\bSigma_{\epsilon,j}^{(k)}}_{\max}^2 = O(T^{-1}).
\end{align*}
Hence we can conclude that $\norm{I_1} = O_P(d_kT^{-1/2})$.

For $I_2$, define $\hat{e}_j := (\qmkcheck^{(m)} - \Umkc1)_j$. Then for $\ell,h\in[d_k]$, the $(\ell,h)$ entry of $I_2$ is given by
\begin{align*}
  (I_2)_{\ell,h} &= \sum_{j_1=1}^{\dmk}\hat{e}_{j_1}\bigg( T^{-1}\a_{\ell,j_1}^\T\Z_{j_1}\G
  \sum_{j_2=1}^{\dmk}(\Umkc1)_{j_2}\Z_{j_2}^\T\a_{h,j_2} - (\Umkc1)_{j_1}\a_{\ell,j_1}^\T\a_{h,j_1}\bigg) =: \sum_{j_1=1}^{\dmk}\hat{e}_{j_1}g_{j_1,\ell,h}.
\end{align*}
By Assumption (RE1) and (E1),
\begin{align*}
  E(g_{j_1,h,\ell}) &= (T^{-1}\tr(\G)-1)(\Umkc1)_{j_1}\a_{\ell,j_1}^\T\a_{h,j_1} = O(T^{-1/2}).
\end{align*}
Also, similar to the treatment of $I_1$ and using (\ref{eqn:quadraticformsquared}),
\begin{align*}
  E(g_{j_1,h,\ell}^2) &= T^{-2}\sum_{j=1}^{\dmk}(\Umkc1)_j^2E(\a_{\ell,j_1}^\T\Z_{j_1}\G\Z_j^\T\a_{h,j})^2\\
  &- 2T^{-1}\tr(\G)(\Umkc1)_{j_1}^2(\a_{\ell,j_1}^\T\a_{h,j_1})^2 + (\Umkc1)_{j_1}^2(\a_{\ell,j_1}^\T\a_{h,j_1})^2\\
  &= \sum_{j=1}^{\dmk}(\Umkc1)_j^2T^{-2}\tr(\G^2)\norm{\a_{h,j}}^2\norm{\a_{\ell,j_1}}^2\\
  &+ (T^{-1}\tr(\G) - 1)^2(\Umkc1)_{j_1}^2(\a_{\ell,j_1}^\T\a_{h,j_1})^2 + T^{-2}\tr(\G^2)(\Umkc1)_{j_1}^2(\a_{\ell,j_1}^\T\a_{h,j_1})^2\\
  &+ T^{-2}(\Umkc1)_{j_1}^2(\nu_4-3)\tr(\diag^2(\G))\a_{h,j_1}^\T\diag(\a_{\ell,j_1}\a_{\ell,j_1}^\T)\a_{h,j_1} = O(T^{-1}),
\end{align*}
so that we can conclude that $\norm{I_2} = O_P\Big(d_k\norm{\qmkcheck^{(m)} - \Umkc1}\sqrt{\frac{\dmk}{T}}\Big) = O_P\Big(d_k^{1/2}\norm{\qmkcheck^{(m)} - \Umkc1}\sqrt{\frac{d}{T}}\Big)$.

For $I_3$, using Theorem 2 of \cite{Latala2005},
\begin{align*}
  \norm{I_3} &\leq \norm{\Q_{2,e}^{(k)}}^2 \norm{T^{-1}\Z_{\epsilon}^{(k)}\G\Z_{\epsilon}^{(k)\T} - \I_d} \leq \norm{\qmkcheck^{(m)} - \Umkc1}^2\max_{j\in[\dmk]}\norm{\bSigma_{\epsilon,j}^{(k)}}\cdot O_P(1 + d/T).
\end{align*}
Hence
\begin{align*}
  \norm{\S_{22,m}} &= \norm{I_0} + \norm{I_1} + 2\norm{I_2} + \norm{I_3}\\
  &= O_P\bigg( \max_{j\in[\dmk]}\norm{\bSigma_{\epsilon,j}^{(k)}} + \frac{d_k}{\sqrt{T}} + \norm{\qmkcheck^{(m)} - \Umkc1}\sqrt{\frac{d_kd}{T}} + \max_{j\in[\dmk]}\norm{\bSigma_{\epsilon,j}^{(k)}}\norm{\qmkcheck^{(m)} - \Umkc1}^2\Big(\frac{d}{T}+1\Big) \bigg)\\
  &=O_P\bigg(\max_{j\in[\dmk]}\norm{\bSigma_{\epsilon,j}^{(k)}}\Big[1 + \norm{\qmkcheck^{(m)} - \Umkc1}^2\frac{d}{T}\Big] + \frac{d_k}{\sqrt{T}} + \norm{\qmkcheck^{(m)} - \Umkc1}\sqrt{\frac{d_kd}{T}}\bigg).
\end{align*}
For $\S_{23,m}$, the treatment is exactly parallel to the treatment of $\S_{12,m}$, so that
\begin{align*}
  \norm{\S_{23,m}} &= \norm{\Q_3^{(k)}}\cdot\sqrt{r_ed_k}\cdot O_P(J_1+J_2)\\
  &= O_P\bigg\{ (S_{\psi}^{(k)})^{1/2}\bigg(\sqrt{\frac{r_ed_k}{T}} + \norm{\qmkcheck^{(m)} - \Umkc1}\sqrt{\frac{r_ed}{T}}\bigg) \bigg\}.
\end{align*}

Finally, by Assumption (RE1) and Theorem 2 of \cite{Latala2005},
\begin{align*}
  \norm{\S_{33,m}} &\leq \norm{\Q_3^{(k)}}^2 O_P\Big(1+\frac{r_e}{T}\Big) = O_P(S_{\psi}^{(k)}).
\end{align*}
This completes the proof of the lemma. $\square$

{\em Proof of Theorem \ref{thm:reiterate_projection_direction}}.
Firstly, (\ref{eqn:lemma2.2}) in Lemma \ref{lem:2}
proves  that $\G_k$ has the $j$-th diagonal element with order $d_k^{\alpha_{k,j}}$ for $j\in[r_k]$ under Assumption (L1), which is the same as those in $\D_k$. Define $\B_k \in \mathbb{R}^{d_k\otimes d_k-1}$ to be an orthogonal compliment of $\U_{k,(1)}$, in the sense that $\V_k := (\U_{k,(1)}, \B_k)$ has $\V_k\V_k^\T = \V_k^\T\V_k = \I_{d_k}$. Then using the SVD of $\A_k$ in (\ref{eqn:SVD_Ak}),
\begin{align*}
  \V^\T\S_{11,m}'\V = \left(
                        \begin{array}{cc}
                          \gmkcheck\G_k & \0 \\
                          \0 & \0 \\
                        \end{array}
                      \right),
\end{align*}
with the $(j,j)$ element of $\gmkcheck\G_k$ of order $\gmkcheck d_k^{\alpha_{k,j}}$, $j\in[r_k]$. By Assumption (L1'), we then have
\[\text{sep}(\U_{k,(1)}^\T\S_{11,m}'\U_{k,(1)}, \B_k^\T\S_{11,m}'B_k) \asymp \gmkcheck d_k^{\alpha_{k,1}},\]
where
\[\text{sep}(\D_1,\D_2) := \min_{\lambda\in\lambda(\D_1),\mu\in\lambda(\D_2)}|\lambda - \mu|.\]
Then by Lemma 3 of \cite{Lametal2011}, which is Theorem 8.1.10 in \cite{GV96}, since $\U_{k,(1)}$ is an eigenvector of $\S_{11,m}'$ and hence the span of $\U_{k,(1)}$ is an invariant subspace for $\S_{11,m}'$,  for the matrix $\E$ in (\ref{eqn:remainder_E}), if
\[\norm{\E} = o_P(\gmkcheck d_k^{\alpha_{k,1}}),\]
then there exists $\check\q_k^{(m+1)}$  which is an eigenvector of $\wt\bSigma_{y,m+1}^{(k)}$ such that
\[\norm{\check\q_k^{(m+1)} - \U_{k,(1)}} = O(\norm{\E}/\gmkcheck d_k^{\alpha_{k,1}}).\]
From the rates proved in Lemma \ref{lem:4} and \ref{lem:5}, with the assumptions $d_k, r_eS_{\psi}^{(k)}, \sqrt{T}S_{\psi}^{(k)} = O(\gmkhat d_{k}^{\alpha_{k,1}}) = O(\gmkcheck d_k^{\alpha_{k,1}})$ since $\gmkhat \asymp_P \gmkcheck$ by (\ref{eqn:gmkhat_rate}) and (\ref{eqn:gmkcheck}), we have
\begin{align}
  \norm{\check\q_k^{(m+1)} - \U_{k,(1)}} &= O_P(\gmkcheck^{-1}d_k^{-\alpha_{k,1}}\norm{\E}) \notag\\
  &= O_P\bigg\{ \sqrt{\frac{r}{T}} + \gmkcheck^{-1/2}d_k^{-\alpha_{k,1}/2}\norm{\qmkcheck^{(m)}-\Umkc1}\sqrt{\frac{rd}{T}} \notag\\
  &+ \gmkcheck^{-1}d_k^{-\alpha_{k,1}} \max_{j\in[\dmk]}\norm{\bSigma_{\epsilon,j}^{(k)}}
  \Big[ 1 + \norm{\qmkcheck^{(m)} - \Umkc1}^2\frac{d}{T} \Big]  \bigg\}. \label{eqn:qmkcheck_inter_rate}
\end{align}
Furthermore, with the assumption $\max_{j\in[\dmk]}\norm{\bSigma_{\epsilon,j}^{(k)}} = O\Big(\prod_{j=1}^{K}d_j^{\alpha_{j,1}}\sqrt{\frac{r}{T}}\Big) = O_P\Big(\gmkcheck d_k^{\alpha_{k,1}}\sqrt{\frac{r}{T}}\Big)$, (\ref{eqn:qmkcheck_inter_rate}) becomes
\begin{align}
  \norm{\check\q_k^{(m+1)} - \U_{k,(1)}} &= O_P\bigg\{ \sqrt{\frac{r}{T}}\Big(1 + \norm{\qmkcheck^{(m)} - \Umkc1}^2\frac{d}{T}\Big) + \gmkcheck^{-1/2}d_k^{-\alpha_{k,1}/2}\norm{\qmkcheck^{(m)} - \Umkc1}\sqrt{\frac{rd}{T}} \bigg\}. \label{eqn:qmkcheck_inter_rate2}
\end{align}
Putting $m=0$ in the above, noting that $\check\q_k^{(0)} = \hat{\q}_{k,pre} = \hat\U_{k,pre,(1)}$ (see the arguments at the beginning of Section \ref{subsec:refiningprojection}) leading to the rate $\norm{\hat\U_{\text{-}k,pre,(1)} - \Umkc1} = O_P(b_k)$ in (\ref{eqn:qmkpre_rate}), we also have $\norm{\qmkcheck^{(0)} - \Umkc1} = O_P(b_k)$, which upon substituting into (\ref{eqn:qmkcheck_inter_rate2}), gives the first statement in Theorem \ref{thm:reiterate_projection_direction}.

By defining $e_{m,k} := \norm{\check\q_k^{(m)} - \U_{k,(1)}}$ and $e_{m,\text{-}k} := \norm{\qmkcheck^{(m)} - \Umkc1}$, we have from an argument similar to the one in (\ref{eqn:qmkpre_rate}) that
\begin{equation}\label{eqn:emk}
e_{m,\text{-}k} = O_P\bigg(\sum_{j=1;j\neq k}^Ke_{m,j}\bigg).
\end{equation}
We see from (\ref{eqn:qmkcheck_inter_rate}) that the coefficient rates of $\norm{\qmkcheck^{(m)} - \Umkc1}$ and $\norm{\qmkcheck^{(m)} - \Umkc1}^2$  are respectively, for each $k\in[K]$,
\[ u_{1k} := \gmkcheck^{-1/2} d_k^{-\alpha_{k,1}/2}\sqrt{\frac{rd}{T}}, \;\;\; u_{2k} := \gmkcheck^{-1}d_k^{-\alpha_{k,1}}\max_{j\in[\dmk]}\norm{\bSigma_{\epsilon,j}^{(k)}}\frac{d}{T}, \]
which are in fact $o_P(K^{-1})$ by the further assumption in Theorem \ref{thm:reiterate_projection_direction}.
From (\ref{eqn:qmkcheck_inter_rate}), defining $u_k := u_{1k} + u_{2k}$,
\begin{align*}
e_{m+n,k} &= O_P\bigg(\sqrt{\frac{r}{T}} + u_{1k}e_{m+n-1,\text{-}k} + u_{2k}e_{m+n-1,\text{-}k}^2\bigg) = O_P\bigg(\sqrt{\frac{r}{T}} + (u_{1k} + u_{2k})e_{m+n-1,\text{-}k}\bigg)\\
&= O_P\bigg(\sqrt{\frac{r}{T}} + u_k\sum_{j=1;j\neq k}^Ke_{m+n-1,j}\bigg).
\end{align*}
Defining $\e_m := (e_{m,1},\ldots,e_{m,K})^\T$, the above becomes
\begin{align*}
\e_{m+n} &= O_P\bigg(\sqrt{\frac{r}{T}}\1_K + \W\e_{m+n-1}\bigg), \;\;\;\text{ where } \;
\W := \left(
         \begin{array}{ccccc}
           0 & u_1 & u_1 & \cdots & u_1 \\
           u_2 & 0 & u_2 & \cdots & u_2 \\
           \vdots & \vdots & \ddots & \vdots & \vdots \\
           u_K & u_K & u_K & \cdots & 0 \\
         \end{array}
       \right),
\end{align*}
with $\norm{\W}_{\infty}\leq K\max_{k\in[K]}u_k = o_P(1)$ by the further assumption in Theorem \ref{thm:reiterate_projection_direction}. Hence iterating the above $n-1$ more times, we have
\begin{align*}
  \e_{m+n} &= O_P\bigg( \sqrt{\frac{r}{T}}\big(\I_K + \W + \cdots + \W^{n-1}\big)\1_K + \W^n\e_{m} \bigg), \; \text{ implying }\\
  \norm{\e_{m+n}}_{\max} &= O_P\bigg(\sqrt{\frac{r}{T}}\big(1 + \norm{\W}_{\infty} + \cdots + \norm{\W}_{\infty}^{n-1}\big) + \norm{\W}_{\infty}^n\norm{\e_m}_{\max}\bigg)\\
  &= O_P\bigg(\sqrt{\frac{r}{T}} + \norm{\W}_{\infty}^n\norm{\e_m}_{\max}\bigg) = O_P\bigg(\sqrt{\frac{r}{T}}\bigg)
\end{align*}
for $n$ large enough. This completes the proof of the theorem. $\square$

{\em Proof of Theorem \ref{thm:reestimation_factor_loading_space}}.
Using the notations in (\ref{eqn:S_11}) and in Lemma \ref{lem:5}, since $\S_{11,m}$ is sandwiched by $\A_k$ and $\A_k^\T$, implying that it is sandwiched by $\U_{k}$ and $\U_{k}^\T$, the span of the columns of $\U_k$ forms an invariant subspace for $\S_{11,m}$. Now Let $\B_k$ be the orthogonal complement of $\U_k$, in the sense that $\U := (\U_k,\B_k)$ is an orthogonal matrix. Then we have
\begin{align*}
\U^\T\S_{11}\U &= \left(
                     \begin{array}{cc}
                       \U_k^\T\S_{11,m}\U_k & \0 \\
                       \0 & \0 \\
                     \end{array}
                   \right), \; \text{ where } \; \text{sep}(\U_k^\T\S_{11,m}\U_k,\0) = \lambda_{r_k}(\U_k^\T\S_{11,m}\U_k), \; \text{ and }\\
  \lambda_{r_k}(\U_k^\T\S_{11}\U_k) &= \lambda_{r_k}\big( \V_k\G_k\V_k^\T
  (\I_{r_k}\otimes \qmkcheck^{(m)\T}\Amk)T^{-1}\Z_f^{(k)}\cA_{f,T}\M_T\cA_{f,T}^\T\Z_f^{(k)\T}(\I_{r_k}\otimes \qmkcheck^{(m)\T}\Amk)^\T\big)\\
  &\geq \lambda_{r_k}(\G_k)\lambda_{r_k}(\gmkcheck\I_{r_k})\lambda_r(T^{-1}\Z_f^{(k)}\cA_{f,T}\M_T\cA_{f,T}^\T\Z_f^{(k)\T})
  \asymp_P(\gmkcheck d_k^{\alpha_{k,r_k}}),
\end{align*}
where we used Theorem 2.8 of \cite{WangPaul2014} (applicable since $r_k = o(T^{1/3})$ by Assumption (L1) and all variables are of bounded fourth moments by Assumption (R1)) to conclude that
\begin{align}
  \lambda_r(T^{-1}\Z_f^{(k)}\cA_{f,T}\M_T\cA_{f,T}^\T\Z_f^{(k)\T}) &\asymp_P (N+1)\bigg(\frac{\tr(\cA_{f,T}\M_t\cA_{f,T}^\T)}{(N+1)T} + \sqrt{\frac{r}{(N+1)T}}\bigg) = 1 + O(T^{-1/2}), \label{eqn:lambda_rZ_f}
\end{align}
where the last equality used Assumption (RE1).
Hence Lemma 3 of \cite{Lametal2011} implies that there exists $\check{\U}_{k}$ with $\check\U_k^\T\check\U_k = \I_{r_k}$ and
\[\norm{\check\U_k - \U_k} = O_P\big\{[\norm{\S_{12,m}} + \norm{\S_{13,m}} + \norm{\S_{22,m}} + \norm{\S_{23,m}} + \norm{\S_{33,m}}]/\text{sep}(\U_k^\T\S_{11,m}\U_k,\0)\big\},\]
such that $\check\U_k$ equals the $r_k$ eigenvectors corresponding to the first $r_k$ largest eigenvalues of $\wt\bSigma_{y,m+1}^{(k)}$ multiplied with an orthogonal matrix, if we can show that the above rate is $o_P(1)$. To this end, by Lemma \ref{lem:5}, with the assumption $d_k, r_e S_{\psi}^{(k)} = O(\gmkhat d_k^{\alpha_{k,1}}) $, we have $\norm{\S_{23,m}} = O_P(\norm{\S_{12,m}})$, and so with $\gmkhat \asymp_P \gmkcheck$ by (\ref{eqn:gmkhat_rate}) and (\ref{eqn:gmkcheck}), we have
\begin{align}
  \norm{\check\U_{k} - \U_k} &= O_P\bigg\{ \gmkcheck^{-1/2}d_k^{\alpha_{k,1}/2 - \alpha_{k,r_k}}\bigg[ \sqrt{rd_k}\bigg( \frac{1}{\sqrt{T}} + \norm{\qmkcheck^{(m)} - \Umkc1}\sqrt{\frac{\dmk}{T}} \bigg) + \sqrt{\frac{rr_eS_{\psi}^{(k)}}{T}}\bigg]\notag\\
  &+ \gmkcheck^{-1}d_k^{-\alpha_{k,r_k}}\bigg[ \max_{j\in[\dmk]}\norm{\bSigma_{\epsilon,j}^{(k)}}\bigg\{1 + \norm{\qmkcheck^{(m)} - \Umkc1}^2\frac{d}{T}\bigg\} + S_{\psi}^{(k)} \bigg]\bigg\}. \label{eqn:Ucheck_rate}
\end{align}
If $m$ is large enough, then following Theorem \ref{thm:reiterate_projection_direction}, we have by (\ref{eqn:emk}) that
\[\norm{\qmkcheck^{(m)} - \Umkc1} = O_P\bigg(\sum_{j=1;j\neq k}^K\norm{\check\q_j^{(m)} - \U_{j,(1)}}\bigg) = O_P\bigg(K\sqrt{\frac{r}{T}}\bigg).\]
Substituting the above rate into (\ref{eqn:Ucheck_rate}) completes the proof, noting that by (\ref{eqn:gmkcheck}) we have $\gmkcheck d_k^{\alpha_{k,1}} \asymp_P g_s$. $\square$

{\em Proof of Theorem \ref{thm:population_eigenvalue_correlation_matrix}}.
First, define
\begin{align}
  \Qm_{1}^{(k)} &:= \diag^{-1/2}(\bSigma_{y,m+1}^{(k)})\A_k(\I_{r_k}\otimes\qmkcheck^{(m)\T}\Amk),\notag\\
  \Qm_{2}^{(k)} &:= \diag^{-1/2}(\bSigma_{y,m+1}^{(k)})(\qmkcheck^{(m)\T}\otimes \I_{d_k})\diag^{1/2}(\bSigma_{\epsilon,1}^{(k)},\ldots,\bSigma_{\epsilon,\dmk}^{(k)}),\label{eqn:Qm}\\
  \Qm_{3}^{(k)} &:= \diag^{-1/2}(\bSigma_{y,m+1}^{(k)})(\qmkcheck^{(m)\T}\otimes\I_{d_k})\bPsi^{(k)},\notag
\end{align}
where $\bSigma_{y,m+1}^{(k)}$ is defined in (\ref{eqn:Sigma_y_population}). Then we have
\[\R_{y,m+1}^{(k)} = \sum_{j=1}^3\Qm_{j}^{(k)}\Qm_{j}^{(k)\T}.\]
Consider
\begin{align}
  \tr(\Qm_{1}^{(k)}\Qm_1^{(k)\T}) &= \tr(\diag^{-1/2}(\bSigma_{y,m+1}^{(k)})\cdot\gmkcheck\A_k\A_k^\T\cdot\diag^{-1/2}(\bSigma_{y,m+1}^{(k)}))\notag\\
  &= \tr(\gmkcheck^{-1/2}\diag^{-1/2}(\A_k\A_k^\T)(1+o(1))^{-1}\cdot\gmkcheck\A_k\A_k^\T\cdot\gmkcheck^{-1/2}\diag^{-1/2}(\A_k\A_k^\T)(1+o(1))^{-1})\notag\\
  &= d_k(1+o(1)), \label{eqn:trace_Qm}
\end{align}
where the second inequality used (\ref{eqn:Sigma_y_diagonal_dominance}). At the same time, for $j\in[r_k]$,
\begin{align*}
  \lambda_j(\Qm_1^{(k)}\Qm_1^{(k)\T}) &= \lambda_j(\diag^{-1/2}(\A_k\A_k^\T)(1+o(1))\A_k\A_k^\T\diag^{-1/2}(\A_k\A_k^\T)(1+o(1))\\
  &= \lambda_j(\U_k^\T\diag^{-1}(\A_k\A_k^\T)\U_k\G_k(1+o(1))), \; \text{ with }\\
  \lambda_j(\G_k)\lambda_{\min}(\diag^{-1}(\A_k\A_k^\T))(1+o(1))&\leq \lambda_j(\U_k^\T\diag^{-1}(\A_k\A_k^\T)\U_k\G_k(1+o(1)))\\ &\leq \lambda_j(\G_k)\lambda_{\max}(\diag^{-1}(\A_k\A_k^\T))(1+o(1)),
\end{align*}
so that by assumption (RE2), there are generic constants $c,C>0$ such that
\begin{equation}\label{eqn:sandwich_Qm}
d_k^{\alpha_{k,j}} \asymp_P c\lambda_j(\G_k) \leq \lambda_j(\Qm_1^{(k)}\Qm_1^{(k)\T}) \leq C\lambda_j(\G_k) \asymp_P d_k^{\alpha_{k,j}}
\end{equation}
in probability, where $\lambda_j(\G_k)$ being asymptotic in probability to $d_k^{\alpha_{k,j}}$ is given by (\ref{eqn:lemma2.2}) in Lemma \ref{lem:2}.
Using (\ref{eqn:sandwich_Qm}), there exists a constant $C>0$ (generic, different from the above) so that
for $j\in[r_k]$,
\begin{align*}
  \frac{\lambda_1(\Qm_1^{(k)}\Qm_1^{(k)\T})}{\lambda_{j}(\Qm_1^{(k)}\Qm_1^{(k)\T})} &\leq Cd_k^{\alpha_{k,1} - \alpha_{k,j}}.
\end{align*}
in probability. Hence using (\ref{eqn:trace_Qm}), in probability, we have
\begin{align*}
  d_k(1+o(1)) \leq \tr(\Qm_1^{(k)}\Qm_1^{(k)\T}) \leq r_k\lambda_1(\Qm_1^{(k)}\Qm_1^{(k)\T})
  \leq r_k\cdot Cd_k^{\alpha_{k,1}-\alpha_{k,j}}\lambda_{j}(\Qm_1^{(k)}\Qm_1^{(k)\T}),
\end{align*}
implying that, in probability, there exists a constant $C>0$ such that as $T,d_k$ are large enough,
\begin{align}
  \lambda_{j}(\R_{y,m+1}^{(k)}) \geq \lambda_{j}(\Qm_1^{(k)}\Qm_1^{(k)\T}) \geq Cd_k^{1-\alpha_{k,1} + \alpha_{k,j}}/r_k > 1, \label{eqn:lower_bound_lambda_j(R)}
\end{align}
since $r_k=o(d_k^{1-\alpha_{k,1} + \alpha_{k,j}})$ by (RE2) for $j\in[r_k]$. For $j\in[d_k]/[r_k]$,
\begin{align*}
  \lambda_j(\R_{y,m+1}^{(k)}) &\leq \lambda_j(\Qm_1^{(k)}\Qm_1^{(k)\T}) + \lambda_1(\Qm_2^{(k)}\Qm_2^{(k)\T}) + \lambda_1(\Qm_3^{(k)}\Qm_3^{(k)\T})\\
  &\leq 0 + \norm{\diag^{-1}(\bSigma_{y,m+1}^{(k)})}\bigg(\sum_{j=1}^{\dmk}(\qmkcheck^{(m)})_j^2\norm{\bSigma_{\epsilon,j}^{(k)}} + \bigg\| \sum_{j=1}^{\dmk}(\qmkcheck^{(m)})_j\bPsi_{j}^{(k)} \bigg\|^2\bigg)\\
  &\leq \gmkcheck^{-1}\norm{\diag^{-1}(\A_k\A_k^\T)(1+o(1))}
  \Big(\max_{j\in[\dmk]}\norm{\bSigma_{\epsilon,j}^{(k)}} + S_{\psi}^{(k)}\Big)\\
  &= O_P\bigg(\prod_{j=1;j\neq k}^Kd_j^{-\alpha_{j,1}}\Big(\max_{j\in[\dmk]}\norm{\bSigma_{\epsilon,j}^{(k)}} + S_{\psi}^{(k)}\Big)\bigg) = o_P(1),
\end{align*}
where the third inequality is in probability after using (\ref{eqn:Sigma_y_diagonal_dominance}), and the last line used (\ref{eqn:gmkcheck}) and Assumption (RE2). This completes the proof of the theorem. $\square$

Before proving Theorem \ref{thm:rkhat_consistency}, define
\begin{align*}
  \S_{y,m+1}^{(k)} := \diag^{-1/2}(\bSigma_{y,m+1}^{(k)})\wt\bSigma_{y,m+1}^{(k)}\diag^{-1/2}(\bSigma_{y,m+1}^{(k)}).
\end{align*}
It is then easy to see that
\[\wh\R_{y,m+1}^{(k)} := \diag^{-1/2}(\S_{y,m+1}^{(k)})\S_{y,m+1}^{(k)}\diag^{-1/2}(\S_{y,m+1}^{(k)}).\]
We state and prove the following lemma.

\begin{lemma}\label{lem:6}
Assume all the assumptions in Theorem \ref{thm:rkhat_consistency} hold. Then for $j\in[r_k]$ and $k\in[K]$,
\begin{align*}
&\bigg| \frac{\lambda_j(\S_{y,m+1}^{(k)})}{\lambda_j(\R_{y,m+1}^{(k)})} - 1 \bigg| = O\bigg(r_kd_k^{\alpha_{k,1}-\alpha_{k,j}-1}\sqrt{\frac{r}{T}}\bigg(d_k^{\alpha_{k,1}} + Kd_k^{\alpha_{k,1}/2}\sqrt{\frac{rd}{T\gmkcheck}} + \frac{K^2r^{1/2}d}{T^{3/2}\gmkcheck}\max_{j\in[\dmk]}\norm{\bSigma_{\epsilon,j}^{(k)}}\bigg)\bigg),\\
&\max_{j\in[d_k]}\bigg| \frac{\lambda_j(\S_{y,m+1}^{(k)})}{\lambda_j(\hat\R_{y,m+1}^{(k)})} - 1 \bigg|
= O_P\bigg( \sqrt{\frac{r}{T}}\bigg(1 + K\sqrt{\frac{rd}{T\gmkcheck}} + \frac{K^2r^{1/2}d}{T^{3/2}\gmkcheck}\max_{j\in[\dmk]}\norm{\bSigma_{\epsilon,j}^{(k)}}\bigg) \bigg),\\
&\max_{j\in[d_k]/[r_k]}|\lambda_j(\S_{y,m+1}^{(k)}) - \lambda_j(\R_{y,m+1}^{(k)})| = O_P\bigg(\gmkcheck^{-1}\bigg\{ \frac{r_eS_\psi^{(k)}}{\sqrt{T}} + \frac{K\sqrt{rr_eS_\psi^{(k)}d}}{T} + \max_{j\in[\dmk]}\norm{\bSigma_{\epsilon,j}^{(k)}}\frac{K^2rd}{T^2}\bigg\}\bigg).
\end{align*}
\end{lemma}

{\em Proof of Lemma \ref{lem:6}}.
Define, using the definition of $\S_{ij,m}$ defined in the proof of Lemma \ref{lem:5},
\[ \S_{ij,r} := \diag^{-1/2}(\S_{ij,m})\S_{ij,m}\diag^{-1/2}(\S_{ij,m}),\;\;\; i,j=1,2,3. \]
Then using (\ref{eqn:Qm}), for $j\in[r_k]$, we can decompose
\begin{align*}
  \bigg| \frac{\lambda_j(\S_{y,m+1}^{(k)})}{\lambda_j(\R_{y,m+1}^{(k)})} - 1 \bigg| &\leq
  \frac{1}{\lambda_j(\R_{y,m+1}^{(k)})}\bigg(\sum_{i=1}^3\norm{\S_{ii,r} - \Qm_i^{(k)}\Qm_i^{(k)\T}} + 2\sum_{i<j}\norm{\S_{ij,r}} + \sum_{i\neq j}\norm{\check\S_{ij}\diag^{-1}(\bSigma_{y,m+1}^{(k)})}\bigg).
\end{align*}
From (\ref{eqn:Sigma_y_diagonal_dominance}) and Assumption (RE2), we have that
\[\lambda_1(\diag^{-1}(\bSigma_{y,m+1}^{(k)})), \;\;\; \lambda_{d_k}(\diag^{-1}(\bSigma_{y,m+1}^{(k)})) \asymp \gmkcheck^{-1} \asymp_P \prod_{j=1;j\neq k}^Kd_j^{-\alpha_{j,1}},\]
where the last order is from (\ref{eqn:gmkcheck}). Hence coupled with the results from Lemma \ref{lem:5} and (\ref{eqn:lower_bound_lambda_j(R)}), and the fact that from an argument similar to (\ref{eqn:qmkpre_rate}) and the result of Theorem \ref{thm:reiterate_projection_direction} that for large enough $m$,
\[ \norm{\qmkcheck^{(m)} - \Umkc1} = O_P\bigg(K\sqrt{\frac{r}{T}}\bigg), \]
we have
\begin{align*}
  \frac{\norm{\S_{11,r} - \Qm_1^{(k)}\Qm_1^{(k)\T}}}{\lambda_j(\R_{y,m+1}^{(k)})}
  &\leq \frac{\norm{\diag^{-1}(\bSigma_{y,m+1}^{(k)})}\norm{\S_{11,m}''}}{\lambda_j(\R_{y,m+1}^{(k)})} = O_P\bigg( r_kd_k^{\alpha_{k,1}-\alpha_{k,j}-1}\gmkcheck^{-1}\cdot \gmkcheck d_k^{\alpha_{k,1}}\sqrt{\frac{r}{T}}  \bigg)\\
  &= O_P\bigg(r_kd_k^{2\alpha_{k,1} - \alpha_{k,j} - 1}\sqrt{\frac{r}{T}}\bigg),\\
  \frac{\norm{\S_{22,r} - \Qm_2^{(k)}\Qm_2^{(k)\T}}}{\lambda_j(\R_{y,m+1}^{(k)})}
  &\leq \frac{\norm{\diag^{-1}(\bSigma_{y,m+1}^{(k)})}\norm{\S_{22,m} - \Qm_2^{(k)}\Qm_2^{(k)\T}}}{\lambda_j(\R_{y,m+1}^{(k)})}\\
  &= O_P\bigg( r_kd_k^{\alpha_{k,1}-\alpha_{k,j}-1}\gmkcheck^{-1}\cdot
  \bigg\{ \max_{j\in[\dmk]}\norm{\bSigma_{\epsilon,j}^{(k)}}\cdot\frac{K^2rd}{T^2} + \frac{d_k}{\sqrt{T}} + \frac{K\sqrt{rd_kd}}{T}
  \bigg\}\bigg),\\
  \frac{\norm{\S_{33,r} - \Qm_3^{(k)}\Qm_3^{(k)\T}}}{\lambda_j(\R_{y,m+1}^{(k)})}
  &= O_P\bigg( r_kd_k^{\alpha_{k,1} - \alpha_{k,j}-1}\gmkcheck^{-1} S_\psi^{(k)} \cdot\frac{r_e}{\sqrt{T}} \bigg) = O_P\bigg( r_kd_k^{\alpha_{k,1} - \alpha_{k,j} -1}\frac{r_eS_\psi^{(k)}}{\gmkcheck\sqrt{T}} \bigg),\\
  \frac{\norm{\S_{12,r}}}{\lambda_j(\R_{y,m+1}^{(k)})} &= O_P\bigg( r_kd_k^{3\alpha_{k,1}/2 - \alpha_{k,j}-1}\gmkcheck^{-1/2}\bigg(\sqrt{\frac{rd_k}{T}} + \frac{Kr\sqrt{d}}{T}\bigg) \bigg),\\
  \frac{\norm{\S_{13,r}}}{\lambda_j(\R_{y,m+1}^{(k)})} &= O_P\bigg( r_kd_k^{3\alpha_{k,1}/2-\alpha_{k,j}-1}\gmkcheck^{-1/2}\sqrt{\frac{rr_eS_\psi^{(k)}}{T}} \bigg),\\
  \frac{\norm{\S_{23,r}}}{\lambda_j(\R_{y,m+1}^{(k)})} &= O_P\bigg( r_kd_k^{\alpha_{k,1} - \alpha_{k,j}-1}\gmkcheck^{-1}(S_\psi^{(k)})^{1/2}\bigg(\sqrt{\frac{r_ed_k}{T}} + \frac{K\sqrt{rr_ed}}{T}\bigg) \bigg),\\
  \frac{\norm{\check\S_{ij}\diag^{-1}(\bSigma_{y,m+1}^{(k)})}}{\lambda_j(\R_{y,m+1}^{(k)})} &= O_P\bigg( r_kd_k^{\alpha_{k,1} - \alpha_{k,j}-1}\gmkcheck^{-1}T^{-1} \bigg).
\end{align*}
These implies that
\begin{align*}
  \bigg| \frac{\lambda_j(\S_{y,m+1}^{(k)})}{\lambda_j(\R_{y,m+1}^{(k)})} - 1 \bigg|
  &= O\bigg(r_kd_k^{\alpha_{k,1}-\alpha_{k,j}-1}\sqrt{\frac{r}{T}}\bigg(d_k^{\alpha_{k,1}} + Kd_k^{\alpha_{k,1}/2}\sqrt{\frac{rd}{T\gmkcheck}} + \frac{K^2r^{1/2}d}{T^{3/2}\gmkcheck}\max_{j\in[\dmk]}\norm{\bSigma_{\epsilon,j}^{(k)}}\bigg)\bigg),
\end{align*}
which is the first statement in the lemma.

For the second statement, for $j\in[\dmk]$ and $k\in[K]$, defining $\u_j$ to be a unit vector with the $j$-th position and 0 elsewhere,
\begin{align*}
  \bigg|\frac{\lambda_j(\S_{y,m+1}^{(k)})}{\lambda_j(\wh\R_{y,m+1}^{(k)})}-1\bigg|
  &= \bigg|\frac{\lambda_j(\S_{y,m+1}^{(k)})}{\lambda_j(\S_{y,m+1}^{(k)}\diag^{-1}(\S_{y,m+1}^{(k)}))}-1\bigg|
  \leq \max_{j\in[\dmk]}|\lambda_j(\diag(\S_{y,m+1}^{(k)}))-1|\\
  &= \max_{j\in[\dmk]}|\lambda_j(\S_{y,m+1}^{(k)} - \R_{y,m+1}^{(k)})| = \max_{j\in[\dmk]}|\u_j^\T(\S_{y,m+1}^{(k)} - \R_{y,m+1}^{(k)})\u_j|\\
  &\leq \max_{j\in[\dmk]}\bigg| \u_j^\T\bigg( \S_{11,r}^{(k)} -\Qm_1^{(k)}\Qm_1^{(k)\T} \bigg)\u_j \bigg|
  +2\max_{j\in[\dmk]}|\u_j^\T\S_{12,r}^{(k)}\u_j| + 2\max_{j\in[\dmk]}|\u_j^\T\S_{13,r}^{(k)}\u_j|\\
  &+ \sum_{i=2}^3|\u_j^\T(\S_{ii,r}^{(k)} - \Qm_i^{(k)}\Qm_i^{(k)\T})\u_j| + 2|\u_j^\T\S_{23,r}^{(k)}\u_j| + \sum_{i\neq j}\norm{\diag^{-1}(\bSigma_{y,m+1}^{(k)})\check\S_{ij}}.
\end{align*}
But similar to the above,
\begin{align*}
  &\max_{j\in[\dmk]}|\u_j^\T(\S_{11,r} - \Qm_1^{(k)}\Qm_1^{(k)\T})\u_j| = O_P(\norm{T^{-1}\Z_f^{(k)}\cA_{f,T}\M_T\cA_{f,T}^\T\Z_f^{(k)\T} - \I_r}) = O_P\bigg(\sqrt{\frac{r}{T}}\bigg),\\
  &\max_{j\in[\dmk]}|\u_j^\T\S_{12,r}^{(k)}\u_j| = O_P\bigg(\gmkcheck^{-1/2}\bigg(\sqrt{\frac{rd_k}{T}} + \frac{Kr\sqrt{d}}{T}\bigg)\bigg), \;\;\; \max_{j\in[\dmk]}|\u_j^\T\S_{13,r}\u_j| = O_P\bigg( \gmkcheck^{-1/2}\sqrt{\frac{rr_eS_\psi^{(k)}}{T}}\bigg),\\
  &\max_{j\in[\dmk]}|\u_j^\T(\S_{22,r} - \Qm_2^{(k)}\Qm_2^{(k)\T})\u_j|
  = O_P\bigg( \gmkcheck^{-1}\bigg\{ \frac{K^2rd}{T^2}\max_{j\in[\dmk]}\norm{\bSigma_{\epsilon,j}^{(k)}}
  + \frac{d_k}{\sqrt{T}} + \frac{K\sqrt{rd_kd}}{T} \bigg\} \bigg),\\
  &\max_{j\in[\dmk]}|\u_j^\T\S_{23,r}^{(k)}\u_j| = O_P\bigg( \gmkcheck^{-1}(S_\psi^{(k)})^{1/2}\bigg(\sqrt{\frac{r_ed_k}{T}} + \frac{K\sqrt{rr_ed}}{T}\bigg)\bigg),\\
  &\max_{j\in[\dmk]}|\u_j^\T(\S_{33,r} - \Qm_3^{(k)}\Qm_3^{(k)\T})\u_j|
  = O_P\bigg(\frac{r_eS_\psi^{(k)}}{\gmkcheck\sqrt{T}}\bigg), \;\;\; \norm{\diag^{-1}(\bSigma_{y,m+1}^{(k)})\check\S_{ij}} = O_P(\gmkcheck^{-1}T^{-1}).
\end{align*}
Hence the above implies that
\begin{align*}
\bigg|\frac{\lambda_j(\S_{y,m+1}^{(k)})}{\lambda_j(\wh\R_{y,m+1}^{(k)})}-1\bigg|
&= O_P\bigg( \sqrt{\frac{r}{T}}\bigg(1 + K\sqrt{\frac{rd}{T\gmkcheck}} + \frac{K^2r^{1/2}d}{T^{3/2}\gmkcheck}\max_{j\in[\dmk]}\norm{\bSigma_{\epsilon,j}^{(k)}}\bigg) \bigg),
\end{align*}
which is the second statement of the lemma.
Finally, consider for $j\in[d_k]/[r_k]$,
\begin{align}
  |\lambda_j(\S_{y,m+1}^{(k)}) - \lambda_j(\R_{y,m+1}^{(k)})| &\leq \bigg|\lambda_j\bigg(\S_{11,r} + \sum_{j=2}^3(\S_{1j,r} + \S_{j1,r} + \Qm_j^{(k)}\Qm_j^{(k)\T})\bigg) - \lambda_j(\R_{y,m+1}^{(k)})\bigg|\notag\\
  &+ 2\norm{\S_{23,r}} + \sum_{j=2}^3\norm{\S_{jj,r} - \Qm_j^{(k)}\Qm_j^{(k)\T}} + \sum_{i\neq j}\norm{\diag^{-1}(\bSigma_{y,m+1}^{(k)})\check\S_{ij}^{(k)}}\notag\\
  &= 2\norm{\S_{23,r}} + \sum_{j=2}^3\norm{\S_{jj,r} - \Qm_j^{(k)}\Qm_j^{(k)\T}} + \sum_{i\neq j}\norm{\diag^{-1}(\bSigma_{y,m+1}^{(k)})\check\S_{ij}^{(k)}}, \label{eqn:diff_ineq}
\end{align}
if we can show that for large enough $T,d_k$, in probability, for $j\in[d_k]/[r_k]$,
\begin{equation}\label{eqn:diff_eq}
\lambda_j\bigg(\S_{11,r} + \sum_{j=2}^3(\S_{1j,r} + \S_{j1,r} + \Qm_j^{(k)}\Qm_j^{(k)\T})\bigg) = \lambda_j(\R_{y,m+1}^{(k)}).
\end{equation}
With (\ref{eqn:diff_ineq}), from the rates obtained in previous arguments, we then have
\begin{align*}
  |\lambda_j(\S_{y,m+1}^{(k)}) - \lambda_j(\R_{y,m+1}^{(k)})| = O_P\bigg(\gmkcheck^{-1}\bigg\{ \frac{r_eS_\psi^{(k)}}{\sqrt{T}} + \frac{K\sqrt{rr_eS_\psi^{(k)}d}}{T} + \max_{j\in[\dmk]}\norm{\bSigma_{\epsilon,j}^{(k)}}\frac{K^2rd}{T^2}\bigg\}\bigg),
\end{align*}
which is the third statement of the lemma. Hence it remains to show (\ref{eqn:diff_eq}). To this end,
define the kernel of $\A_k^\T\diag^{-1/2}(\bSigma_{y,m+1}^{(k)})$ to be
\begin{align*}
  \cN := \{ \w:  \A_k^\T\diag^{-1/2}(\bSigma_{y,m+1}^{(k)})\w=\0\} \equiv \{\w:\A_k^\T\w=\0\}
\end{align*}
in probability by (\ref{eqn:Sigma_y_diagonal_dominance}) and assumption (RE2), and define $\cN^\perp$ the corresponding row space, which has $\cN^{\perp} = \text{span}(\A_k)$ in probability by the above argument. Hence by Assumption (L1) that $\A_k$ is of full rank, dimension of $\cN^\perp$ is exactly $r_k$ (in probability), and the rank-nullity theorem implies that, in probability, the dimension of $\cN$ is exactly $d_k-r_k$. Consider, by (\ref{eqn:lambda_rZ_f}) and Assumption (L1), for some constant $c>0$, we have in probability that
\begin{align*}
  \lambda_{r_k}(\S_{11,r}) &\geq \gmkcheck\min_{\w\in\cN^\perp}\norm{\w^\T\diag^{-1/2}(\bSigma_{y,m+1}^{(k)})\A_k}^2
  \lambda_r(T^{-1}\Z_f^{(k)}\cA_{f,T}\M_T\cA_{f,T}^\T\Z_f^{(k)\T}) \geq cd_k^{\alpha_{k,r_k}}.
\end{align*}
At the same time, we have
\begin{align*}
  \norm{\Qm_2^{(k)}\Qm_2^{(k)\T} + \Qm_3^{(k)}\Qm_3^{(k)\T}} &= O_P(\gmkcheck^{-1}(\max_{j\in[\dmk]}\norm{\bSigma_{\epsilon,j}^{(k)}} + S_\psi^{(k)})),
\end{align*}
and earlier calculations shown that
\begin{align*}
  \norm{\S_{12,r}} &= O_P\bigg(\gmkcheck^{-1/2}d_k^{\alpha_{k,1}/2}\bigg(\sqrt{\frac{rd_k}{T}} + \frac{Kr\sqrt{d}}{T}\bigg)\bigg),\;\;\;
  \norm{\S_{13,r}} = O_P\bigg( \gmkcheck^{-1/2}d_k^{\alpha_{k,1}/2}\sqrt{\frac{rr_eS_\psi^{(k)}}{T}} \bigg).
\end{align*}
Hence in probability, the eigenvalues of $\S_{11,r}$ are dominating those in
\[\G := \sum_{j=2}^3(\S_{1j,r} + \S_{j1,r} + \Qm_j^{(k)}\Qm_j^{(k)\T}).\]
Hence the $r_k$ eigenvectors corresponding to the largest $r_k$ eigenvalues of $\G + \S_{11,r}^{(k)}$ coincide with those of $\S_{11,r}$'s for large enough $d_k$, which are all necessarily in $\cN^\perp$. This means that the $(r_k+1)$-th largest eigenvalue of $\G+\S_{11,r}$ and beyond will have eigenvectors in $\cN$ since the dimension of $\cN^{\perp}$ is $r_k$. Take $\w \in \cN$, then it is easy to see that
\[ \w^\T(\G + \S_{11,r})\w =  \w^\T(\Qm_2^{(k)}\Qm_2^{(k)\T} + \Qm_3^{(k)}\Qm_3^{(k)\T})\w = \w^\T\R_{y,m+1}^{(k)}\w,\]
showing that $\lambda_j(\G+\S_{11,r}) = \lambda_j(\R_{y,m+1}^{(k)})$ for $j\in[d_k]/[r_k]$, which is (\ref{eqn:diff_eq}). This completes the proof of the lemma. $\square$

{\em Proof of Theorem \ref{thm:rkhat_consistency}}. For $j\in[r_k]$ and $k\in[K]$, we can decompose
\begin{align*}
  \lambda_j(\wh\R_{y,m+1}^{(k)}) &\leq \lambda_j(\R_{y,m+1}^{(k)})\bigg\{1 + \bigg|\frac{\lambda_j(\wh\R_{y,m+1}^{(k)})}{\lambda_j(\S_{y,m+1}^{(k)})}-1\bigg|
  +\bigg|\frac{\lambda_j(\S_{y,m+1}^{(k)})}{\lambda_j(\R_{y,m+1}^{(k)})}-1\bigg|
  \\ &+ \bigg|\frac{\lambda_j(\wh\R_{y,m+1}^{(k)})}{\lambda_j(\S_{y,m+1}^{(k)})}-1\bigg|
  \cdot\bigg|\frac{\lambda_j(\S_{y,m+1}^{(k)})}{\lambda_j(\R_{y,m+1}^{(k)})}-1\bigg|\bigg\}\\
  &= \lambda_j(\R_{y,m+1}^{(k)}) (1 + O_P(r_kd_k^{\alpha_{k,1}-\alpha_{k,j}-1}a_T(\delta_{k,1}) + a_T(0)))\\
  &\succeq_P r_kd_k^{1-\alpha_{k,1} + \alpha_{k,j}}(1 + O_P(r_kd_k^{\alpha_{k,1}-\alpha_{k,j}-1}a_T(\delta_{k,1}) + a_T(0))),
\end{align*}
where the second last line used the results in Lemma \ref{lem:6}, and the last line used the result from Theorem \ref{thm:population_eigenvalue_correlation_matrix}. Similarly, for $j\in[d_k]/[r_k]$,
\begin{align*}
  \lambda_j(\wh\R_{y,m+1}^{(k)}) &\leq \lambda_j(\R_{y,m+1}^{(k)})\bigg( 1 +
  \bigg|\frac{\lambda_j(\wh\R_{y,m+1}^{(k)})}{\lambda_j(\S_{y,m+1}^{(k)})}-1\bigg| \bigg)
  + |\lambda_j(\S_{y,m+1}^{(k)}) - \lambda_j(\R_{y,m+1}^{(k)})|\\
  &+
  |\lambda_j(\S_{y,m+1}^{(k)}) - \lambda_j(\R_{y,m+1}^{(k)})|\cdot \bigg|\frac{\lambda_j(\wh\R_{y,m+1}^{(k)})}{\lambda_j(\S_{y,m+1}^{(k)})}-1\bigg|\\
  &= \lambda_j(\R_{y,m+1}^{(k)}) + O_P(a_T(0)) \leq 1 + O_P(a_T(0)),
\end{align*}
where the second last line used Assumption (RE2) to conclude that
\[\max_{j\in[d_k]/[r_k]}|\lambda_j(\S_{y,m+1}^{(k)}) - \lambda_j(\R_{y,m+1}^{(k)})| = O_P(a_T(0)),\]
and the last inequality used a result in Theorem \ref{thm:population_eigenvalue_correlation_matrix}. This completes the proof of the theorem. $\square$

\bibliographystyle{apalike}
\bibliography{tensorrank}
	
\end{document}